\documentclass[a4paper,12pt,openright,twoside]{scrbook}

\usepackage{amstext,amsmath,amssymb,amsfonts,bbm}
\usepackage{amsthm}
\usepackage[utf8]{inputenc}
\usepackage[T1]{fontenc}
\usepackage[pdftex]{graphicx}
\usepackage[svgnames]{xcolor}
\usepackage{bm}

\usepackage{bookmark,hyperref}

\usepackage[a4paper,heightrounded,bindingoffset=0.5cm]{geometry}
\usepackage{caption}

\def\be{\begin{equation}}
\def\ee{\end{equation}}

\newcommand{\bra}[1]{\ensuremath{\langle#1|}}
\newcommand{\ket}[1]{\ensuremath{|#1\rangle}}
\newcommand{\bk}[2]{{\langle#1\,|\,#2\rangle}}
\newcommand{\bek}[3]{{\langle#1\,|\,#2\,|\,#3\rangle}}
\newcommand{\id}{1\!\! 1}
\renewcommand{\sl}{SL(2,\mathbb{C})}
\newcommand*{\mathgraphics}[2][]{\vcenter{\hbox{\includegraphics[#1]{#2}}}}
\def\tl{\widetilde}
\def\pp{\partial}
 
 \def\w{\wedge}  \def\i{\iota}
 
\def\f{\frac}
\def\dag{^\dagger}

\def\Tr{\mathrm{Tr}}
\def\d{\mathrm{d}}

\theoremstyle{definition}
\theoremstyle{definition}
\theoremstyle{definition}\newtheorem*{rem}{Remark}
\newenvironment{calc}{\vspace{11pt}\noindent\rule{\textwidth}{.1ex}\small\newline}{\newline\noindent\rule{\textwidth}{.1ex}\vspace{11pt}}

\begin{document}
\frontmatter
\begin{titlepage}
\begin{center}
{\Large Alma Mater Studiorum $\cdot$ Universit\`a di
Bologna}
 
{\rule[8pt]{.8\textwidth}{1pt}}

{\Large\sffamily Dottorato di ricerca in Fisica Teorica\\Ph.D. in Theoretical Physics}
\end{center}
\vspace{15mm}

\begin{center}
\Large 
 ciclo XXIII\\
settore scientifico disciplinare FIS/02
\end{center}

\vspace{1cm}

\begin{center}\rule[10pt]{\textwidth}{1pt}
{\Huge\sffamily The relation between Geometry and Matter in Classical and Quantum Gravity and Cosmology}
\rule{\textwidth}{1pt} 
\end{center}
\vspace{.8cm}
\begin{center}
 \Large Daniele Regoli
\end{center}


\vfill
\begin{flushleft}
{\large relatore: Alexandre Kamenchtchik\\[1ex]
coordinatore: Fabio Ortolani}
\end{flushleft}\vspace{.6ex}
\centering{\rule{\textwidth}{.1mm}}
\begin{center}
{\large esame finale 2011}
\end{center}
\end{titlepage}
\clearpage{\pagestyle{plain}\cleardoublepage}
\thispagestyle{empty}

\vspace*{0.4\textheight}

\hfill
\begin{flushright}
{\large\itshape
Quaerendo invenietis}\\
\large J.~S.~Bach\end{flushright}
\vfill

\clearpage{\pagestyle{plain}\cleardoublepage}

\currentpdfbookmark{Contents}{Contents}
\tableofcontents

\mainmatter
\chapter*{Introduction\markboth{Introduction}{Introduction}}
\addcontentsline{toc}{chapter}{Introduction}

Cosmology and Quantum Gravity are the two main 
areas of Physics  the research collected in this dissertation is about.

Cosmology is the study of the universe as a dynamical system. It is a rather peculiar chapter of Physics for more than one reason. The main one, I believe, is that we have but one universe (someone may not agree on this) and there is no way to do experiments on the universe, in the proper sense of the word. We can at best \emph{observe} it, and that's what astronomers do very well. Another reason, at least for myself, is on purely intellectual basis: Cosmology is one of the  areas (if not \emph{the} area) of Science that most touches in deep the questions Science was born for. One more argument, which is somehow mixed with the preceding one, is that Cosmology is in a certain sense a rather young science: Einstein's General Relativity (1916) is the first theoretical framework that has allowed a scientific and mathematical approach to the problem of the study of the universe as a single system. Before that, questions like `where does the universe come from?' and `what is the fate of the universe?' where mere philosophy, with no scientific attempt to be answered\footnote{Maybe it is superfluous to remark that they are not answered even nowadays. The point are not the answers, which will probably never come, but the scientific and rigorous attempt to give them.}.  

Since its birth, Cosmology has taken gigantic steps, both on the phenomenological and theoretical viewpoints. The Friedmann models (see section~\ref{sec:cosmo}) are the first theoretical models comprehending the dynamical \emph{expansion} of the universe, together with the Big Bang initial singularity, both compatible with observations and accepted from the scientific community\footnote{The Big Bang, being a singularity of the theory, cannot be \emph{observed}, strictly speaking. What is accepted, is the existence, some 15 billion years ago, of a phase of the universe in which everything (spacetime itself) was contracted in a tiny and extremely hot bubble. But on the nature of that bubble, on its real size and so on, the debate is very far from being closed.}. 

I dare say that expansion and initial moment of the universe have somehow given birth  (for what it is possible) to two macro lines of research.
Observations of the Cosmic Microwave Background have indeed fostered the research on the nature of the universe in its very beginning. We talk then about ``early-time Cosmology'', referring to the study of primordial perturbations amplified by some (quantum) mechanism (e.g. inflation), giving birth to the large-scale structures we observe nowadays, such as galaxies, clusters and so on. 
On the other hand we have the discovery of the cosmic acceleration~\cite{Riess,Perlmutter}, that has guided many theoreticians in the study of cosmological models able to reproduce this (and others to it related) experimental evidence.

The first part of this collection is focused on the second of this problem, the one regarding the entire evolution of the universe, its expansion, its acceleration and so on; while I will not touch the cosmological perturbations or amplification mechanisms such as inflation.\\[14pt]

Quantum Gravity is not a self-consistent theory, yet. It is a work in progress attempt to find a coherent set of tools in order to formulate some kind of predictions at physical scales where both Quantum Mechanics and General Relativity should hold. And we know that taken as they are, Quantum Mechanics and General Relativity \emph{cannot} be both valid. Thus, we cannot properly  speak of a theory, since a theory is something self-contained (at least at some degree) that has the possibility of making scientific predictions. 

Loop Quantum Gravity is one of these attempts, so one usually says that it is a \emph{candidate} for Quantum Gravity. The keystone of Loop Quantum Gravity, namely the feature that identifies it among other candidates, is to quantize (canonically) the metric tensor (or, better, an appropriate manipulation of the metric tensor) \emph{without assuming the existence of a somehow fixed classical background}, around which fluctuates the ``quantum part'' of the metric. This is known  as \emph{background independence}. 

Since its birth in 1986, LQG has reached amazing and important results, the main of which -- I dare say -- is the suggestive prediction of the `granular' nature of space: LQG indeed predicts the existence of a minimum length in space. 
However, much is still to be done; and the main drawback, which is actually common to all the candidates Quantum Gravity theories, is the lack (more or less total) of \emph{testable} predictions.

A decade ago a new ``spin-off'' theory was born in the framework of LQG: spinfoam theory. It is an attempt to define quantum gravitational amplitudes between space-geometries (given by LQG) in terms of a sum-over-histories. The histories are called `spinfoams', a kind of bubble-like representation of spacetime at fundamental level. Many problems are still open, and it is a very stimulating field of research, particularly for young researchers, since there are entire lines still to be explored and thousands of bridges with other theoretical frameworks still to be built.\\[14pt]

The main results of the research here presented are the following:
\begin{itemize}
 \item we analyzed in depth two-field cosmological models, with one scalar and one phantom scalar field. The procedure of reconstruction of two-field potentials reproducing a dynamical evolution (specifically an evolution with a crossing of the phantom divide line) has been worked out: we have discovered that two-field models have a huge freedom with respect to single-field models, namely there is an infinite number of potentials which give, with a specific choice of initial conditions, the wanted dynamics for the model. Moreover, a thorough analysis of the phase space of two specific two-field models is carried out, showing that inside a single model, varying the initial conditions on the fields, qualitatively different families of evolutions are present, with different cosmological singularities as well.
\item
In line with the preceding result, we have coupled our two-field models with cosmic magnetic fields, which are experimentally observed quantities. We have shown the sensitivity of the amplification of the magnetic fields to the change of underlying cosmological model, but keeping the \emph{same background evolution of the universe}. This, in principle, is a way to experimentally discriminate between cosmological models with different potentials.
\item
The greatest problem of phantom scalar fields is quantum instability. We have shown that in the framework of $PT$~symmetric Quantum Theory (somehow adapted to Cosmology) this instability can be cured: we have an \emph{effective} phantom scalar field, with stable quantum fluctuations.
\item
We analyzed one-field cosmological models comprehending a peculiar version of the Big Bang and Big Crunch singularity, namely singularity with finite and \emph{non-zero} radius. We performed a detailed analysis of the phase space and reveal the presence of different classes of evolutions of the universe.
\item
In the framework of spinfoam theory for Quantum Gravity, we proposed a fixing of the face amplitude of the spinfoam sum, somewhat neglected until now, motivated essentially by a form of ``unitarity'' of gravitational evolution. 

\end{itemize}

I list here the publications in which the above said research has been collected (chronologically from the latest) :
\begin{itemize}
\item E.~Bianchi, D.~Regoli and C.~Rovelli,
  \emph{``Face amplitude of spinfoam quantum gravity''}
  Classical and Quantum Gravity {\bfseries CQG}  27, 185009 (2010)\\
  (arXiv:1005.0764 [gr-qc]).

\item A.~A.~Andrianov, F.~Cannata, A.~Y.~Kamenshchik and D.~Regoli,
  \emph{``Phantom Cosmology based on $PT$-symmetry''}, International  Journal of Modern Physics  D {\bfseries IJMPD}  19, 97 (2010).

  A.~A.~Andrianov, F.~Cannata, A.~Y.~Kamenshchik and D.~Regoli,
  \emph{``Cosmology of non-Hermitian $(C)PT$-invariant scalar matter''},
  {\bfseries J. Phys. Conf. Ser.}  171, 012043 (2009).

\item F.~Cannata, A.~Y.~Kamenshchik and D.~Regoli,
\emph{``Scalar field cosmological models with finite scale factor singularities''}, {\bfseries Physics Letters B} 670 (2009) 241-245
  (arXiv:0801.2348v1 [gr-qc]).

\item A.~A.~Andrianov, F.~Cannata, A.~Y.~Kamenshchik and D.~Regoli,
  \emph{``Two-field cosmological models and large-scale cosmic magnetic fields''}, Journal of Cosmology and  Astroparticle Physics ({\bfseries JCAP}) 10 (2008) 019 \\
  (arXiv:0806.1844v1 [hep-th]).

\item A.~A.~Andrianov, F.~Cannata, A.~Y.~Kamenshchik and D.~Regoli,
  \emph{``Reconstruction of scalar potentials in two-field cosmological models''}, Journal of Cosmology and  Astroparticle Physics ({\bfseries JCAP}) 02 (2008) 015\\ 
  (arXiv:0711.4300 [gr-qc]).
\end{itemize}

This work  is thus intended to pursue two targets:
\begin{itemize}
 \item to provide, for the fields of interest, a sufficient framework of notions and results that are the current background on which research is performed and are necessary to understand (or closely related to) the subject of the following goal
\item to resume and explain the results of the research I have been working on during this three-year PhD course, under the supervision of Dr.~Alexander~Kamenshchik.
\end{itemize}

\emph{Gravity} is the main character of the play. \emph{Matter} is the other one. `Matter' of course in a cosmological sense, i.e. everything that is embedded in spacetime. Talking in field theory jargon, we have nothing but fields: the gravitational field (the metric) and all the others: the matter fields (including those whose (quantum) excitations are the well-known elementary particles, and possibly others, which should model/explain such things as the dark matter, dark energy and,  generically, other types of exotic matter/energy). All living on the spacetime manifold.

We know that if we stay away from the quantum regime, General Relativity is the stage on which our characters play\footnote{There are of course many proposals to modify somehow GR, e.g. the Ho\v{r}ava-Lifshitz proposal~\cite{modifiedgrav1}, the $f(R)$ models~\cite{modifiedgrav2,modifiedgrav3}, and others, but I will not consider them here at all.}.  GR, in the intuitive rephrasing by Wheeler~\cite{MTW}, is simply encoded in the following aphorism-like sentence: matter teaches spacetime how to curve, spacetime teaches matter how to move.
The ``how'', is encoded in Einstein's equation and in general in the Einstein-Hilbert action for GR.

Classical Cosmology is precisely the study of this interplay between spacetime geometry and matter, on the scale of the entire universe. The idea is to build models with various kinds of matter, possibly encoded in terms of fields, that are able to reproduce the observed data on the dynamical evolution of the universe.
More specifically for what concerns this thesis, the key empirical observation is the cosmic acceleration \cite{Riess,Perlmutter}, and the possibility of a kind of ``super-acceleration'', that has to be somehow explained theoretically, which we tried to, for our (little) part.

When one enters the quantum regime, attention must be paid. We know that there is a typical scale, the \emph{Planck scale}, under which we expect both GR and QM to be valid. However there is no accepted framework for this scale. No complete Quantum Gravity theory exists, yet. Actually, it is the Holy Grail of contemporary theoretical physics.

The second (shorter) part of the present thesis is entirely focused on one specific candidate of Quantum Gravity: loop quantum gravity. I will have a more appropriate occasion to discuss this thoroughly~\ref{sec:lqg_intro}, but for the moment let me stress that I believe it fundamental to study what happens to spacetime under the Planck scale in order to properly catch the nature of the relation between geometry and matter. I think Quantum Gravity is going to revolutionize our concept of the interplay between gravity and matter as much as Einstein's theory has drastically changed the idea of matter fields top of a passive, fixed background. My personal contribution in this respect, is in the framework of spinfoam models for quantum gravity: a kind of path-integral formulation of quantum gravity, based and founded on the results of canonical Loop Quantum Gravity.

What follows is divided in:
\begin{itemize}
 \item Part \ref{part:1}, dedicated to classical Cosmology: it starts with an introductory chapter intended to provide the necessary concepts of classical Cosmology; this is followed by four `research'-chapters, each dedicated to one of the four publications about Cosmology listed before (or, equivalently, see \cite{mio_2field},\cite{mio_PT},\cite{mio_magnetic}, \cite{mio_soft}). The second of these four chapters, namely the one dedicated to ``$PT$-phantom Cosmology'', includes also a short review of $PT$-symmetric Quantum Mechanics, whose ideas  play a key role in the research explained in (what remains of) that chapter.     
\item Part \ref{part:2}, focused on loop quantum gravity and spinfoam models: here I dedicate more time to properly introduce the basics of these theories. Namely, four entire chapters are devoted to a review of fundamental concepts of this theoretical framework. Indeed loop quantum gravity stands on much more advanced and sophisticated basis than classical Cosmology, and, since the work I have done is about a rather technical aspect of spinfoam models, I thought it absolutely necessary to properly review all the underlying framework, trying to be as self contained as possible. 

The last chapter of this second part resumes the results and explains the content of the paper \cite{mio_faccia}, about a proposal for fixing the face amplitudes of spinfoam models.
\end{itemize}

A brief conclusion~\ref{sec:concl}  is intended to give a global view \emph{ex post} on all the issues discussed.
\part{Cosmology and Dark Energy models}\label{part:1}
\chapter{Cosmology: the basics}\label{sec:cosmo}

Cosmology is the study of the universe as a dynamical system. At a classical level the tools to handle this kind of study are given by the Einstein's General Relativity theory, \emph{in nuce} by equations
\be
R_{\mu\nu}-\f12g_{\mu\nu}R=\f{8\pi G}{c^4}T_{\mu\nu}\ ,
\label{EE}
\ee
where $R_{\mu\nu}$ is the Ricci tensor, $g_{\mu\nu}$ is the metric tensor, $R=g^{\mu\nu}R_{\mu\nu}$ is the scalar curvature and $T_{\mu\nu}$ is the stress-energy tensor.
As for the constants, $G$ is Newton's gravitational constant and $c$ is the speed of light\footnote{The reader can find detailed informations in every good textbook on General Relativity. Let me just say that for general GR framework I really like the book by Hawking and Ellis \cite{HawkingEllis} while for classical Cosmology (and for much more) I find very useful the book by Landau and Lifshitz~\cite{LL}.}. 

Of course, to (hope to) resolve these equations -- in order to catch the dynamics of the universe, encoded in the solution $g_{\mu\nu}(x)$ -- one has to know the source term $T_{\mu\nu}$ point by point in spacetime. This can be achieved only by drastic exemplifications on the nature of the source and of the spacetime. Namely one assumes a rather strict -- but also very reasonable at this large scale -- symmetry: roughly speaking the assumption is homogeneity and isotropy of space. More precisely one assumes that the spacetime 4d manifold admits a foliation is 3d spaces that are homogeneous and isotropic, i.e. one admits the existence of a family of observers that see uniform space-like surfaces. This can be rephrased in a more suggestive and simple way by saying that we exclude the existence of special regions in space. This is known as \emph{cosmological principle} and, as we will see in a moment, it drastically exemplifies equations \eqref{EE}. Of course this principle is just a device through which one tempts to extract physics from a general model which would be by far too complicated. One can relax in many ways this principle, for example in studying cosmological perturbations or by admitting anisotropies in some ways~\cite{Mukhanov-book}. However, for our purposes, it won't be necessary to abandon this principle.

The assumption of such a principle results in the following form of the line element in spacetime (the Friedmann-Lem\`aitre-Robertson-Walker line element, or simply FLRW):
\be
\d s^2=dt^2-a^2(t)\d l^2\ ,
\label{FRWline}
\ee
where 
\be
\d l^2 =\f{\d r^2}{1-k r^2}-r^2(\d \theta^2-\sin^2\theta\d\phi^2)\ .
\ee
Recall that $k$ is the curvature parameter and equals 1, 0 or -1 for a closed, open-flat and open-hyperbolic universe respectively. In all what follows we shall take $k=0$, i.e. we assume a (spatially-)flat universe. This is in accordance with all the observational data\footnote{Actually it is true that $k=0$ is coherent with all the data, but some argue this is only a proof that the space is locally flat, and still it may have a positive or negative global curvature. I am not going to discuss this (important) issue here.}.

In equation \eqref{FRWline} the only dynamical variable is the \emph{scale factor} $a(t)$, the only dynamical degree of freedom in classical (uniform) Cosmology. As is clear from \eqref{FRWline}, it represents the `size' of the universe at a certain moment, and thus it gives information on the expansion or contraction of the universe itself.

Recall that for a non-dissipative fluid one has $T_{\mu\nu}=(p+\varepsilon)u_\mu u_\nu-pg_{\mu\nu}$ with $p$ and $\varepsilon$ the pressure and the energy density of the fluid, respectively, and $u_\mu=\d x_\mu(s)/\d s$ is the 4-velocity of the fluid. With this in mind Einstein's equations become much more simple than \eqref{EE}, namely
\begin{align}
&3\left(\f{\dot{a}}{a}\right)^2+3\f{k}{a^2}=8\pi G\varepsilon\ ,\\\label{fri1}
&\f{2a\ddot{a}+a^2+k}{a^2}=-8\pi Gp\ .
\end{align}
These are the well-known Friedmann equations. Actually one usually replaces the second of these equations with the simpler requirement of conservation of energy, i.e.
\be
\dot\varepsilon=3\ \f{\dot a}{a}\ (p+\varepsilon)\ ,
\label{fri2}
\ee
and thus uses equations \eqref{fri1} and \eqref{fri2} as the set of two independent equations that governs the dynamics of the universe\footnote{Intuitively, one for determining the source (say $\varepsilon$) in terms of the scale factor, and then the other to determine the scale factor itself.}.

\begin{figure}\centering
 \includegraphics{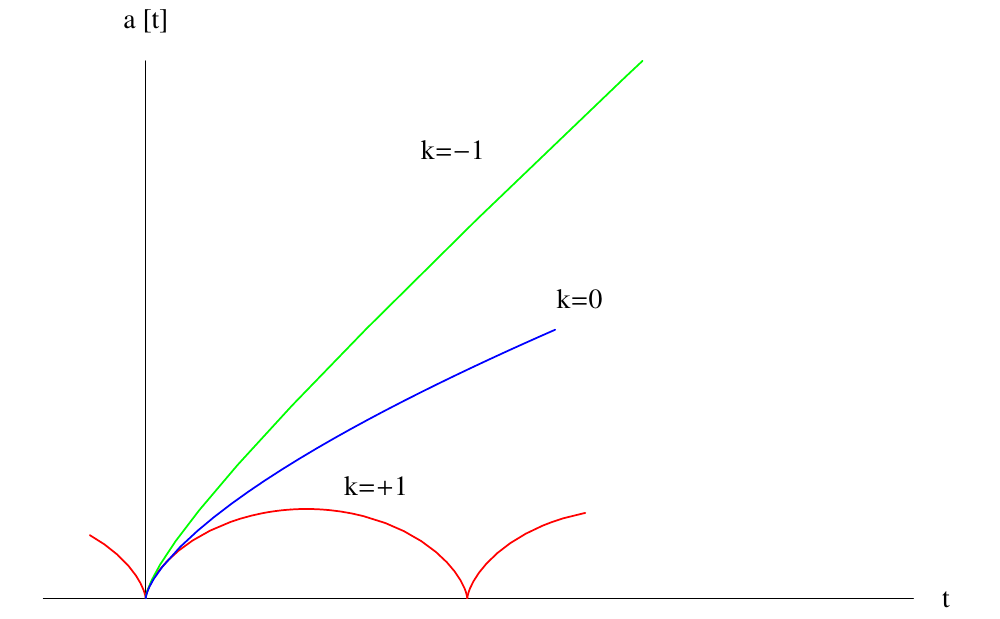}
\caption{}\label{fig:fri}
\end{figure}

Solving the system \eqref{fri1}, \eqref{fri2} for a dust-like fluid (a fluid with zero pressure, which is a good approximation for visible matter like galaxies, clusters of galaxies and so on) one obtains the well-known Friedmann models, summarized in figure \ref{fig:fri}.

\begin{rem}
These results, I want to stress it, are striking. Doing research in this field one is often `obliged' to investigate technical (and often exotic) details somewhat losing the importance of the Friedmann models: with a couple of amazingly simple equations one is able to obtain a qualitative (past and future) evolution of the universe. Of course these results must be corrected in order to fit observations introducing different kind of matter/energy, must be generalized to embrace anisotropies or inhomogeneities and must be supported by some quantum gravity theory to investigate what physics is inside the initial (and possibly final) singularity. But much of  classical Cosmology is already inside these simple models. 
\end{rem}

In what follows we shall make a huge usage of the \emph{Hubble parameter}
\be
h(t)=\f{\dot a(t)}{a(t)}\ ,
\label{Hubble}
\ee
in terms of which the Friedmann equations \eqref{fri1}, \eqref{fri2} become\footnote{I use here, and everywhere in what follows about Cosmology, the convention $8\pi G/3=1$.} 
\begin{align*}
&h^2=\varepsilon -\frac{k}{a^2}\ ,\\
&\dot\varepsilon= -3h(\varepsilon+p)\ ,
\end{align*}
and, for the flat case,
\begin{subequations}
\begin{align}\label{fri3}
&h^2=\varepsilon\ ,\\ 
&\dot\varepsilon= -3h(\varepsilon+p)\ .
\end{align}
\end{subequations}
and thus can be written in the compact form
\be
\dot h=-\f32(\varepsilon+p)\ .
\label{fri}
\ee

For sake of completeness, let me recall the form of the curvature in terms of the Hubble parameter:
\be
R=-6\left(\dot h+h^2+\f{k}{a^2}\right)\ .
\ee
which is important to understand how $h$ and $a$ determines curvature singularities.

\section{Accelerated expansion}

Up till here we have talked about a universe filled with galaxies, and we have seen its large scale evolution, which is an expansion starting from the Big Bang singularity and ending eventually either in a future singularity after a contraction (the closed case) or in an endless cold expansion.

However it is now well-known that this is not the case. The cosmic expansion is accelerated \cite{Riess,Perlmutter}. This is the empirical observation that acted like a spark for the explosion of a large amount of cosmological models trying to incorporate this effect \cite{dark1}-\cite{dark6}, \cite{quintessence1,quintessence2}, \cite{darkmodel1}-\cite{darkmodel11} in the decade after it was first discovered (1998). The first part of this thesis lays in this line.

Now I shall introduce the general ideas and the framework for these models as well as the technical jargon necessary to understand them.

Take a generic cosmic fluid, and write its equation of state as follows
\begin{equation}\label{state_equation}
p=w\varepsilon\ ,
\end{equation}
The $w$ parameter is crucial to discriminate between different kind of expansions, as well as of fluids, of course.
Indeed, Friedmann equations tell us that in order to have acceleration ($\ddot a>0$) one has to have
\begin{equation}
w<-\frac{1}{3}\ .
\end{equation}
Such a fluid is generically known as \emph{Dark Energy} (DE)~\cite{dark1}-\cite{dark5}.
The  $w=-1$ (constant) case is the $\Lambda$CDM model of Cosmology\footnote{Obviously we are here talking of the DE side of the story. In every decent cosmological model there must be also a baryonic sector as well as a dark matter sector, which, when talking specifically about DE, are often understood.}, and is actually the best candidate model to fit observational data, as well as the simplest DE model.
As is well-known, at a sufficiently large time scale (when the baryonic and dark matter contribution becomes negligible) the evolution of $\Lambda$CDM is de~Sitter-like.

There are some motivations to investigate DE models \emph{different} from the simple $\Lambda$CDM. Some more practical and some more ``philosophical''. Let me spend some words on this issue. The not interested reader of course can skip the entire following subsection.

\subsection{\texorpdfstring{$\Lambda$CDM}{LambdaCDM}: ups and downs}

Let me remark one thing: $\Lambda$CDM  is in accordance with all observational data. This, as contemporary theoretical physics teaches us, does not prevent theoreticians to investigate alternatives, if there are some motivations supporting this research. 

A practical (and I think more common-sense) motivation is the following: it is true that observations do not exclude $\Lambda$CDM, but there are other possibilities not excluded as well. Moreover, some observations indicate a best fit for $w$ which is \emph{less then} -1~\footnote{See~\cite{phant-obs1}-\cite{phant-obs7} and \cite{obs-new1}-\cite{obs-new3}.} (we shall see that this is not a painless difference), and contemplate the possibility of ``crossing'' this benchmark $w=-1$, thus with a non-constant value of $w$. In my humble opinion I think this is enough for  researchers to explore alternatives, even if exotic; obviously keeping clear in mind that \emph{there is} a good candidate model in accordance with data.

Theoretical, or ``philosophical'', motivations have been put out as well (I draw arguments mainly by \cite{bianchirovellilambda}). They are generically of two types. The first is the `coincidence problem': observations tell us that the dark energy density is about 2.5 times bigger than baryonic + dark matter(dama) density (about 74\% for DE and 26\% for the rest). Since the baryonic and dark matter contributions dilute in time (as $a^{-1/3}$) while the cosmological constant does not, it will eventually dominate the evolution of the universe and its density should be much greater than the observed + dama sector, starting from some cosmic era on. The argument is that it is not likely that we are just in that era in which the DE density is comparable to the observed+dama one, it would be too much of a coincidence. Think of it as a kind of cosmological principle in time. Thus, the argument claims, $\Lambda$CDM must be wrong. This is a rather poor argument, in my point of view. First, because probability arguments are to be taken very carefully, always. Second, because it is quite clear that in a universe dominated by the cosmological constant, matter and thus us, wouldn't exist. This is a kind of anthropic (counter)argument, but -- I think -- of a very light and reasonable kind.

The second type is based on QFT reasonings: the cosmological constant (i.e. the DE in $\Lambda$CDM) is a kind of vacuum energy; QFT `predicts' the existence of a energy of the vacuum, but if we compare the QFT calculation with the (tiny) observed value for $\Lambda$ they differ by 120 orders of magnitude. So $\Lambda$CDM must be wrong. 
For this argument there are counter-arguments as well: it is true that QFT predicts a vacuum energy, but it is actually a huge amount of energy, and it is not at all observed. If QFT vacuum energy was really there, any region of space with a quantum field would have a huge mass, and it would certainly be observed. The Casimir effect reveals only the effect of a difference (or, better, a change) in vacuum energy. Thus it is likely that there would be some unknown mechanism in QFT that prevents the vacuum energy to gravitate, or protect it from huge radiative corrections, rather than $\Lambda$CDM is `wrong'.
  
Thus, I do think it is worthwhile exploring alternatives of $\Lambda$CDM, but mainly for practical and observational reasons. And still I consider the $\Lambda$CDM model as a good standpoint for classical Cosmology.

\subsection{The case \texorpdfstring{$-1<w<-1/3$}{-1<w<1/3}}

Let us briefly sketch the behavior of this class, i.e. the class of models with state parameter $w$ of DE type but greater than -1.
The calculations starting from the evolution equation \eqref{fri} are straightforward
\be
h(t)=\frac{2}{3(w+1)(t-t_0)}\ .
\ee
with a Big Bang initial singularity, in the sense that
\be
a(t)\xrightarrow[t\rightarrow t_0^+]{}0\ ,
\ee
indeed
\be
a(t)=(t-t_0)^{2/3(w+1)},\quad\dot a=\frac{2}{3(w+1)}(t-t_0)^{-\frac{3w+1}{w+1}}.
\ee
But differently from the Friedmann models -- where $\dot a$ is increasingly high going towards the initial instant -- here $\dot a$ goes to zero as well.

One could as well tempt to analyze cases where $w$ varies with time. One remarkable example is the so-called Chaplygin gas~\cite{darkmodel5}, characterized by the following state equation
\be
p=-\frac{A}{\varepsilon},
\ee
with $A$ a positive constant. This fluid gives a sort of interpolation between a dust-like era and a de~Sitter era.

\subsection{The case \texorpdfstring{$w<-1$}{w<-1}: the Big Rip type singularity}\label{sec:rip}

I usually call the evolution of the models with $w$ constant and less than -1 ``super-acceleration''. Indeed they are characterized by
\be
\dot h>0\ ,
\ee
that is 
\be
\frac{\ddot a}{a}>\frac{\dot a^2}{a^2}\ .
\ee
The dynamical evolution reads
\be
 a(t)=(t-t_0)^{-2/3|w+1|}\ ,\quad \dot a=\frac{2}{3|w+1|}(t-t_0)^{-\frac{3w+1}{w+1}}\ .
\ee
Notice that now the exponents are both negative, thus
\be
a\xrightarrow[t\rightarrow t_0^+]{}\infty\ ,\quad\dot a\xrightarrow[t\rightarrow t_0^+]{}-\infty\ .
\ee
These features -- scale factor and its velocity both infinite at a finite instant -- define the \emph{Big Rip} singularity, which is typical of the super-accelerated models~\cite{Rip1,Rip2}. Sometimes, the fluid responsible of such an exotic evolution is dubbed \emph{phantom energy}~\cite{phantom1}-\cite{phantom10}.

I hope it is now clear to the reader that the benchmark $w=-1$ (somehow represented by the $\Lambda$CDM model) discriminates between dynamical evolutions qualitatively different: cosmic acceleration with a Big Bang and a cold infinity expansion versus a ``super-acceleration'' with a different cosmological singularity.
The line $w=-1$ has deserved a name, the \emph{phantom divide line}.

\begin{rem}[\emph{the phantom stability problem}]
Let me anticipate here the big drawback of the phantom energy. A kind of energy with state parameter $w<-1$ violates all the energy conditions prescribed by General Relativity (see e.g. \cite{HawkingEllis}). Indeed, when a phantom energy is modeled by a scalar field (see section \ref{sec:21}) it requires a negative kinetic term. Its hamiltonian becomes, obviously, unbounded and such a field is plagued by manifest quantum instability: its quantum fluctuations grow exponentially with time~\cite{phantomstability1,phantomstability2}. This is actually the main critic done to the use of such things as phantom fields. And, I have to admit, it is a quite strong and reasonable critic. However, ghost fields like these, were studied long before the phantom field was first introduce; and in chapter \ref{sec:PT} (or equivalently in \cite{mio_PT}) I will describe a mechanism that could fix this plague, paying the price of enlarging the framework of Quantum Theory to its $PT$~symmetric version. 
\end{rem}
\chapter{Two-field cosmological models}\label{sec:2field}

\emph{This chapter is devoted to reviewing and explaining the content of \cite{mio_2field}. The idea is to work with  two scalar fields, one of which phantom, and try with this model to reproduce an evolution from a Big Bang to a Big Rip singularity. The procedure of reconstruction of the model starting by a specif evolution is carried out in detail. Then, the dynamical phase space of two of such models is studied, analyzing the dynamical classes included in it by varying the initial conditions on the fields. }

\section{Introduction}\label{sec:21}
As we have seen in the previous section, the discovery of cosmic acceleration \cite{Riess,Perlmutter} has
stimulated the construction of a class of dark energy models
\cite{dark1}-\cite{dark6}, \cite{quintessence1,quintessence2}, \cite{darkmodel1}-\cite{darkmodel11} describing this effect. In what follows we shall deal with cosmological models based on scalar fields, i.e. the matter content of the universe is modeled by means of scalar fields. Notice that the cosmological models based on  scalar fields were considered long before the observational discovery of cosmic acceleration~\cite{before1}-\cite{before3}.

According to some authors, the analysis of observations permits the existence of the moment when the universe changes the value of the parameter $w$ from $w > -1$ to $w < -1$ \cite{phant-obs1}-\cite{phant-obs7}. This transition is called ``crossing of the phantom divide line''.
The most recent investigations  have shown that the phantom divide line crossing is still not excluded by the
data \cite{obs-new1}-\cite{obs-new3}.

It is easy to see that the standard minimally coupled scalar field cannot give rise to the phantom dark energy, because in this model
the absolute value of  energy density is always greater than that of pressure, i.e. $|w| < 1$. A possible  way out of this situation
is the consideration of the scalar field models with the negative kinetic term, i.e. \emph{phantom} field models, as I will briefly review in a moment.
Thus, the important problem arising in connection with the phantom energy is  the crossing of the phantom
divide line. The general belief is that while this crossing is not admissible
in simple minimally coupled models its  explanation
 requires  more complicated models such as ``multifield'' ones
or models with non-minimal coupling between scalar field and gravity
(see e.g. \cite{divide1}-\cite{divide6}).

Some authors of \cite{mio_2field} described the phenomenon of the change
of sign of the kinetic term of the scalar field implied by the Einstein
equations\cite{crossing,cusps}. It was shown that such a change is possible only when the potential of the
scalar field possesses some cusps and, moreover, for some very special
initial conditions on the time derivatives and values of the
considered scalar field approaching to the phantom divide line.
At the same time, two-field models  including one standard scalar field and
a phantom field can describe the phenomenon of the (de-)phantomization under very general conditions and using rather
simple potentials \cite{2f1}-\cite{2f4}.

In the  paper under consideration \cite{mio_2field} we have focused  to the drastic
difference between two- and one-field models.

The reconstruction procedure of the (single) scalar field potential models
is well-known \cite{Star}, \cite{Barrow1,Barrow2}, \cite{we-tach}, \cite{Yurov}, \cite{Guo1,Guo2}, \cite{Zhuk} and \cite{Szydlowski1}-\cite{Szydlowski4}. Let me recapitulate it briefly.

If the matter is represented by a spatially homogeneous minimally coupled scalar field, then
the energy density and the pressure are given by the formul\ae
\begin{equation}
\varepsilon = \frac12\dot{\phi}^2 + V(\phi)\ ,
\label{2:energy}
\end{equation}
\begin{equation}
p = \frac12\dot{\phi}^2 - V(\phi)\ ,
\label{2:pressure}
\end{equation}
where $V(\phi)$ is a scalar field potential.
Friedmann equations thus read
\begin{align}
&h^2=\epsilon=\tfrac{1}{2}\dot\phi^2+V(\phi)\ ,\\
&\ddot\phi+3h\dot\phi+\frac{\d V}{\d\phi}=0\ ,\label{2:KGscalar}
\end{align}
Where the second is clearly the Klein-Gordon equation on a FRLW background.
Combining these equations we have
\be
V = \frac{\dot{h}}{3} + h^2\ ,
\label{2:poten}
\ee
and
\be
\dot{\phi}^2 = -\frac23 \dot{h}\ .
\label{2:phi-dot}
\ee
Equation \eqref{2:poten} represents the potential as a function of time $t$. Integrating equation \eqref{2:phi-dot}
one can find the scalar field as a function of time. Inverting this dependence we can obtain the time parameter as
a function of $\phi$ and substituting the corresponding formula into equation \eqref{2:poten} one arrives to the uniquely
reconstructed potential $V(\phi)$. It is necessary to stress that this potential reproduces a given cosmological evolution
only for some special choice of initial conditions on the scalar field and its time derivative \cite{crossing,we-tach}.
Below I shall show that in the case of two scalar fields one has an enormous freedom in the choice of
the two-field potential providing the same cosmological evolution. This freedom is connected with the fact
that the kinetic term has now two contributions.

Notice that equation \eqref{2:phi-dot} immediately implies one thing: in order to have (at least a phase of) super-acceleration (which, I recall, is defined by $\dot h>0$) with an energy density produced by a scalar field, one has to have a negative sign of the kinetic term. This is even clearer if one take the specific Hubble parameter
\be
h(t)=\f{2}{3(w+1)t}\ ,
\ee
i.e. the Hubble parameter for a fluid with $w$ = constant, and tries to reconstruct the scalar field model.  Integrating \eqref{2:phi-dot} one has
\be
\phi(t)=\pm\sqrt{\frac{4}{9(w+1)}}\ln t\ ,
\ee
which clearly implies $w>-1$.
Had we chose a scalar field with a negative kinetic term, we would have been able to reproduce the $w<-1$ case.

In order to examine the problem of phantom divide line crossing we shall be interested in the case of one standard
scalar field and one phantom field $\xi$ ,
whose kinetic term has a negative sign. In this case the total energy density and pressure will be given by
\begin{align}\label{2:energy1}
\varepsilon &= \frac12\dot{\phi}^2 - \frac12\dot{\xi}^2 + V(\phi,\xi)\ ,\\ \label{2:pressure1}
p &= \frac12\dot{\phi}^2 - \frac12\dot{\xi}^2 - V(\phi,\xi)\ .
\end{align}
The relation \eqref{2:poten} expressing the potential as a function of $t$ does not change in form, but instead  of
equation \eqref{2:phi-dot} we have
\be
\dot{\phi}^2 - \dot{\xi}^2  = -\frac23 \dot{h}\ .
\label{2:phi-dot1}
\ee
Now, one has rather a wide freedom in the choice of the time dependence of one of two fields. After that the time dependence
of the second field can be found from equation \eqref{2:phi-dot1}. However, the freedom is not yet exhausted. Indeed, having two representations for the time parameter $t$ as a function of $\phi$ or $\xi$, one can construct an infinite number of potentials
$V(\phi,\xi)$ using the formula \eqref{2:poten} and some rather loose consistency conditions. It is rather difficult to characterize
all the family of possible two-field potentials, reproducing given cosmological evolution $h(t)$. In the following, I describe
some general principles of construction of such potentials and then consider some concrete examples.

\section{The system of equations for two-field cosmological models}
The system of equations, which we  study contains (\ref{2:poten}) and (\ref{2:phi-dot1}) and two Klein-Gordon equations
\begin{align}
\ddot{\phi} + 3h\dot{\phi} + \frac{\partial V(\phi,\xi)}{\partial \phi} =& 0\ ,
\label{2:KG}\\
\ddot{\xi} + 3h\dot{\xi} - \frac{\partial V(\phi,\xi)}{\partial \xi} =& 0\ .
\label{2:KG1}
\end{align}
From equations \eqref{2:KG} and \eqref{2:KG1} we can find the partial derivatives $\frac{\partial V(\phi,\xi)}{\partial \phi}$
and $\frac{\partial V(\phi,\xi)}{\partial \xi}$ as functions of time $t$. The consistency relation
\be
\dot{V} = \frac{\partial V(\phi,\xi)}{\partial \phi}(t)\dot{\phi} + \frac{\partial V(\phi,\xi)}{\partial \xi}(t)\dot{\xi}
\label{2:consistency}
\ee
is respected.

Before starting the construction of potentials for particular cosmological evolutions, it is useful to
consider some  mathematical aspects of the problem of reconstruction of a function of two variables in general terms.

\subsection{Reconstruction of the function of two variables, which in turn depends on a third parameter}
Let us consider the function of two variables $F(x,y)$  defined on a curve, parameterized by $t$. Suppose that we know
the function $F(t)$ and its partial derivatives  as functions of $t$:
\be
F(x(t),y(t)) = F(t)\ ,
\label{2:function}
\ee
\be
\frac{\partial F(x(t),y(t))}{ \partial x} = \frac{\partial F}{ \partial x}(t)\ ,
\label{2:partial}
\ee
\be
\frac{\partial F(x(t),y(t))}{ \partial y} = \frac{\partial F}{ \partial y}(t)\ .
\label{2:partial1}
\ee
These three functions should satisfy the consistency relation
\be
\dot{F}(t) = \frac{\partial F}{ \partial x}(t)\dot{x} + \frac{\partial F}{ \partial y}(t)\dot{y}\ .
\label{2:consistency1}
\ee

As a simple example we can consider the curve
\begin{equation}
x(t) = t,\ \ y(t) = t^2\ ,
\label{2:curve}
\end{equation}
while
\begin{equation}
F(t) = t^2\ ,
\label{2:example}
\end{equation}
\begin{equation}
\frac{\partial F}{\partial x} = t^2\ ,
\label{2:example1}
\end{equation}
\begin{equation}
\frac{\partial F}{\partial y} = t\ ,
\label{2:example2}
\end{equation}
and equation (\ref{2:consistency1}) is satisfied.

Thus, we would like to reconstruct the function $F(x,y)$ having explicit expressions in right-hand side of equations (\ref{2:function})--
(\ref{2:partial1}). This reconstruction is not unique. We shall begin the reconstruction process taking such simple ansatzes
as
\begin{equation}
F_1(x,y) = G_1(x) + H_1(y)\ ,
\label{2:sum}
\end{equation}
\begin{equation}
F_2(x,y) = G_2(x)H_2(y)\ ,
\label{2:product}
\end{equation}
\begin{equation}
F_3(x,y) = (G_3(x) + H_3(y))^{\alpha}\ .
\label{2:power}
\end{equation}

The assumption (\ref{2:sum})  immediately implies
\begin{equation}
\frac{\partial F_1}{\partial x} = \frac{\partial G_1}{\partial x}\ ,
\label{2:partial2}
\end{equation}
\begin{equation}
\frac{\partial F_1}{\partial y} = \frac{\partial H_1}{\partial y}\ .
\label{2:partial3}
\end{equation}
Therefrom  one obtains
\begin{equation}
G_1(x) = \int^x \frac{\partial F_1}{\partial x'}(t(x')) dx'\ ,
\label{2:G}
\end{equation}
\begin{equation}
H_1(y) = \int^y \frac{\partial H_1}{\partial y'}(t(y')) dy'\ .
\label{2:H}
\end{equation}
Hence
\begin{equation}
F_1(x,y) = \int^x \frac{\partial F_1}{\partial x'}(t(x')) dx' + \int^y \frac{\partial H_1}{\partial y'}(t(y')) dy'\ .
\label{2:result}
\end{equation}
For an example given by equations (\ref{2:curve})--(\ref{2:example2}) the function $F_1(x,y)$ is
\begin{equation}
F_1(x,y) = \int^x t^2(x')dx' + \int^y t(y')dy' = \int^x x'^2 dx' \pm \int^y \sqrt{y'} dy'\ .
\label{2:example3}
\end{equation}
Explicitly
\begin{equation}
F_1(x,y)=\left\{
\begin{aligned}
\frac{1}{3}\left( x^3+2y^{3/2}\right)\ , && x>0\ ,\\
\frac{1}{3}\left( x^3-2y^{3/2}\right)\ , && x<0\ .\end{aligned}\right.\label{2:ex_somma}
\end{equation}

Similar reasonings give for the assumptions (\ref{2:product}),and  (\ref{2:power})  correspondingly
\begin{equation}
F_2(x,y) = \exp\left\{\int \left(\frac1F\frac{\partial F}{\partial x}\right)(t(x))dx
+ \int \left(\frac1F\frac{\partial F}{\partial y}\right)(t(y))dy\right\}\ ,
\label{2:F2}
\end{equation}
\begin{equation}
F_3(x,y) = \left\{\int \frac{dx}{\alpha} \left(F^{\frac{1}{\alpha}-1}\frac{\partial F}{\partial x}\right)(t(x))
+ \int \frac{dy}{\alpha} \left(F^{\frac{1}{\alpha}-1}\frac{\partial F}{\partial y}\right)(t(y))\right\}^{\alpha}\ .
\label{2:F3}
\end{equation}
For our simple example (\ref{2:curve})--(\ref{2:example2}) the functions $F_2(x,y)$ and $F_3(x,y)$ have the form
\begin{equation}
F_2(x,y) = xy\ ,
\label{2:exampleF2}
\end{equation}
\begin{equation}
F_3(x,y)=
\left\{\begin{aligned}
\left[\frac{1}{3}\left(x^{3/\alpha} +2y^{3/2\alpha}\right)\right]^{\alpha}\ , && x>0\ ,\\
\left[\frac{1}{3}\left(x^{3/\alpha}(-)^{3/\alpha}2y^{3/2\alpha}\right)\right]^{\alpha}\ , && x<0\ .\end{aligned}\right.
\end{equation}
Thus, we have seen that the same input of ``time'' functions (\ref{2:example})--(\ref{2:example2}) on the curve (\ref{2:curve})
produces quite different functions of variables $x$ and $y$.

Naturally, one can introduce many other assumptions for reconstruction of $F(x,y)$. For example, one can consider
linear combinations of $x$ and $y$ as functions of the parameter $t$ and decompose the presumed function $F$ as a sum
or a product of the functions of these new variables.

Now I  present a way of constructing  the whole family of solutions starting from a given one. Let us suppose that we have
a function $F_0(x,y)$ satisfying all the necessary conditions. Let us take an arbitrary function
\begin{equation}
f(x,y) = f\left(\frac{t(x)}{t(y)}\right)\ ,
\label{2:f-small}
\end{equation}
which depends only on the ratio $t(x)/t(y)$. We require also
\begin{equation}
f(1) = 1\ ,\quad f'(1) = 0\ ,
\end{equation}
i.e. the function reduces to unity and its derivative vanishes on the curve $(x(t),y(t))$.
Then it is obvious that
the function
\begin{equation}
F(x,y) = F_0(x,y)f\left(\frac{t(x)}{t(y)}\right)
\label{2:F-new}
\end{equation}
is also a solution. This permits us to generate a whole family of solutions, depending on a choice of the
function $f$.
Moreover, one can construct other solutions, adding to the function $F(x,y)$ a term proportional to
$(t(x)-t(y))^2$.

\subsection{Cosmological applications, an evolution ``Bang to Rip''}\label{sec:BtR}
To show how this procedure works in Cosmology, we consider a relatively simple cosmological evolution, which
nevertheless is of  particular interest, because it describes the phantom divide line crossing.
Let us suppose that the Hubble variable for this evolution behaves as
\begin{equation}
h(t) = \frac{A}{t(t_R - t)}\ ,
\label{2:Hubble-bang-to-rip}
\end{equation}
 where $A$ is a positive constant. At the beginning of the cosmological evolution, when $t \rightarrow 0$ the universe
is born from the standard Big Bang type cosmological singularity, because $h(t) \sim 1/t$. Then, when $t \rightarrow t_R$,
the universe is super-accelerated, approaching the Big Rip singularity $h(t) \sim 1/(t_R - t)$.
Substituting the function (\ref{2:Hubble-bang-to-rip}) and its time derivative into equations. (\ref{2:phi-dot1}) and (\ref{2:poten})
we come to
\begin{equation}
\dot{\phi}^2 - \dot{\xi}^2 = - \frac{2A(2t-t_R)}{3t^2(t_R-t)^2}\ ,
\label{2:BtR}
\end{equation}
\begin{equation}
V(t) = \frac{A(2t-t_R+3A)}{3t^2(t_R-t)^2}\ .
\label{2:poten1}
\end{equation}
For convenience let me
choose also the parameter $A$ as
\begin{equation}
A = \frac{t_R}{3}\ ,
\label{2:special-choice}
\end{equation}
Then,
\begin{equation}
h(t) = \frac{t_R}{3t(t_R-t)}\ ,
\label{2:Hubble-simple}
\end{equation}
and
\begin{equation}
V(t) = \frac{2t_R}{9t(t_R-t)^2}\ .
\label{2:poten2}
\end{equation}

Let us consider now a  special choice of functions $\phi(t)$ and $\xi(t)$ used already\footnote{Notice that the origin of two scalar fields has been associated in \cite{2f1}-\cite{2f4} with a non-hermitian complex scalar field theory and there a classical solution was found as a saddle point in  "double" complexification. I will discuss in more detail the link between non-hermiticity and phantom fields in chapter \ref{sec:PT}.} in \cite{2f1}-\cite{2f4},
\begin{equation}
\phi(t) = -\frac43 {\rm arctanh} \sqrt{\frac{t_R-t}{t_R}}\ ,
\label{2:phi1}
\end{equation}
\begin{equation}
\xi(t) = \frac{\sqrt{2}}{3} \ln \frac{t}{t_R-t}\ .
\label{2:xi1}
\end{equation}
The derivatives of the potential with respect to the fields $\phi$ and $\xi$ could be found from the Klein-Gordon equations
(\ref{2:KG}) and (\ref{2:KG1}):
\begin{equation}
\frac{\partial V}{\partial \xi} = \ddot{\xi} + 3h\dot{\xi} = \frac{\sqrt{2}}{3}\frac{2t_R}{t(t_R-t)^2}\ ,
\label{2:der-pot}
\end{equation}
\begin{equation}
\frac{\partial V}{\partial \phi} = -\ddot{\phi} - 3h\dot{\phi} = -\frac{\sqrt{t_R}}{t(t_R-t)^{3/2}}\ .
\label{2:der-pot1}
\end{equation}
We can obtain also the time parameter as a function of $\phi$ or $\xi$:
\begin{equation}
t(\phi) = \frac{t_R}{\cosh^2(-3\phi/4)}\ ,
\label{2:time-phi}
\end{equation}
\begin{equation}
t(\xi) = \frac{t_R}{\exp(-3\xi/\sqrt{2})+1}\ .
\label{2:time-xi}
\end{equation}

Now we can make a hypothesis about the structure of the potential $V(\xi,\phi)$:
\begin{equation}
V_2(\xi,\phi) = G(\xi)H(\phi)\ .
\label{2:hypo}
\end{equation}
Applying the technique described in the subsection A, we can get $G(\xi)$:
\begin{align}
\ln G(\xi) &= \int \left(\frac1V \frac{\partial V}{\partial \xi}\right)(t(\xi)) d\xi  \\\nonumber
&= \int \frac{9t(\xi)(t(\xi)-t_R)^2}{2t_R}\frac{\sqrt{2}}{3}\frac{2t_R}{t(\xi)(t_R-t(\xi))^2}d\xi\\\nonumber
& = 3\sqrt{2}\xi\ ,
\label{2:G-hypo}
\end{align}
 and
\begin{equation}
G(\xi) = \exp(3\sqrt{2}\xi)\ .
\label{2:G-hypo1}
\end{equation}
To find $H(\phi)$ one can use the analogous direct integration, but we preferred to implement a formula
\begin{equation}
H(\phi) = \frac{V(t(\phi)}{G(\xi(t(\phi)))}\ ,
\label{2:another}
\end{equation}
which gives
\begin{equation}
H(\phi(t)) = \frac{2t_R}{9t^3}\ ,
\label{2:H-new}
\end{equation}
and hence,
\begin{equation}
H(\phi) = \frac{2}{9t_R^2}\cosh^6(-3\phi/4)\ .
\label{2:H-new1}
\end{equation}
Finally,
\begin{equation}
V_2(\xi,\phi) = \frac{2}{9t_R^2}\cosh^6(-3\phi/4)\exp(3\sqrt{2}\xi)\ .
\label{2:V-2new}
\end{equation}
Here we have reproduced the potential studied in \cite{2f1}-\cite{2f4}.

Making the choice
\begin{equation}
V_1(\xi,\phi) = G(\xi) + H(\phi),
\label{2:choice1}
\end{equation}
we derive
\begin{eqnarray}
V_{1}(\xi,\phi) = \frac{2}{3t_R^2}\left[-\frac13e^{-3\xi/\sqrt{2}} + 3\sqrt{2}\xi
+2e^{3\xi/\sqrt{2}}
+ \frac{1}{3} e^{6\xi/\sqrt{2}} \right. \nonumber \\
\left.+\frac{\sinh^4(-3\phi/4) + \sinh^2(-3\phi/4) -1}{\sinh^2(-3\phi/4)}
+ \ln \sinh^4(-3\phi/4)\right]\ .
\label{2:poten3}
\end{eqnarray}

Now, we can make another choice of the field functions $\phi(t)$ and $\xi(t)$, satisfying  the condition
(\ref{2:BtR}):
\begin{equation}
\phi(t) = \frac{\sqrt{2}}{3} \ln \frac{t}{t_R-t}\ ,
\label{2:phi-new3}
\end{equation}
\begin{equation}
\xi(t) = \frac{4}{3} {\rm arctanh} \sqrt{\frac{t}{t_R}}\ .
\label{2:xi-new3}
\end{equation}
The time parameter $t$ is a function of fields is
\begin{equation}
t(\phi) = \frac{t_R}{\exp(-3\phi/\sqrt{2})+1}\ ,
\label{2:time-phi1}
\end{equation}
\begin{equation}
t(\xi) = \frac{t_R}{\tanh^2(3\xi/4)}\ .
\label{2:time-xi1}
\end{equation}

Looking for the potential as a sum of functions of two fields as in equation (\ref{2:choice1}) after lengthy but
straightforward calculations one comes to the following potential:
\begin{eqnarray}
V_1(\xi,\phi) = &\frac{2}{3t_R^2}\left[1 + \frac23 \sinh^4 (3\xi/4) + 3\sinh^2 (3\xi/4)
- \frac{1}{3\sinh^2 (3\xi/4)} \right. \nonumber \\
&+ 2 \ln \sinh^2 (3\xi/4) + \frac23 \exp (-3\phi/\sqrt{2}) - 3\sqrt{2}\phi \nonumber \\
&\left. - 2\exp(3\phi/\sqrt{2})-\frac13\exp(\phi/\sqrt{2})\right]\ .
\label{2:poten5}
\end{eqnarray}

Similarly for the potential designed as a product of functions of two fields (\ref{2:hypo})
we obtain
\begin{equation}
V_2(\xi,\phi) = \frac{2}{9t_R^2}\sinh^2(3\xi/4)\cosh^2(3\xi/4)\exp(-3\sqrt{2}\phi)\ .
\label{2:poten6}
\end{equation}

One can make also other choices of functions $\phi(t)$ and $\xi(t)$ generating other potentials, but I shall not do it here,  concentrating instead on the qualitative and numerical analysis of two toy cosmological models described by
potentials (\ref{2:poten5}) and (\ref{2:V-2new}).

\section{Analysis of cosmological models}
It is well known \cite{Bel-Khal}  that for the qualitative analysis of the system of cosmological
equations it is convenient to present it as a dynamical system, i.e. a system of first-order differential equations.
Introducing the new variables $x$ and $y$ we can write
\begin{equation}\left\{
\begin{array}{ll}
\dot{\phi} &= x\ ,  \\
\dot{\xi} &= y\ ,  \\
\dot{x}& =-3\ {\rm sign}(h)\ x \sqrt{\frac{x^2}{2} - \frac{y^2}{2} + V(\xi,\phi)} - \frac{\partial V}{\partial \phi}\ ,\\
\dot{y} &=-3\ {\rm sign}(h)\ y \sqrt{\frac{x^2}{2} - \frac{y^2}{2} + V(\xi,\phi)} - \frac{\partial V}{\partial \xi}\ .

\end{array}\right.\label{2:system}
\end{equation}
Notice that the reflection
\begin{eqnarray}
&&x \rightarrow -x\ , \nonumber \\
&&y \rightarrow -y\ , \nonumber \\
&&t \rightarrow -t
\label{inversion}
\end{eqnarray}
transforms the system into one describing the cosmological evolution with the opposite sign of the Hubble parameter.
The stationary points of the system (\ref{2:system}) are given by
\begin{equation}
x = 0\ ,\ y = 0\ ,\ \frac{\partial V}{\partial \phi} = 0\ ,\ \frac{\partial V}{\partial \xi} = 0\ .
\label{2:stationary}
\end{equation}

\subsection{Model I}\label{sec:model1}
In this subsection I shall analyze the cosmological model with two fields -- standard scalar and phantom -- described by
the potential (\ref{2:poten5}).
For this potential the system of equations (\ref{2:system})  reads

\begin{equation}\left\{
\begin{array}{ll}
\dot\phi=&x\ ,\\
\dot\xi=&y\ ,\\
\dot x=&-3\ {\rm sign}(h)\ x\sqrt{\frac{x^2}{2}-\frac{y^2}{2}+\frac{2}{9t_R^2}\sinh^2(3\xi/4)\cosh^6(3\xi/4)e^{-3\sqrt{2}\phi}} \\
&+\frac{2\sqrt{2}}{3t_R^2}\sinh^2(3\xi/4)\cosh^6(3\xi/4)e^{-3\sqrt{2}\phi}\ , \\ 
\dot y=&-3\ {\rm sign}(h)\ y\sqrt{\frac{x^2}{2}-\frac{y^2}{2}+\frac{2}{9t_R^2}\sinh^2(3\xi/4)\cosh^6(3\xi/4)e^{-3\sqrt{2}\phi}} \\&+\frac{\sinh(3\xi/4)\cosh^5(3\xi/4)}{3t_R^2}\left[\frac{1}{3}\cosh^2(3\xi/4)+\sinh^2(3\xi/4)\right]e^{-3\sqrt{2}\phi}\ .
\end{array}\right.\label{2:sistV4}
\end{equation}

It is easy to see that there are stationary points
\begin{equation}
\phi = \phi_0\ ,\  \xi = 0\ ,\  x = 0\ ,\  y = 0\ ,
\label{2:stat}
\end{equation}
where $\phi_0$ is arbitrary.
For these points the potential and hence the Hubble variable vanish. Thus, we have a static cosmological solution.
We should study the behavior of our system in the neighborhood of the point~(\ref{2:stat}) in linear approximation:
\begin{equation}\left\{
\begin{array}{ll}
\dot\phi=&x\ ,\\
\dot\xi=&y\ ,\\
\dot x=&0\ ,\\
\dot y=&+\frac{\xi}{4t_R^2}e^{-3\sqrt{2}\phi_0}\ .
\end{array}\right.\label{2:sistV4_li}
\end{equation}

One sees that the dynamics of $\phi$ in this approximation is frozen and hence we can focus on the study
of the dynamics of the variables $\xi,y$. The eigenvalues of the corresponding subsystem of two equations
are
\begin{equation}
\lambda_{1,2} = \mp \frac{e^{-3\phi_0/\sqrt{2}}}{2t_R}\ .
\label{2:saddle}
\end{equation}
These eigenvalues are real and have opposite signs, so one has a saddle point in the plane $(\xi,y)$ and
this means that the points (\ref{2:stat}) are  unstable.

One can make another qualitative observation. Freezing the dynamics of $\xi$ independently of $\phi$,
namely choosing $y = 0, \xi =0$, which implies also $\dot{y} = \ddot{\xi} = 0$, one has the following equation of motion for
$\phi$:
\begin{equation}
\ddot{\phi} + 3h\phi = 0\ .
\label{2:massless}
\end{equation}
Equation (\ref{2:massless}) is nothing but the Klein-Gordon equation for a massless scalar field on the Friedmann background,
whose solution is
\begin{equation}
\phi(t) = \frac{\sqrt{2}}{3}\ln t\ ,
\end{equation}
and which gives a Hubble variable
\begin{equation}
h(t) = \frac{1}{3t}\ .
\end{equation}
This is an evolution of the flat Friedmann universe, filled with stiff matter with the equation of state $p =\varepsilon$.
It describes a universe born from the Big Bang singularity and infinitely expanding. Naturally, for the opposite sign
of the Hubble parameter, one has the contracting universe ending in the Big Crunch cosmological singularity.

\begin{figure}[htp]\centering
\includegraphics{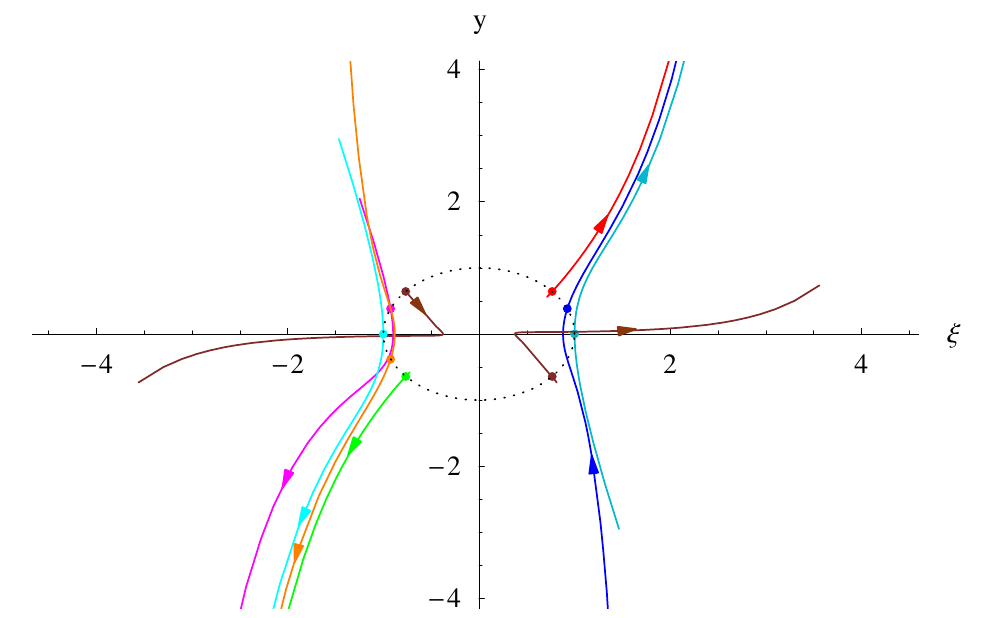}
\caption{An example of section of the 4d phase space obtained with numerical calculations. We can see a series of trajectories corresponding to different choices of the initial conditions. Every initial condition in the graphic is chosen on the dashed ``ellipse'' centered in the origin and is emphasized by a colored dot. A ``saddle point like'' structure of the set  of the trajectories clearly emerges.}\label{2:figure1}
\end{figure}
\begin{figure}[htp]\centering
\includegraphics[scale=.7]{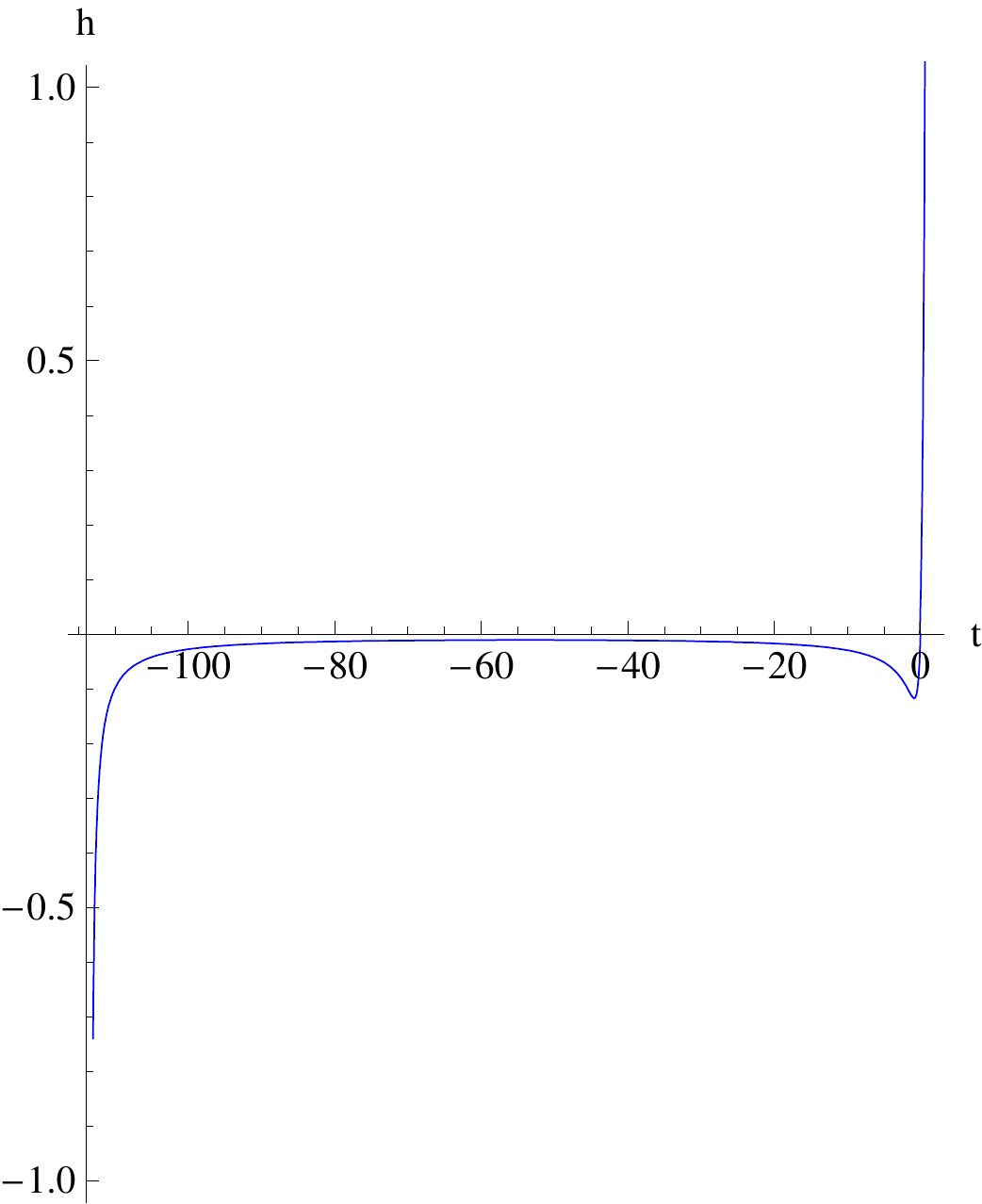}
\caption{Typical behavior of the Hubble parameter for the model I trajectories. The presence of two stationary points (namely a maximum followed by a minimum) indicates the double crossing of the phantom divide line. Both the initial and final singularities are  characterized by a Big Rip behavior, the first is a  contraction and the second is an expansion.}\label{2:figure2}
\end{figure}

Now, I describe some results of numerical calculations to have an idea about the structure of the set 
of possible cosmological evolutions coexisting in the model under consideration. 
We have carried out two kinds of simulations. First, we have considered neighborhood of the plane 
$y,\xi$ with the initial conditions on the field $\phi$ such that $\phi(0) = 0, \dot\phi (0) = 0$ (see figure \ref{2:figure1}). The initial conditions 
for the phantom field were chosen in such a way that the sum of absolute values of the kinetic and potential energies were 
fixed. Then, running the time back and forward we have seen that the absolute majority of the cosmological evolutions began 
at the singularity of the ``anti-Big Rip'' type (figure \ref{2:figure2}). Namely, the initial cosmological singularity were characterized by an 
infinite value of the cosmological radius and an infinite negative value of its time derivative (and also of the Hubble variable).
Then the universe squeezes, being dominated by the phantom scalar field $\xi$. At some moment the universe passes the phantom divide line and the universe continues squeezing but with $\dot{h} < 0$. Then it achieves the minimal value of the cosmological radius 
and an expansion begins. At some moment the universe undergoes the second phantom divide line crossing and its expansion becomes
super-accelerated culminating in an encounter with a Big Rip singularity. Apparently this scenario is very different from the standard cosmological scenario and from its phantom version Bang-to-Rip, which has played a role of an input in the construction 
of our potentials.
The second procedure, which we have used is the consideration of trajectories close to our initial trajectory of the Bang-to-Rip type. The numerical analysis shows that this trajectory is unstable and the neighboring trajectories again have anti-Big Rip --
double crossing of the phantom line -- Big Rip behavior described above.  
However, it is necessary to emphasize that a small subset of the trajectories of the Bang-to-Rip type exist, being not 
in the vicinity of our initial trajectory.

\subsection{Model II}\label{sec:model2}
In this subsection we shall study the cosmological model with the potential (\ref{2:V-2new}). Now the system of equations
(\ref{2:system}) looks like
\begin{equation}\left\{
\begin{array}{ll}
\dot\phi=&x\ ,\\
\dot\xi=&y\ , \\
\dot x=&-3\ {\rm sign}(h)\ x\sqrt{\frac{x^2}{2}-\frac{y^2}{2}+\frac{2}{9t_R^2}\cosh^6(3\phi/4) \exp\{3\sqrt{2}\xi\}} \\
&-\frac{1}{t_R^2}\cosh^5(3\phi/4)\sinh(3\phi/4)e^{3\sqrt{2}\xi}\ , \\
\dot y=&-3{\rm sign}(h)y\sqrt{\frac{x^2}{2}-\frac{y^2}{2}+\frac{2}{9t_R^2}\cosh^6(3\phi/4) \exp\{3\sqrt{2}\xi\}}\\&
+\frac{2\sqrt{2}}{3t_R^2}\cosh^6(3\phi/4)e^{3\sqrt{2}\xi}\ .
\end{array}\right.\label{2:2sist}
\end{equation}
Notice that the potential (\ref{2:V-2new}) has an additional reflection symmetry
\begin{equation}
V_2(\xi,\phi) = V_2(\xi,-\phi)\ .
\label{2:reflection}
\end{equation}
This provides the symmetry with respect to the origin in the plane $(\phi,x)$. The system (\ref{2:2sist})
has no stationary points. However, there is an interesting point
\begin{equation}
\phi=0\ ,\  x = 0\ ,
\label{2:interest}
\end{equation}
which freezes the dynamics of $\phi$ and hence, permits to consider independently the dynamics of $\xi$ and $y$, described by
the subsystem
\begin{equation}\left\{
\begin{array}{ll}
\dot\xi= &y\ , \\
\dot y=&-3\ {\rm sign}(h)\ y\sqrt{-\frac{y^2}{2}+\frac{2}{9t_R^2}e^{3\sqrt{2}\xi}}+\frac{2\sqrt{2}}{3t_R^2}e^{3\sqrt{2}\xi}\ .
\end{array}\right.\label{2:sis2dim}
\end{equation}
Apparently, the evolution of the universe is driven now by the phantom field and is subject to super-acceleration.

In this case the qualitative analysis of the differential equations for $\xi$ and $y$, confirmed by the numerical simulations 
gives a predictable result: being determined by the only phantom scalar field the cosmological evolution is characterized by 
the growing positive value of $h$. Namely, the universe begins its evolution from the anti-Big Rip singularity ($ h = -\infty$)
then $h$ is growing  passing at some moment of time the value $h = 0$ (the point of minimal contraction of the cosmological radius $a(t)$) and then expands ending  its evolution in the Big Rip cosmological singularity ($h = +\infty$).

\begin{figure}[htp]\centering
\includegraphics{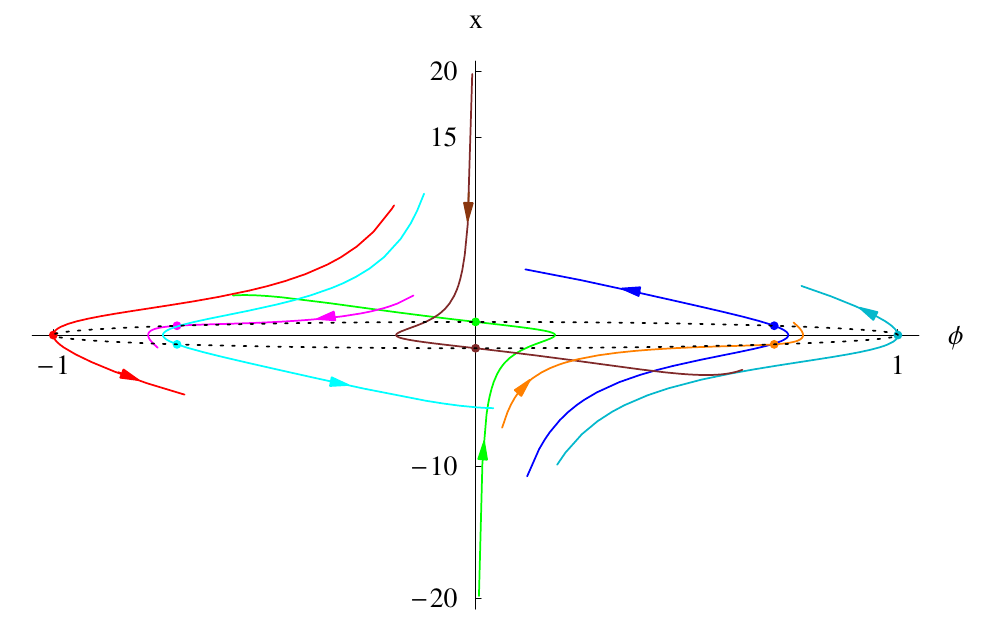}
\caption{An example of section of the 4D phase space for the model II obtained with numerical calculations. We can see a family of different trajectories corresponding to different choices of the initial conditions. Every initial condition in the graphic is chosen on the dashed ``ellipse''. The origin represents the point of freezing.}\label{2:figure3}
\end{figure}

Another numerical simulation can be done by fixing initial conditions for the phantom field as $\xi(0) = 0, y(0) = 0$ (see figure~\ref{2:figure3}). 
Choosing various values of the initial conditions for the scalar field $\phi(0),x(0)$ around the point of freezing $\phi=0,
x=0$ we found two types of cosmological trajectories:\\
1. The trajectories starting from the anti-Big Rip singularity and ending in the Big Rip after the double crossing of the phantom 
divide line. These trajectories are similar to those discussed in the preceding subsection for the model I.\\
2. The evolutions of the type Bang-to-Rip. 

Then we have carried out the numerical simulations of cosmological evolutions, choosing the initial conditions around 
the point of the phantomization point with the coordinates 
\begin{align}
\phi(0) &=0\ , \nonumber \\
x(0) &= \frac{\sqrt{2}}{3}\ ,\nonumber \\
\xi(0) &= \frac43{\rm arctanh} \frac{1}{\sqrt{2}}\ , \nonumber \\
y(0) &= \frac{\sqrt{2}}{3}\ .
\label{2:phant-point}
\end{align}

This analysis shows that in contrast to the model I, here the standard phantomization trajectory is stable and 
the trajectories of the type Bang-to-Rip are not exceptional, though less probable then those of the type 
anti-Big Rip to Big Rip. 

\section{Conclusions}
In the paper \cite{mio_2field} we have considered the problem of reconstruction of the potential in a theory with two scalar fields
(one standard and one phantom) starting with a given cosmological evolution. It is known ( see e.g. \cite{Star,we-tach}) 
that in the
case of the only scalar field this potential is determined uniquely as well as the initial conditions for the scalar field,
reproducing the given cosmological evolution.  Changing the initial conditions, one can find a variety of 
cosmological evolutions, sometimes qualitatively different from the ``input'' one (see e.g. \cite{we-tach}).
In the case of two fields the procedure of reconstruction becomes much more involved. As we have shown here, there is a huge variety
of different potentials reproducing the given cosmological evolution (a very simple one in the case, which we have
explicitly studied here). Every potential entails  different cosmological evolutions, depending on the initial
conditions.


It is interesting that the existence of different dynamics of scalar fields corresponding to the same evolution of 
the Hubble parameter $h(t)$ can imply some observable consequences connected with the possible interactions of 
the scalar fields with other  fields. In this case, the time dependence of the scalar fields considered above 
can directly affect physically observable quantities.
Indeed, some results in this direction are presented in chapter \ref{sec:magnetic}, devoted to the explanation of the paper \cite{mio_magnetic}.


\chapter{Two-field models and cosmic magnetic fields}\label{sec:magnetic}

\emph{The present chapter is intended to present the content of the paper \cite{mio_magnetic}. Starting from the models studied in the preceding chapter, we couple one of the fields to a cosmic electromagnetic field, whose existence is experimentally confirmed. The idea is to try and find a discrimination between cosmological models presenting the \emph{same} background evolution, by means of the effects of the coupling with other (observable) fields. }

\section{Introduction to the content of the paper~\texorpdfstring{\cite{mio_magnetic}}{on cosmic magnetic fields}}

In the preceding chapter I have shown the procedure of reconstruction of the potential in two-field models. It was shown that there exists a huge variety of potentials and time dependences of the fields realizing the same cosmological evolution. Some concrete examples were considered, corresponding to the evolution beginning with the standard Big Bang singularity and ending in the Big Rip singularity \cite{Rip1,Rip2}.

One can ask oneself: what is the sense of  studying  different potentials and scalar field dynamics if they imply 
the same cosmological evolution?    The point is that the scalar and phantom field can interact with other  fields 
and   influence not only  the global cosmological evolution but also  other observable quantities.  

One of the possible effects of the presence of normal and phantom fields could be their influence on the dynamics of cosmic magnetic fields. The problem of the origin 
and of possible amplification of cosmic magnetic fields is widely discussed in the literature \cite{magnetic1,magnetic2,magnetic3}. 
In particular, the origin of such fields can be attributed to primordial quantum fluctuations~\cite{quantum1,quantum2,quantum3} 
and their further evolution can be influenced by hypothetic interaction with pseudo-scalar fields breaking 
the conformal invariance of the electromagnetic field \cite{break1}-\cite{break5}, \cite{Sorbo}. In the  paper under consideration we analyzed the evolution 
of magnetic fields created as a result of quantum fluctuations, undergoing the inflationary period 
with unbroken conformal invariance and beginning the interaction with pseudo-scalar or pseudo-phantom fields 
after exiting the inflation and entering the Big Bang expansion stage, which is a part of the Bang-to-Rip scenario 
described in the preceding chapter. We used different field realizations of this scenario and we
shall see how the dynamics of the field with negative parity influences the dynamics of cosmic magnetic fields. 

 What follows is so organized: in section~\ref{sec:m1} I very briefly recall  the Bang-to-Rip scenario and the two models that will be considered (the same of chapter \ref{sec:2field}; in section~\ref{sec:m2} I introduce the interaction of the fields (phantom or normal) with an electromagnetic field 
 and write down the corresponding equations of motion; in section~\ref{sec:m3} I describe the numerical simulations of the evolution 
 of magnetic fields and present the results of these simulations; the last section \ref{sec:m4} is devoted to concluding remarks.

\section{Cosmological evolution and (pseudo)-scalar fields}\label{sec:m1}
I shall consider a spatially flat Friedmann universe with the FLRW metric~\eqref{FRWline}.

Notice that the physical distance is 
obtained by multiplying $dl$ by the cosmological radius $a(t)$. 
We would like to consider the cosmological evolution characterized by the following time dependence of the 
Hubble variable $h(t)$, where, I recall, ``dot'' denotes the differentiation with respect to 
the cosmic time $t$\footnote{Cfr. equation \eqref{2:Hubble-simple}.}:
\begin{equation}
h(t) = \frac{t_R}{3t(t_R-t)}\ .
\label{m:BtR}
\end{equation}
I called this kind of scenario Bang-to-Rip in section \ref{sec:BtR}: at small values of $t$ the universe expands 
according to power law: $a \sim t^{1/3}$ while at $t \rightarrow t_R$ the Hubble variable explodes and one encounters 
the typical Big Rip type singularity. (The factor one third in (\ref{m:BtR}) was chosen for calculation simplicity). 

In the paper we are considering \cite{mio_magnetic} we kind of continued the work presented in the preceding chapter, or equivalently in \cite{mio_2field}.
Specifically we considered precisely the two models of the preceding chapter (see sections \ref{sec:model1}, \ref{sec:model2}) and coupled them with a magnetic field.

We have seen in chapter \ref{sec:2field} that, analyzing the Friedmann equation~\eqref{fri3}\footnote{I use here, in accordance with \cite{mio_2field,mio_magnetic}, the following system of units $\hbar = 1, c =1$ and 
$8\pi G = 3$. In this system the Planck mass $m_P$, the Planck length $l_P$ and the Planck time $t_P$ are equal to $1$. Then when we need to make the transition to the ``normal'', say, cgs units, we should simply express the Planck units in terms of the cgs units.
In all that follows we tacitly assume that all our units are normalized by the proper Planck units. Thus, the scalar field entering 
as an argument into the dimensionless expressions should be divided by the factor $\sqrt{m_P/t_p}$.},
for two-field models there is huge variety of potentials $V(\phi,\xi)$ 
realizing a given evolution, in contrast to models with one scalar field. Moreover, besides the freedom in the choice of the potential, one can choose different dynamics of the fields 
$\phi(t)$  and $\xi(t)$ realizing the given evolution. 
For simplicity I repeat here the main ingredients of the two models of \ref{sec:model1}, \ref{sec:model2}:  
The first potential is (cfr. with \eqref{2:poten6})
\begin{equation}
V_{I}(\xi,\phi) = \frac{2}{9t_R^2}\cosh^6(-3\phi/4)\exp(3\sqrt{2}\xi)\ .
\label{m:V-I}
\end{equation}
and the fields
\begin{equation}
\phi(t) = -\frac43 {\rm arctanh} \sqrt{\frac{t_R-t}{t_R}}\ ,
\label{m:phi1}
\end{equation}
\begin{equation}
\xi(t) = \frac{\sqrt{2}}{3} \ln \frac{t}{t_R-t}\ .
\label{m:xi1}
\end{equation}
(The expression for the potential should be multiplied by the factor $m_P/(c l_P)$, while the expressions for the fields $\phi(t)$ and $\xi(t)$ should be multiplied by $\sqrt{m_P/t_P}$. For the relation between Planck units and cgs ones see e.g. \cite{Mukhanov-book}).   
If we would like to substitute one of these  two fields by the pseudo-scalar field, conserving the correct parity of 
the potential, we can choose only the field $\phi$ because the potential $V_{I}$ is even with respect to $\phi$, but not with respect to $\xi$. In what follows I shall call the model with the potential (\ref{m:V-I}), the pseudo-scalar 
field (\ref{m:phi1})  and the scalar phantom (\ref{m:xi1}) model~$I$.  

Consider another potential (cfr. with \eqref{2:V-2new})
\begin{equation}
V_{II}(\xi,\phi) = \frac{2}{9t_R^2}\sinh^2(3\xi/4)\cosh^2(3\xi/4)\exp(-3\sqrt{2}\phi)\ .
\label{m:V-II}
\end{equation}
with the fields 
\begin{equation}
\phi(t) = \frac{\sqrt{2}}{3} \ln \frac{t}{t_R-t}\ ,
\label{m:phi-new3}
\end{equation}
\begin{equation}
\xi(t) = \frac{4}{3} {\rm arctanh} \sqrt{\frac{t}{t_R}}\ .
\label{m:xi-new3}
\end{equation}
This potential is even with respect to  the field  $\xi$. Hence the model $II$ is based on the potential
(\ref{m:V-II}), the pseudo-phantom field (\ref{m:xi-new3}) and the scalar field (\ref{m:phi-new3}). 
They will be the fields with the negative parity which couple to the magnetic field.

\section[Interaction between magnetic and pseudo-scalar/phantom field]{Post-inflationary evolution of a magnetic field  interacting with a pseudo-scalar or pseudo-phantom fields}\label{sec:m2}
The action of an electromagnetic field interacting with a pseudo-scalar or pseudo-phantom field $\phi$ is
\begin{equation}
S = -\frac14\int d^4x \sqrt{-g} ( F_{\mu\nu}F^{\mu\nu} + \alpha \phi F_{\mu\nu} \tilde{F}^{\mu\nu})\ ,
\label{m:action} 
\end{equation}
where $\alpha$ is an interaction constant and the dual electromagnetic tensor $\tilde{F}^{\mu\nu}$ is defined as
\begin{equation}
\tilde{F}^{\mu\nu} \equiv \frac12 E^{\mu\nu\rho\sigma}F_{\rho\sigma}\ ,
\label{m:dual}
\end{equation}
where
\begin{equation}
E_{\mu\nu\rho\sigma} \equiv \sqrt{-g}\ \epsilon_{\mu\nu\rho\sigma}\ ,\quad 
E^{\mu\nu\rho\sigma} \equiv -\frac{1}{\sqrt{-g}}\ \epsilon^{\mu\nu\rho\sigma}\ , 
\label{m:dual1}
\end{equation}
with the standard Levi-Civita symbol
\begin{equation}
\epsilon_{\mu\nu\rho\sigma} = \epsilon_{[\mu\nu\rho\sigma]},\ \epsilon_{0123} = +1\ .
\label{m:LC}
\end{equation}

Variating the action (\ref{m:action}) with respect to the field $A_{\mu}$ we obtain the field equations
\begin{equation}
\nabla_{\mu}F^{\mu\nu} = -\alpha\partial_{\mu}\phi\tilde{F}^{\mu\nu}\ ,
\label{m:Maxwell}
\end{equation}
\begin{equation}
\nabla_{\mu}\tilde{F}^{\mu\nu} = 0\ .
\label{m:Maxwell1}
\end{equation}
The Klein-Gordon equation for the pseudo-scalar field is 
\begin{equation}
\nabla^{\mu}\nabla_{\mu}\phi + \frac{\partial V}{\partial \phi} = - \alpha F_{\mu\nu} \tilde{F}^{\mu\nu}\ .
\label{m:KG} 
\end{equation}
The Klein-Gordon equation for the pseudo-phantom field (which is the one that couples with the magnetic field in the model $II$) differs from equation (\ref{m:KG}) by change of sign in front of the kinetic term. 
In what follows we shall neglect the influence of magnetic fields on the cosmological evolution, i.e. we will discard the electromagnetic coupling in equation (\ref{m:KG}). 

If one wants to rewrite these formul\ae\  in terms of the three-dimensional quantities (i.e. the electric and magnetic fields) one can find the expression of the electromagnetic tensor in a generic curved background, starting from a locally flat reference frame --- in which it is well known the relation between electromagnetic fields and $F$ --- and using a coordinate transformation. It is easy to see that we have, for the metric (\eqref{FRWline}):
\begin{equation}
 F^{\mu\nu}=\frac{1}{a^2}\left(\begin{array}{cccc}0 & -aE_1 & -aE_2 & -aE_3\\
             aE_1 & 0 & B_3 & -B_2 \\
             aE_2 & -B_3 & 0 & B_1 \\
             aE_1 & B_2 & -B_1 & 0     
            \end{array}\right)\ .
\end{equation}
The field equations (\ref{m:Maxwell}), (\ref{m:Maxwell1}) and (\ref{m:KG}) rewritten in terms of $\vec{E}$ and $\vec{B}$ become
\begin{subequations}\begin{equation}
\vec{\nabla}\cdot\vec{E} = - \alpha \vec{\nabla}\phi\cdot\vec{B}\ .
\label{m:Maxwell2}
\end{equation}
\begin{equation}
\partial_0(a^2\vec{E}) - \vec{\nabla}\times(a\vec{B}) = - \alpha[\partial_0\phi(a^2\vec{B}) - \vec{\nabla}\phi\times(a\vec{E})]\ ,
\label{m:Maxwell3}
\end{equation}
\begin{equation}
\partial_0(a^2\vec{B}) - \vec{\nabla}\times(a\vec{E}) = 0\ ,
\label{m:Maxwell4}
\end{equation}
\begin{equation}
\vec{\nabla}\cdot\vec{B} = 0\ .
\label{m:Maxwell5}
\end{equation}\end{subequations}

For a spatially homogeneous pseudo-scalar field  equations (\ref{m:Maxwell2}) and (\ref{m:Maxwell3}) look like
\begin{subequations}\begin{equation}
\vec{\nabla}\cdot\vec{E} = 0\ ,
\label{m:Maxwell6}
\end{equation}
\begin{equation}
\partial_0(a^2\vec{E}) - \vec{\nabla}\times(a\vec{B}) = -\alpha\partial_0\phi(a^2\vec{B})\ .
\label{m:Maxwell7}
\end{equation}\end{subequations}
Taking the curl of (\ref{m:Maxwell7}) and substituting into it 
the value of  $\vec{E}$ from (\ref{m:Maxwell4}) we obtain
\begin{equation}
\partial^2_0(a^2\vec{B})+h(t)\partial_0(a^2\vec{B})-\frac{\Delta^{(3)}(a^2\vec{B})}{a^2}-\frac{\alpha}{a}{\partial_0\phi\vec{\nabla}\times(a^2\vec{B})}=0\ ,
\label{m:Maxwell8}
\end{equation}
where $\Delta^{(3)}$ stands for the three-dimensional Euclidean Laplacian operator.

Let me introduce 
\begin{equation}
\vec{F}(\vec{x},t) \equiv a^2(t)\vec{B}(\vec{x},t)
\label{m:F-def}
\end{equation}
and its Fourier transform
\begin{equation}
\vec{F}(\vec{k},t) = \frac{1}{(2\pi)^{3/2}} \int e^{-i\vec{k}\cdot\vec{x}} \vec{F}(\vec{x},t)d^3x\ .
\label{m:Fourier}
\end{equation}
Here the field $\vec{B}$ is an observable magnetic field entering into the expression for the Lorentz force.  
The field equation for  $\vec{F}(\vec{k},t)$ is 
\begin{equation}
 \ddot{\vec{F}}(\vec{k},t)+h(t)\dot{\vec{F}}(\vec{k},t)+\left[\left(\frac{k}{a}\right)^2-\frac{i\alpha}{a}\dot{\phi}\vec{k}
\times\right]
\vec{F}(\vec{k},t)=0\ ,
\label{m:Max-Four}
\end{equation}
where ``dot'' means the time derivative.
This last equation can be further simplified: assuming $\vec{k}=(k,0,0)$ and defining the functions $F_\pm\equiv (F_2\pm iF_3)/\sqrt{2}$ one arrives to
\begin{equation}
 \ddot{F}_{\pm} + h\dot{F}_{\pm}  +\left[\left(\frac{k}{a}\right)^2\pm\alpha\frac{k}{a}\dot{\phi}\right]F_{\pm}=0\ ,
\label{m:Max-Four1}
\end{equation}
where I have omitted the arguments $k$ and $t$.  

Assuming that the electromagnetic field has a quantum origin (as all the fields in the cosmology of the early universe 
\cite{initial})  the modes of this field are represented by harmonic oscillators. Considering their vacuum 
fluctuations responsible for their birth we can neglect the small breakdown of the conformal symmetry and treat them as free.
In conformal coordinates $(\eta,\vec{x})$ such that the Friedmann metric has the form 
\begin{equation}
ds^2 = a^2(\eta)(d\eta^2 - dl^2)\ ,
\label{m:conformal}
\end{equation}
the electromagnetic potential $A_i$ with the gauge choice $A_0=0$, $\partial_jA^j=0$ satisfies the standard harmonic oscillator equation of motion
 \begin{equation}
 \ddot{A}_i + k^2 A_i = 0\ .
\label{m:harmonic}
 \end{equation}
Hence the initial amplitude of the field $A_i$ behaves as $A_i = 1/\sqrt{2k}$, while the initial amplitude of the 
 functions $F$ is $\sqrt{k/2}$. The evolution of the field $F$ during the inflationary period was described in 
\cite{Sorbo}, where it was shown that the growing solution at the end of inflation is amplified by some factor depending 
on the intensity of the interaction between the pseudo-scalar field and magnetic field. 

Here we are interested in the evolution of the magnetic field interacting with the pseudo-scalar field after inflation, 
where our hypothetic  Bang-to-Rip scenario takes place. More precisely, I would like to see how different types 
of scalar-pseudo-scalar potentials and field dynamics providing the same cosmological evolution could be distinguished by their influence 
on the evolution of the magnetic field. We do not take into account the breaking of the conformal invariance during 
the inflationary stage 
and all the effects connected with this breakdown will be revealed only after the end of inflation and the beginning of 
the Bang-to-Rip evolution. This beginning is such that  the value of the Hubble parameter, 
characterizing this evolution is equal to that of the inflation, i.e. 
\begin{equation}
h(t_0) = \frac{t_R}{3t_0(t_R-t_0)} = h_{\rm inflation} \simeq 10^{33} {\rm s}^{-1}\ .
\label{m:beginning}
\end{equation}
In turn, this implies that we begin evolution at the time moment of the order of $10^{-33}{\rm s}$.

I shall consider both the components $F_{+}$ and $F_{-}$ and we shall dwell on the scenarios $I$ and $II$ 
described in the preceding section~\ref{sec:m1}. 
Anyway, our assumption regarding the initial conditions for equation (\ref{m:Max-Four1}) can be easily modified in order to 
account for the previous possible amplification of primordial magnetic fields as was discussed by \cite{Sorbo}. 
Thus, all estimates for the numerical values of the magnetic fields in today's universe  should be multiplied by some 
factor corresponding to the amplification of the magnetic field during the inflationary stage.  Hence, our results refer more to 
differences between various models of a post inflationary evolution than of the real present values of magnetic fields, whose amplification might be also combined effect of different mechanism  \cite{magnetic1,magnetic2,magnetic3}, \cite{break1}-\cite{break5}, \cite{quantum1,quantum2,quantum3}, \cite{Sorbo}.

\section{Generation of magnetic fields: numerical results}\label{sec:m3}

In this section I present the results of numerical simulations for the two models $I$ and $II$.
In these models, the equations of motion for the modes $F_\pm(k,t)$ (\ref{m:Max-Four1}) reads\footnote{The reader can easily verify that this equation is obtained imposing the normalization $a|_{\rm today} = 1$, where today-time is taken to be near the crossing of the phantom divide line, i.e. at $t\simeq t_R/2$. This implies in turn that at the beginning of the ``Bang-to-Rip'' evolution the cosmological radius is $a(t_0) \simeq 10^{-17}$.}:
\begin{equation}
 \ddot{F}_{\pm} + \frac{t_R\ \dot{F}_{\pm}}{3t(t_R-t)} +
 \left[k^2\left(\frac{t_R-t}{t}\right)^{2/3}\pm\alpha\left(\frac{t_R-t}{t}\right)^{1/3} k\dot{\Phi}\right]F_{\pm}=0\ ,
\label{m:edm}
\end{equation}
where $\Phi$ stands for the scalar field $\phi$ in the model~$I$ and for the phantom $\xi$ in the model~$II$, so that
\begin{equation}
\dot\Phi_I=\dot\phi= \frac23 \frac{\sqrt{t_R}}{t \sqrt{t_R - t}};\quad \dot\Phi_{II}=\dot\xi= \frac23 \frac{\sqrt{t_R}}{\sqrt{t} (t_R - t)}\ .
\label{m:dots}
\end{equation}

Equation (\ref{m:edm}) is solved for different values of the wave number $k$ and the coupling parameter $\alpha$.  
(The parameter $\alpha$ has the dimensionality inverse with respect to that of the scalar field; the wave number $k$ has the dimensionality of inverse length; the time $t_R = 10^{17}{\rm s}$).  Qualitatively let me remark that in (\ref{m:Max-Four1}) the coupling term influence becomes negligible after some critical 
period. After that the magnetic fields in our different scenarios evolve as if the parameter $\alpha$ in  (\ref{m:Max-Four1}) 
had been put equal to zero. Indeed, it can be easily seen that the interaction term vanishes with the growth of the cosmological radius $a$. Then the distinction between the two models is to be searched in the early time behavior of the field evolution. 

Noting that in both our models $I$ and $II$ the time derivative $\dot{\Phi}$ is positive, by inspection of the linear term in equation (\ref{m:Max-Four1}) we expect the amplification 
to be mainly given for the mode $F_{-}$  provided the positive sign for $\alpha$ is chosen; so we will restrict our attention on $F_{-}$.
We can also argue that the relative strength of the last two terms in the left-hand side of (\ref{m:edm})  is crucial for determining the behavior of the solution: when the coupling term prevails (I remark that we are talking about $F_-$ so this term is \emph{negative} in our models) then we expect an amplification, while when the first term dominates we expect an oscillatory behavior.
For future reference it is convenient to define
\begin{equation}
{\bf A}(t;\Phi)\equiv \frac{k}{a(t)\alpha\dot{\Phi}}=\frac{k}{\alpha\dot\Phi}\left(\frac{t_R-t}{t}\right)^{1/3}\ ,
\label{m:A}
\end{equation}
which is just the ratio between the last two terms in the left-hand side of equation~(\ref{m:edm}).

Indeed our numerical simulations confirm these predictions.
Let us consider the model $I$ with $\alpha=1(t_P/m_P)^{1/2}$ and $k=10^{-55}l_P^{-1}$, where $l_P$ is the Planck length. Such a value of the wave 
number $k$ corresponds to the wave length of $1 kpc$ at the present moment.  
We obtain an early-time amplification of about 2 orders of magnitude, with the subsequent oscillatory decay. Notice that the parameter ${\bf A}$ in this model at the beginning is very 
small: this corresponds to the dominance of the term proportional to $\dot{\Phi}$ and, hence, to the amplification of the field 
$F_{-}$. At the time scale of the order of $10^{51} t_P$, where $t_P$ is the Planck time, this regime turns to that with big values of ${\bf A}$ where the 
influence of the term  proportional to $\dot{\Phi}$ is negligible.   

For the same choice of the parameters $\alpha$ and $k$ in the model $II$  
the amplification is absent. 

\begin{figure}[htp]\centering
\includegraphics{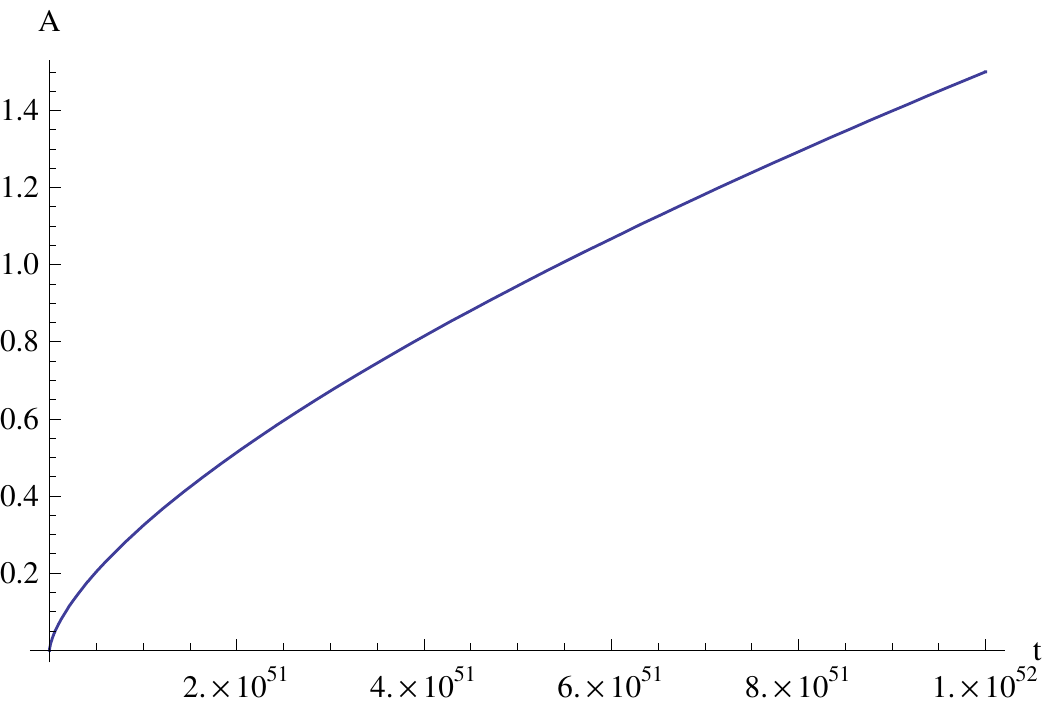}
\caption{Plot of the ratio ${\bf A}$ for the model $I$, with the parameter choice $\alpha=1 (t_P/m_P)^{1/2}$, $k=10^{-55}l_P^{-1}$. It can be easily seen that at a time scale of order $10^{51} t_P$ the ratio becomes greater than 1.}\label{mag1}
\end{figure}

\begin{figure}[htp]\centering

\includegraphics{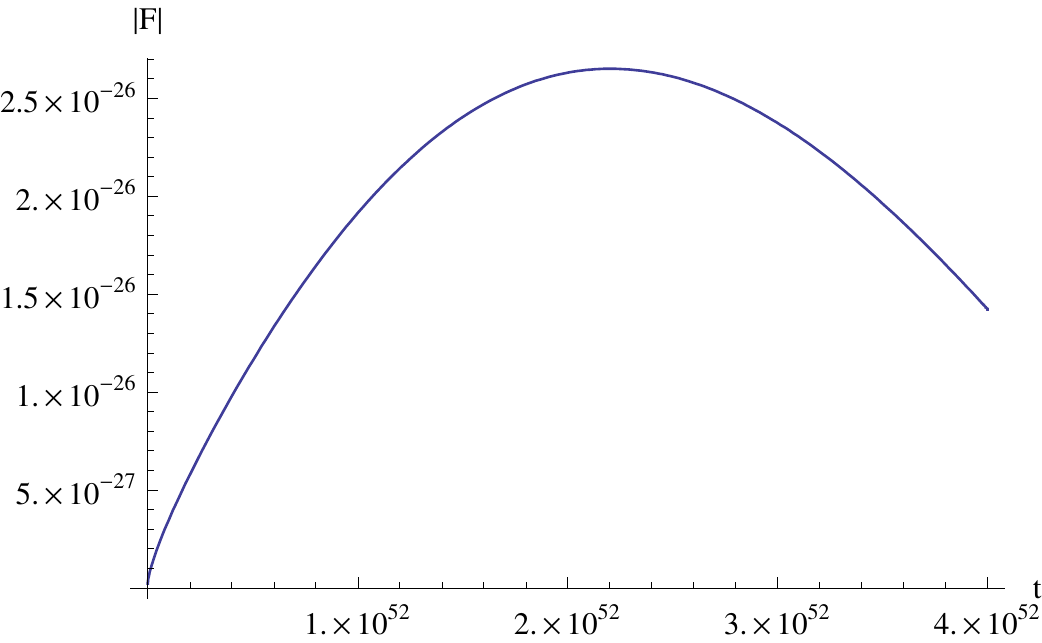}

\caption{Plot of the time evolution of the absolute value of the complex field $F$ (given in Planck units) in model $I$ with the parameter choice $\alpha=1 (t_P/m_P)^{1/2}$, $k=10^{-55}l_P^{-1}$. The behavior, as said above in the text, consists in an  amplification till a time of order $10^{51} t_P$, after which the oscillations begin.} \label{mag2}
\end{figure}

\begin{figure}[htp]\centering

\includegraphics{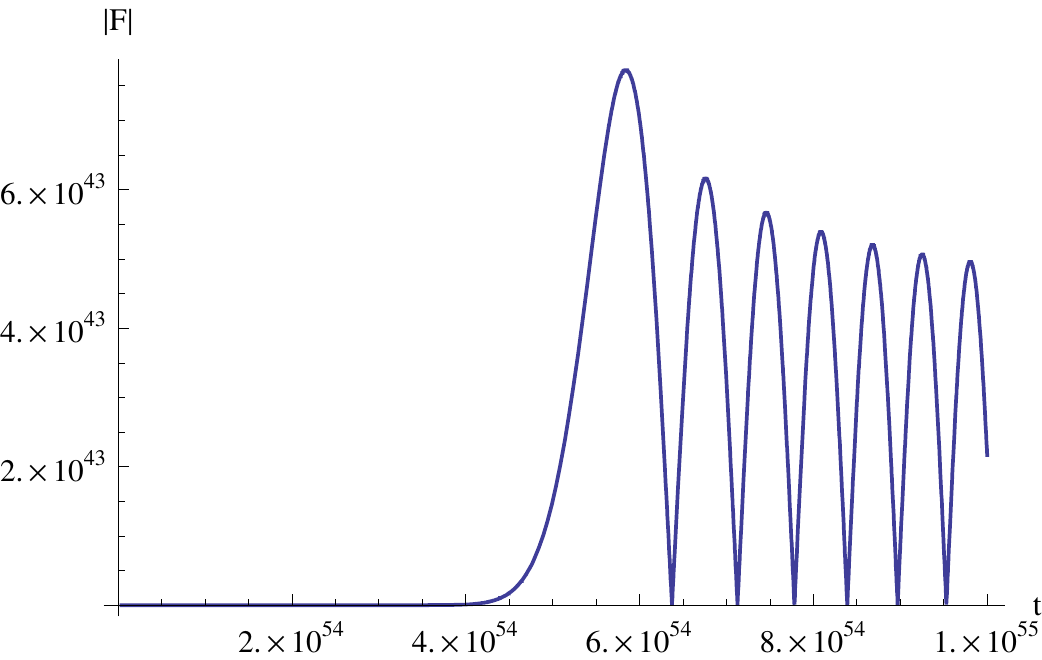}

\caption{Plot of the time evolution of the absolute value of the complex field $F$ (given in Planck units) in model $I$ with the parameter choice $\alpha=100 (t_P/m_P)^{1/2}$, $k=10^{-55}l_P^{-1}$. The behavior, consists in an  amplification till a time of order $10^{54} t_P$, after which the oscillations begin.} \label{mag3}
\end{figure}

In figure~\ref{mag1}  you see the time dependence of the 
function ${\bf A}$ for the model $I$  for the values of $\alpha=1 (t_P/m_P)^{1/2}$ and $k=10^{-55}l_P^{-1}$ chosen above.
The figure~\ref{mag2} manifests the amplification of the magnetic field in the model~$I$. 

Naturally the effect of amplification of the magnetic field grows with the  coupling constant $\alpha$ and diminishes when the  wave number $k$ increases. In figure~\ref{mag3} are displayed the results for the case of $\alpha = 100 (t_P/m_P)^{1/2}$, which is admittedly extreme and possibly non realistic, but good for illustrative purposes. Here the 
amplification is more evident and extends for a longer time period. 
 
In \cite{mio_magnetic}, we also tried to make some estimates of the cosmic magnetic fields in the universe today, using the correlation functions.
The correlation function for the variable $F$ is defined as the quantum vacuum average  
\begin{equation}
G_{ij}(t,\vec{x}-\vec{y}) = \langle 0| F_i(t,\vec{x}) F_j(t,\vec{y})|0\rangle
\label{m:corr}
\end{equation}
and can be rewritten as 
\begin{equation}
G_{ij}(t,\vec{x}-\vec{y}) = \int \frac{d^3k}{(2\pi)^3} e^{-i\vec{k}\cdot (\vec{x}-\vec{y})} F_i(t,\vec{k})F_j^*(t,\vec{k}).
\label{m:corr1}
\end{equation}
Integrating over the angles, we come to 
\begin{equation}
G_{ij}(t,\vec{x}-\vec{y}) =  \int \frac{k^2 dk}{2\pi^2} \frac{\sin(k|\vec{x}-\vec{y}|)}{k|\vec{x}-\vec{y}|}F_i(t,k)F_j^*(t,k).
\label{m:corr2}
\end{equation}
To estimate the integral (\ref{m:corr2}) we notice that the main contribution to it comes from the region where $k \approx 1/|\vec{x}-\vec{y}|$ (see e.g.\cite{Mukhanov-book}) and it is of order
\begin{equation}
\frac{1}{L^3} |F_i(t,1/L)|^2\ ,
\label{m:corr3}
\end{equation}
where $L = |\vec{x}-\vec{y}|$.
In this estimation the amplification factor is 
\begin{equation}
\left|\frac{F(1/L)}{\sqrt{1/L}}\right|\ ,
\label{m:amplific}
\end{equation}
where the subscript $i$ is not present since we have taken the trace over polarizations.

Now we are in a position to give numerical values for the magnetic fields at different scales in the model $I$ for different values 
of the coupling parameter $\alpha$. These values (see Table~\ref{tab}) correspond to three values of the coupling parameter $\alpha$ (1,10 and 100 $(t_P/m_P)^{1/2}$) \footnote{It is useful to remark that at this length scales values of $\alpha$ less than 1 make the coupling with the cosmological evolution negligible.} and 
to two spatial scales $L$ determined by the values of the wave number $k$. We did not impose some physical restrictions on the value 
of $\alpha$. It is easy to see that the increase of $\alpha$ implies the growth of the value of the magnetic field $B$.  

Let me stress once again that we ignored the effects of amplification of the magnetic fields 
during inflation to focus on seizable effects during evolution. 

\begin{table}[h]
\caption{The Table displays the values of the magnetic field $B$ \\corresponding to the chosen values of $\alpha$ and $k$.
The length $L$\\ refers to the present moment when $a = 1$.}
\renewcommand{\arraystretch}{1.5}
\begin{tabular}{@{}l|ccc}

&$\alpha=1 \sqrt{t_P/m_P}$&$\alpha= 10 \sqrt{t_P/m_P}$ &$\alpha = 100 \sqrt{t_P/m_P}$\\ \hline

&&&\\
$k=10^{-55}l_P^{-1}\ (L=1kpc)$&$B \sim 10^{-67} G$&$B \sim	 10^{-60} G$&$B \sim 100 G$\\ 
&&&\\
$k=10^{-50}l_P^{-1}\ (L = 10^{-2}pc)$&$B\sim10^{-55} G$&$B \sim 10^{-49} G$&$B \sim 10^{13} G$	
\end{tabular}
\label{tab}
\end{table}

Finally notice that our quantum ``initial'' conditions correspond to physical magnetic fields which for presented values of $k$ are 
$B_{in} \sim k^2/a^2$ is equal to $10^{-34} G$ for $k = 10^{-55}l_P^{-1}$ and $B_{in} \sim 10^{-24} G$ for 
$k = 10^{-50}l_P^{-1}$.

\section{Conclusions}\label{sec:m4}
We have seen that the evolution of the cosmic magnetic  fields interacting with a pseudo-scalar (pseudo-phantom) field 
is quite sensitive to the concrete form of the dynamics of this field in two-field models
where different scalar field dynamics and potentials realize the same cosmological evolution. 

The sensitivity of the evolution of the magnetic field with respect to its helicity is confirmed,   
given the sign of the coupling constant $\alpha$ and that the $\Phi$ is a monotonic function of time 
(as it is really so in our models).
We gave also some numerical estimates of the actual magnetic fields up to the factor of amplification 
of such fields during the inflationary period. The toy model of the Bang-to-Rip evolution studied in this paper \cite{mio_magnetic},
cannot  be regarded as the only responsible for the amplification of cosmic magnetic fields 
implying their present observable values.
It rather complements some other mechanisms acting before. 
However, the difference  between cosmic magnetic fields arising  in various models 
(giving the same expansion law after the inflation) is essential. 
It may provide a discriminating test for such models.
\chapter{Phantom without phantom in a PT-symmetric background}\label{sec:PT}

{\it This chapter is devoted to presenting the content of the paper \cite{mio_PT}. \emph{In nuce}, the idea is that if we put ourselves in the framework of $PT$~symmetric Quantum Theory we can have an \emph{effective} phantom field which is stable to quantum fluctuations. ``Effective'' in the usual sense, i.e. the real field is a standard scalar field, but becomes  a phantom field once we `sit' on a (specific) classical solution.

Before going into the details of the model presented in \cite{mio_PT}, let me briefly introduce the key ideas and concepts of $PT$~symmetric Quantum Mechanics.}

\section{PT-symmetric Quantum Mechanics: a brief introduction}

$PT$~symmetric Quantum Mechanics was born in 1998 in the seminal paper \cite{PTstart} by Carl Bender and Stefan Boettcher. The motivating idea is very simple: how to make sense of non-hermitian hamiltonians. Indeed it had been  known for long (since late 50's) that some non-hermitian hamiltonians come out naturally in some systems \cite{non-hermit1}-\cite{non-hermit5}, but everybody where rather skeptical about their physical sense.

However, since the 80's, there where hints in the following direction: some non-hermitian hamiltonians do have a real and bounded spectrum \cite{PT1,PT2}. You guess the point: Bender and collaborators rigorously proved since the very beginning of the theory in 1998 that it is not necessary to require hermiticity in order to have a real and bounded spectrum. We can weaken that axiom of QM with the more physical requirement: the hamiltonian operator must be $PT$~symmetric.

A $PT$ transformation is a combined transformation made of parity
\be
P\left\{\begin{aligned}x&\rightarrow -x\\
        p&\rightarrow -p
       \end{aligned}\right.
\ee
and time reversal
\be
T\left\{\begin{aligned}
        x&\rightarrow x\\
	p&\rightarrow -p\\
	i&\rightarrow -i
       \end{aligned}\right.
\ee
thus
\be
PT\left\{\begin{aligned}
                    x&\rightarrow -x\\
		    p&\rightarrow p\\
		    i&\rightarrow -i
                   \end{aligned}\right.
\ee
we can think of it as a spacetime reflection.

$PT$~symmetric hamiltonians do have real and bounded spectrum, thus they have the usual physical interpretation  as operators whose eigenvalues are the energy levels of the system under investigation\footnote{Actually it is only a sub-class of $PT$~symmetric operators that have this property. This sub-class is said to have an \emph{un-broken} $PT$~symmetry. See \cite{PTreview1,PTreview2} for more details.}.

As Gell-Mann said (and as QM teaches us) ``everything that is not forbidden is compulsory''. This is the sort of principle that guided -- together with the \emph{need} of dealing with non-hermitian hamiltonians -- the flourishing of the $PT$~symmetric approach in the last decade or so.

Remarkably Bender and collaborators proved also another (even more) fundamental result \cite{PTunitarity}: in $PT$~symmetric QM it is possible to have \emph{unitary} evolution. This is not at all a trivial result. Indeed the self-adjointness of the hamiltonian is `responsible' not only for the reality of the energy levels, but also -- and I dare say most importantly -- for the unitarity of time evolution, i.e. the conservation of probability in time. Indeed in standard QM unitarity is completely due to the hermeticity of the hamiltonian\footnote{I am using here hermeticity and self adjointness as synonimous. Strictly speaking, they are not, because of domain subtleties.}: the evolution operator is $U(t)=e^{-iHt}$ and is unitary as long as $H=H\dag$. I am not going in details in this respect, I only say that it has been proved that it can be defined a proper inner product with respect to which $PT$~symmetric QM does have a unitary evolution \cite{PTunitarity,PTunitarity1,PTunitarity2}.

\begin{rem}One key point must be stressed: non-hermitian hamiltonians classically mean \emph{complex} forces. This implies that we have to consider \emph{complex dynamical variables} $x$ and $p$. This is completely acceptable in the quantum world, once you understand that this only means that the position and momentum operators \emph{are not observables} in $PT$~symmetric quantum theory. Observables \emph{must} have real spectrum, this has to be true indeed in any theory.
 
However, this is a problem in the corresponding classical limit. One has to include somehow complex valued trajectories. Here, I must admit, I see a drawback of this approach. Frankly, I am not aware of any reasonable interpretation about it, and it seems to me that in the literature this point is quite often underestimated. 
\end{rem}

One more thing to say is that, at present, no empirical clear and definitive evidence of the existence of systems with complex hamiltonian has been found.

It is far beyond the scope of this thesis to discuss the details of this approach, let alone the many insights that have been reached in these years in many respects.
However, in the following section I present a class of very well studied and archetypal complex $PT$~symmetric hamiltonians, which I find particularly instructive and also very useful to understand the mechanism that plays a central role in the model studied in~\cite{mio_PT}.

\section{PT-symmetric oscillator}\label{sec:PTosc}
Take the hamiltonian
\be
H_\epsilon=p^2+x^2(ix)^\epsilon\ ,
\label{PTosc}
\ee
this is complex for $\epsilon\neq 0$ and $PT$~symmetric\footnote{For $\epsilon<0$ this hamiltonian is said to have a \emph{broken} $PT$~symmetry, and the reality of the spectrum is not guaranteed anymore~\cite{PTreview1,PTreview2}. Thus we restrict to positive (or null) value of $\epsilon$.}.
This is a generalization of the standard harmonic oscillator. We shall call this class the $PT$~symmetric oscillator, as is sometimes done in the literature.

I will review here mainly the \emph{classical} analysis of the hamiltonians \eqref{PTosc}, i.e. the study of the trajectories of a point particle subject to the complex potential $x^2(ix)^\epsilon$. Notice that the richness of the model arises precisely from considering it in the complex $x$ plane, as it is natural to do since we have complex forces.
The equation of motion is
\be
\f{\d^2x}{\d t^2}=2i(2+\epsilon)(ix)^{1+\epsilon}\ ,
\ee
that can be integrated, e.g. using the energy first integral $H=E$, to give
\be
\f12\f{\d x}{\d t}=\pm\sqrt{E+(ix)^{2+\epsilon}}\ ,
\ee
where $E$ is the constant value of the energy of the particle. We set for simplicity $E=1$; obviously this does not spoil the generality of the following considerations.
Let us analyze separately the cases $\epsilon=0,1,2$.

\subsection*{$\epsilon=0$}
\begin{figure}\centering
 \includegraphics[scale=1]{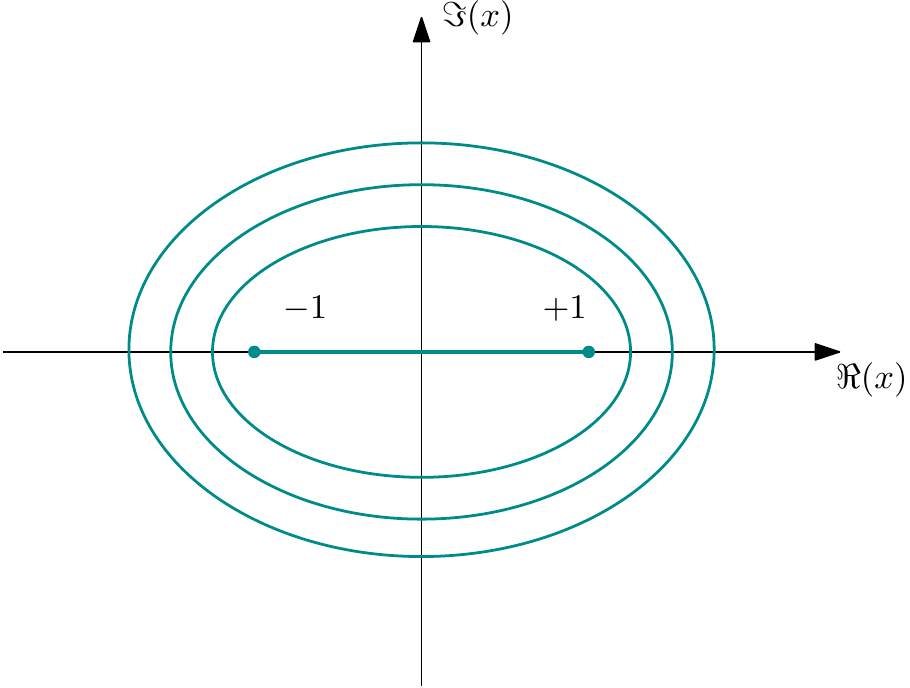}
\caption{Qualitative plot of the trajectories in the complex $x$ plane of the system with hamiltonian \eqref{PTosc} with $\epsilon=0$.}\label{fig:e0}
\end{figure}

This is the standard harmonic oscillator. It has two turning points $x=-1,+1$, both on the real axis. The trajectory depends on the choice of $x(0)$. Here we can choose an initial condition wherever we want on the complex $x$ plane. The solutions are all periodic of period $2\pi$, and are all oscillating. If we start on the real axis, then -- as we know from the standard oscillator -- we remain on the real axis, otherwise we oscillate on ellipses on the complex plane (see figure \ref{fig:e0}).
Notice that the trajectories are $P$ invariant (reflection with respect to the origin) $T$ invariant (reflection with respect to the real axis) and $PT$ invariant (reflection about the imaginary axis).

\subsection*{$\epsilon=1$}
\begin{figure}\centering
 \includegraphics[scale=1]{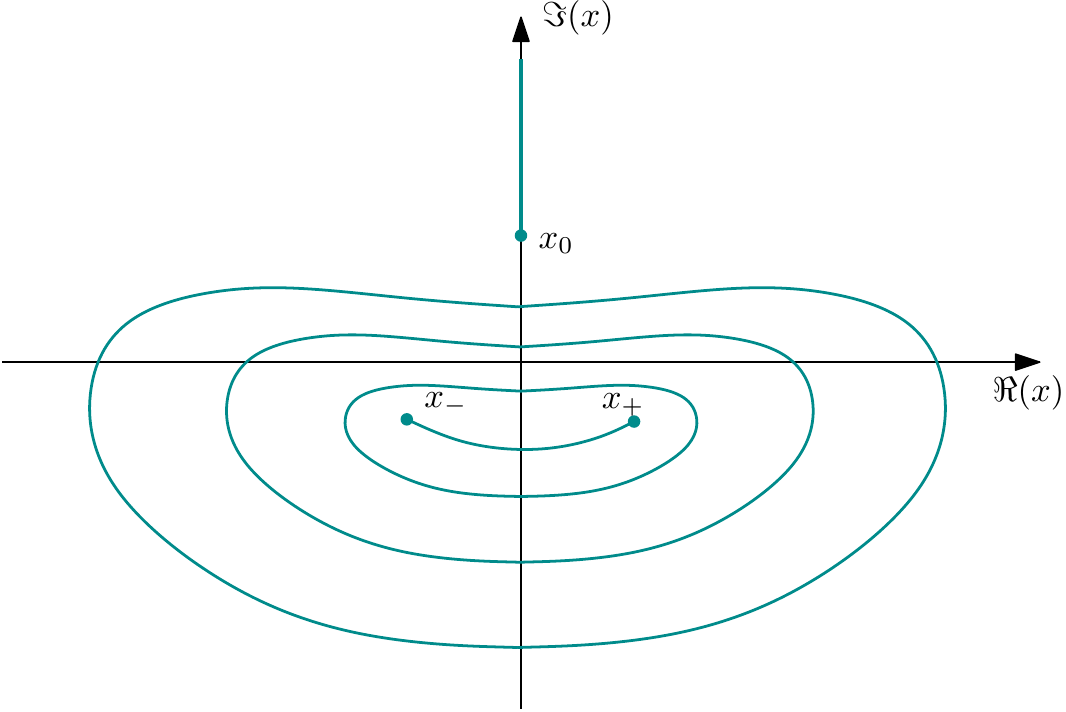}
\caption{Qualitative plot of the trajectories for the case $\epsilon=1$.}\label{fig:e1}
\end{figure}

In this case there are three turning points, namely the solution of the equation
\be
ix^3=1\ ,
\ee
i.e. $x_-=e^{-5i\pi/6}$, $x_+=e^{-i\pi/6}$ and $x_0=i$ (see figure \ref{fig:e1}). In this case we have two classes of trajectories: periodic and non-periodic. Actually the non-periodic trajectories are a null-measure subset of all the trajectories. Namely, trajectories starting in $x(0)=i\tl{x}$ with $\tl{x}$ real and greater than or equal to 1, goes up to to infinity on the imaginary axis. In other words, trajectories starting at the $x_0$ turning point or somewhere upper but always on the imaginary axis, are not periodic. All the other are periodic and have the same period, by virtue of  Cauchy's theorem. Notice that the trajectories are manifestly $PT$-invariant, but $P$ and $T$ symmetries taken separately are lost.

\subsection*{$\epsilon=2$}
\begin{figure}\centering
 \includegraphics[scale=1]{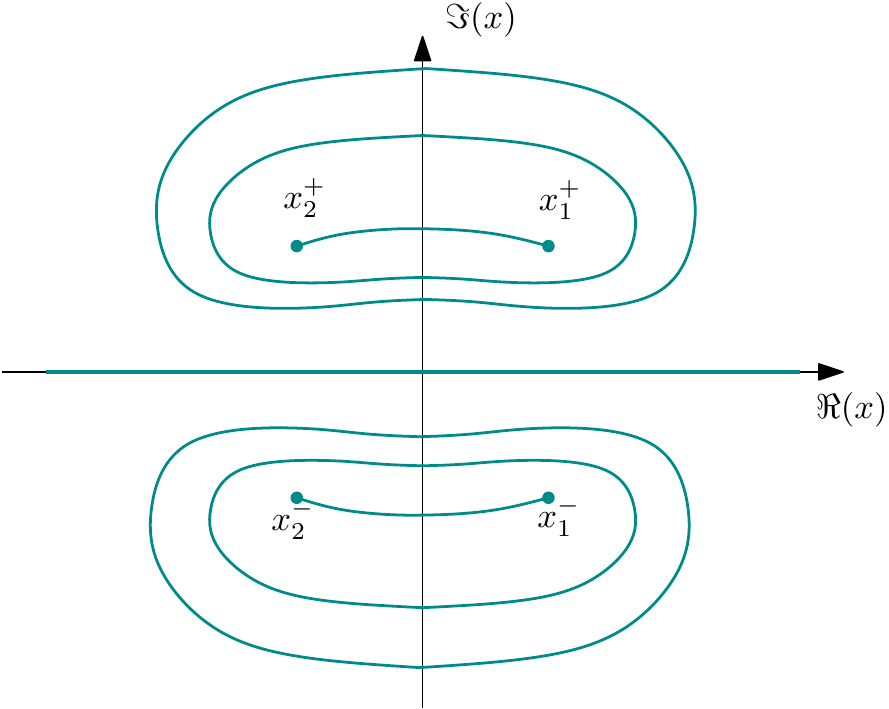}
\caption{Qualitative plot of the trajectories for the case $\epsilon=2$}\label{fig:e2}
\end{figure}

This is the most interesting case. Notice that the hamiltonian \eqref{PTosc} now reads
\be
H_2=p^2-x^4\ ,
\label{osc4}
\ee
which is actually a real but unbounded hamiltonian. Being it real we can analyze it on real trajectories, and we know that the system is completely unstable: the particle will roll down the maximum in $x=0$ going away at infinity. The energy is doomed to fall to $-\infty$. 
However, we can also expand our horizon and see what happens in the complex plane. There are four (complex) turning points, solutions of  $x^4=-1$
\be
x^{\pm}_1=e^{\pm i\pi/4}\ ,\quad x^{\pm}_2=e^{\pm 3i\pi/4}\ ,
\ee
two above and two under the real axis (see figure \ref{fig:e2}). The trajectories are here divided in three families: one encircling the top turning points $x_{1,2}^+$, one around the bottom turning points $x_{1,2}^-$ and one  stuck on the real axis -- again of null measure on the whole set. The latter is of course the one we talked about previously. Indeed it is unstable, in the sense that it is not periodic and goes to infinity. So, here you are: while on standard basis you look to hamiltonian \eqref{osc4} and think `unstable, good for nothing', in the complex plain you'd better  bet on the periodicity/stability of the same system described by \eqref{osc4}.

This last example shows the key point: a manifest instability can be `absorbed' by the `complexity'. Or better (but less suggestively), what is manifestly unstable in a ``real'' world, might not be so in a ``complex'' world.\\[14pt]

With these brief examples I end this introduction to $PT$~symmetric QM. Actually, this was more a presentation of the \emph{ideas} behind $PT$~symmetry rather than of the theory itself. Indeed there is a huge amount of work, both in the foundations of the theory and in more border-line issues. The interested reader is encouraged to look to the very good reviews by Bender \cite{PTreview1,PTreview2}.

\section{Introduction and resume of the paper~\texorpdfstring{\cite{mio_PT}}{on PT~cosmology}}
As I pointed out at the beginning of the chapter, complex (non-hermitian) hamiltonians with $PT$~symmetry have been vigorously investigated in Quantum Mechanics and Quantum Field Theory~\cite{PT1,PT2}. A possibility of applications to quantum cosmology has been pointed out in~\cite{PT-cosm}.  In~\cite{mio_PT}  we mainly focused attention on complex  field theory. We explored the use of a particular  complex scalar field lagrangian, whose  solutions of the classical equations of motion provide us with  real physical observables and well-defined geometric characteristics.

In the said paper we proposed a cosmological model inspired by $PT$~symmetric Quantum Theory, choosing
potentials so that the equations of motion have classical phantom solutions for homogeneous and isotropic
universe. Meanwhile quantum fluctuations have positive energy density and thus ensuring the stability around a
classical background configuration.

We considered the complex extension of matter lagrangians requiring the reality of all the physically measurable
quantities and the well-definiteness of geometrical characteristics. It is worthwhile  to underline here that we
considered only \emph{real} space-time manifolds. Attempts to use complex manifolds for studying the problem of dark
energy in cosmology can be found here~\cite{complex-man1,complex-man2,complex-man3}.

In~\cite{mio_PT}, we start with the model of two scalar fields with positive kinetic terms. The potential of the
model is additive. One term of the potential is real, while the other is complex and $PT$~symmetric. We find a
classical complex solution of the system of the two Klein-Gordon equations together with the Friedmann equation.
The solution for one field  is real (the ``normal'' field) while the solution for the other field is purely
imaginary, realizing classically the phantom behavior. Moreover, the effective lagrangian for the linear
perturbations has the correct potential signs for both the fields, so that the problem of stability does not
arise. However, the background (homogeneous Friedmann) dynamics is determined by an effective action including
two real fields one normal and one phantom. As a byproduct, we notice that the phantom phase in the cosmological
evolution is inevitably transient. The number of phantom divide line crossings, (i.e. events such that the
ratio $w$ between pressure and energy density passes through the value $-1$)  can be only even and the Big Rip
never occurs. The avoidance of Big Rip singularity  constitutes an essential difference between our model and
well-known quintom models, including one normal and one phantom fields\cite{two-exp1,two-exp2}. The other differences will
be discussed in more detail later.

What follows is so organized: in sections~\ref{sec:PT1} and \ref{sec:PT2}  the cosmological model we considered is described, together with
a brief explanation of the ``interplay'' between $PT$~symmetric Quantum Mechanics and two-field models; in section~\ref{sec:PT3} I present the results of the qualitative
analysis and of the numerical simulations for the dynamical system under consideration; conclusions and
perspectives are presented in the last section~\ref{sec:PT4}.

\section{Phantom and stability}\label{sec:PT1}

As I pointed out at the end of section~\ref{sec:rip}, the great problem of phantom fields is the quantum instability~\cite{phantomstability1,phantomstability2}. Their hamiltonian is unbounded from below, thus its quantum fluctuations grow exponentially. This should remind the reader of the $PT$~oscillator case~\ref{sec:PTosc}, which indeed I will again discuss, in a slightly more specific context. The idea is to exploit the $PT$~symmetric framework to ``go around'' (in the complex plane) this kind of instability.

We shall study the flat Friedmann cosmological model described by the (usual) FLRW metric~\eqref{FRWline}.

Let us consider the matter represented by scalar fields with complex potentials. Namely, we shall try to find a
complex potential possessing the solutions of classical equations of motion which guarantee the reality of all
observables. Such an approach is of course inspired by the quantum theory of the $PT$~symmetric non-hermitian hamiltonians,
whose spectrum is real and bounded from below. Thus, it is natural for us to look for lagrangians which have
consistent counterparts in the quantum theory.

Let me elucidate how the phantom-like classical dynamics arises in such lagrangians. For this purpose, at risk of being repetitive, choose
the one-dimensional $PT$ symmetric potential of an-harmonic oscillator $V^{(2+\epsilon)}(q) =  \lambda q^2
(iq)^\epsilon,\ 0 <\epsilon \leq 2 $ which has been rapidly analyzed above (\ref{sec:PTosc}). For illustration, let me take here the most interesting case of $\epsilon = 2, V^{(4)}(q) =
-\lambda q^4$. As said above, the classical dynamics for {\it real} coordinates $q(t)$ offers the infinite motion with
increasing speed and energy or, in the quantum mechanical language, indicates the absence of bound states and
unboundness of energy from below. However, just there is a more consistent solution which, at the quantum level,
provides the real discrete energy spectrum, certainly, bounded from below. It has been proven, first, by means of
path integral~\cite{PT1} and, further on, by means of the theory of ordinary differential equations~\cite{buslaev}. In fact, this classically ``crazy'' potential on a curve in the {\it complex} coordinate plane
generates the same energy spectrum as a two-dimensional quantum an-harmonic oscillator $V^{(4)}(\vec q) =\lambda
(q_1^4 + q_2^4)/4$ with {\it real} coordinates $q_{1,2}$ in the sector of zero angular momentum~\cite{PT1}.
Although the superficially unstable an-harmonic oscillator is well defined on the essentially complex coordinate
contour  any calculation in the style of perturbation theory (among them the semi-classical expansion) proceeds
along the contour with a fixed complex part (corresponding to a "classical" solution) and varying unboundly in
real direction. In particular, the classical trajectory for $V^{(4)}(q)$ with keeping real the kinetic, $(\dot
q)^2$ and potential, $- \lambda q^4(t)$ energies (as required by its incorporation into a cosmological scenario)
can be chosen imaginary,
$$q = i\xi,\quad \ddot \xi = -2\lambda \xi^3,\quad \dot\xi^2 = C - \lambda \xi^4,\quad C> 0 $$
which   obviously represents a bounded, finite motion with $|\xi| \leq (C/\lambda)^{1/4}$. Such a motion supports
the quasi-classical treatment of bound states with the help of Bohr's quantization. Evidently, the leading,
second variation of the lagrangian around this solution, $q(t) = i \xi(t) + \delta q (t)$ gives a positive
definite energy,
$${\cal L}^{(2)} = p(t)\delta\dot q(t)  - H,\ H = \frac14 p^2(t) + 12 \xi^2(t) (\delta q (t))^2$$ realizing the perturbative
stability of this an-harmonic oscillator in the vicinity of imaginary classical trajectory. It again reflects the
existence of positive discrete spectrum for this type of an-harmonicity. However the {\it classical} kinetic
energy $-\dot\xi^2$ is negative,  i.e. it is {\it phantom-like}.

In a more general, Quantum Field Theory setting let us consider a non-hermitian (complex) lagrangian of a scalar
field
\begin{equation}
L = \frac12\partial_{\mu}\phi\partial^{\mu}\phi^* - V(\phi,\phi^*)\ , \label{PT:Lagrange}
\end{equation}
with the corresponding action,
\begin{equation}
S[\phi, g] =  \int \, d^4x \sqrt{-g} \left(L + \frac16 R(g)\right)\ , \label{PT:action}
\end{equation}
where $g$ stands for the determinant of a metric $g^{\mu\nu}$ and $R(g)$ is the scalar curvature term and
the Newton gravitational constant is as usual normalized to $3/8\pi$ to simplify the Friedmann equations.

We employ potentials $V(\Phi,\Phi^*)$ satisfying the invariance condition
\begin{equation}
(V(\Phi,\Phi^*))^* = V(\Phi^*,\Phi)\ , \label{PT:condition}
\end{equation}
while the condition
\begin{equation}
(V(\Phi,\Phi^*))^* = V(\Phi,\Phi^*)\ , \label{PT:condition1}
\end{equation}
is not satisfied. This condition represents a generalized requirement of $(C)PT$~symmetry.

Let's  define two real fields,
\begin{equation}
\phi \equiv  \frac12(\Phi + \Phi^*)\ ,\quad \chi \equiv  \frac{1}{2i}(\Phi - \Phi^*)\ . \label{PT:split2}
\end{equation}
Then, for example, such a potential can have a form
\begin{equation}
V(\Phi,\Phi^*) = \tilde V(\Phi+\Phi^*, \Phi-\Phi^*) = \tilde V (\phi, i\chi)\ , \label{PT:factor}
\end{equation}
where $\tilde V (x,y)$ is a real function of its arguments. In the last equation one can recognize the link to
the so called $(C)PT$ symmetric potentials if to supply the field $\chi$ with a discrete charge or negative
parity. When keeping in mind the perturbative stability we impose also the requirement for the second variation
of the potential to be a positive definite matrix which, in general, leads to its $PT$~symmetry~\cite{2f1,2f2}.

Here, the functions $\phi$ and $\chi$ appear as the real and the imaginary parts of the complex scalar field
$\Phi$, however, in what follows, we shall treat them as independent spatially homogeneous variables depending
only on the time parameter $t$ and, when necessary, admitting the continuation to {\it complex} values.

\section{Cosmological solution with classical phantom field}\label{sec:PT2}

It appears that among known $PT$~symmetric hamiltonians (lagrangians) possessing the real spectrum one which is
most suitable for our purposes is that with the exponential potential. It is connected with the fact that the
properties of scalar field based cosmological models with exponential potentials are well studied\cite{Barrow1,we-tach,power-law1,power-law2}. 
In particular, the corresponding models have some exact solutions providing a universe
expanding according to some power law $a(t) = a_0 t^{q}$. We shall study the model with two scalar fields and the
additive potential. Usually, in cosmology the consideration of models with two scalar fields (one normal and one
phantom, i.e. with the negative kinetic term) is motivated by the desire of describing the phenomenon of the so
called phantom divide line crossing. At the moment of the phantom divide line crossing the equation of
state parameter $w = p/\varepsilon$ crosses the value $w = -1$ and (equivalently) the Hubble variable $h$ has an
extremum. Usually the models using two fields are called ``quintom models''~\cite{2f1}-\cite{2f4}. As a matter of fact
the phantom divide line crossing phenomenon can be described in the models with one scalar field, provided some particular
potentials are chosen\cite{crossing,cusps} or in the models with non-minimal coupling between the  scalar field and
gravity~\cite{non-minimal}. However, the use of two fields make all the considerations more simple and natural. In the framework we are considering here, the necessity of using two scalar fields follows from other requirements. We would like to implement
a scalar field with a complex potential to provide the effective phantom behavior of this field on some classical
solutions of equations of motion. Simultaneously, we would like to have the standard form of the effective
Hamiltonian for linear perturbations of this field. The combination of these two conditions results in the fact
the background contribution of both the kinetic and potential term in the energy density, coming from this field
are negative. To provide the positivity of the total energy density which is required by the Friedmann equation
(\ref{fri3}) we need the other normal scalar field. Thus, we shall consider the two-field scalar lagrangian
with the complex potential
\begin{equation}
L = \frac{\dot{\phi}^2}{2} + \frac{\dot{\chi}^2}{2} - A e^{\alpha \phi} + B e^{i\beta \chi}\ , \label{PT:lagrangian}
\end{equation}
where $A$ and $B$ are real, positive constants. This lagrangian is the sum of two terms. The term representing
the scalar field $\phi$ is a standard one, and it can generate a power-law cosmological expansion~\cite{Barrow1,we-tach,power-law1,power-law2}. The kinetic term of the scalar field $\chi$ is also standard, but its potential is complex. Notice that the exponential potential has been analyzed in great detail in~\cite{exponential1,exponential2}. The most important feature of this potential is that the spectrum of the corresponding
Hamiltonian is real and bounded from below, provided correct boundary conditions are assigned.

Inspired by this fact we are looking for a classical complex solution of the system, including two Klein-Gordon
equations for the fields $\phi$ and $\chi$:
\begin{equation}
\ddot{\phi} + 3 h \dot{\phi} + A\alpha e^{\alpha\phi} = 0\ , \label{PT:KG}
\end{equation}
\begin{equation}
\ddot{\chi} + 3 h \dot{\chi} - iB\beta e^{i\beta\chi} = 0\ , \label{PT:KG1}
\end{equation}
and the Friedmann equation
\begin{equation}
h^2 = \frac{\dot{\phi}^2}{2} + \frac{\dot{\chi}^2}{2} + A e^{\alpha \phi} - B e^{i\beta \chi}\ . \label{PT:Friedmann}
\end{equation}
The classical solution which we are looking for should provide the reality and positivity of the right-hand side
of the Friedmann equation (\ref{PT:Friedmann}). The solution where the scalar field $\phi$ is real, while the scalar
field $\chi$ is purely imaginary
\begin{equation}
\chi = i\xi,\quad\xi\ {\rm real}\ , \label{PT:xi}
\end{equation}
uniquely satisfies this condition. Moreover, the lagrangian (\ref{PT:lagrangian}) evaluated on this solution is real
as well. This is remarkable because on homogeneous solutions the lagrangian coincides with the pressure, which
indeed should be real.

Substituting the equation~(\ref{PT:xi})  into the Friedmann equation (\ref{PT:Friedmann}) we shall have
\begin{equation}
h^2 = \frac{\dot{\phi}^2}{2} - \frac{\dot{\xi}^2}{2} + A e^{\alpha \phi} - B e^{-\beta \xi}\ . \label{PT:Friedmann1}
\end{equation}
Hence, effectively we have the Friedmann equation with two fields: one ($\phi$) is a standard scalar field, the
other ($\xi$) has a phantom behavior, as we pointed out above. In the next section we shall study the
cosmological dynamics of the (effective) system, including (\ref{PT:Friedmann1}), (\ref{PT:KG}) and
\begin{equation}
\ddot{\xi} + 3 h \dot{\xi} - B\beta e^{-\beta\xi} = 0\ . \label{PT:KG2}
\end{equation}

The distinguishing feature of such an approach rather than the direct construction of phantom lagrangians becomes
clear when one calculates the linear perturbations  around the classical solutions. Indeed the second variation
of the action for the field $\chi$ gives the quadratic part of the effective lagrangian of perturbations:
\begin{equation}
L_{eff} = \frac12 \dot{\delta \chi}^2  - B\beta^2 e^{i\beta\chi_0} (\delta\chi)^2\ , \label{PT:perturb}
\end{equation}
where $\chi_0$ is a homogeneous purely imaginary solution of the dynamical system under consideration. It is easy
to see that on this solution, the effective lagrangian (\ref{PT:perturb}) will be real and its potential term has a
sign providing the stability of  the background solution with respect to linear perturbations as the related
Hamiltonian is positive,
\begin{equation}
H_{eff}^{(2)} = \frac12 {\delta \pi}^2  + B\beta^2 e^{-\beta\xi_0} (\delta\chi)^2 ,\quad \delta \pi
\Leftrightarrow \delta\dot\chi . \label{PT:perturbham}
\end{equation}

Let us list the main differences between our model and quintom models, using two fields (normal scalar and
phantom) and exponential potentials\cite{two-exp1,two-exp2}. First, we begin with two normal (non-phantom) scalar fields,
with normal kinetic terms, but one of these fields is associated to a complex ($PT$~symmetric) exponential
potential. Second, the (real) coefficient multiplying this exponential potential is negative. Third, the
background classical solution of the dynamical system, including two Klein-Gordon equations and the Friedmann
equation,  is such that the second field is purely imaginary, while all the geometric characteristics are
well-defined. Fourth, the interplay between transition to the purely imaginary solution of the equation for the
field $\chi$ and the negative sign of the corresponding potential provides us with the effective lagrangian for
the linear perturbations of this field which have correct sign for both the kinetic and potential terms: in such
a way the problem of stability of the our effective phantom field is resolved. Fifth, the qualitative analysis of
the corresponding differential equations, shows that in contrast to the quintom models in our model the Big Rip
never occurs. The numerical calculations confirm this statement.

In the next section we shall describe the cosmological solutions for our system of equations.

\section{Cosmological evolution}\label{sec:PT3}
First of all notice that our dynamical system  permits the existence of cosmological trajectories which cross
the phantom divide line. Indeed, the crossing point is such that the time derivative of the Hubble parameter
\begin{equation}
\dot{h} = -\frac32(\dot{\phi}^2 - \dot{\xi}^2) \label{PT:PDL}
\end{equation}
is equal to zero. We always can choose $\dot{\phi} = \pm \dot{\xi}$, at  $t = t_{PDL}$ provided the values of the
fields $\phi(t_{PDL})$ and $\xi(t_{PDL})$ are chosen in such a way, that the general potential energy $A
e^{\alpha \phi} - B e^{-\beta \xi}$ is non-negative. Obviously, $t_{PDL}$ is the moment of  the phantom divide line crossing.
However, the event of the phantom divide line crossing cannot happen only once. Indeed, the fact that the universe has crossed
phantom divide line means that it was in effectively phantom state before or after such an event, i.e. the
effective phantom field $\xi$ dominated over the normal field $\phi$. However, if this dominance lasts for a long
time it implies that not only the kinetic term $-\dot{\xi}^2/2$ dominates over the kinetic term $\dot{\phi}^2/2$
but also the potential term $-B\exp(-\beta\xi)$ should dominate over $A\exp(\alpha\phi)$; but it is impossible,
because contradicts to the Friedmann equation (\ref{PT:Friedmann1}). Hence, the period of the phantom dominance
should finish and one shall have another point of phantom divide line crossing. Generally speaking, only the regimes with even
number of phantom divide line crossing events are possible. Numerically, we have found only the cosmological trajectories with
the double phantom divide line crossing. Naturally, the trajectories which do not experience the crossing at all also exist and
correspond to the permanent domination of the normal scalar field. Thus, in this picture,  there is no place for
the Big Rip singularity as well, because such a singularity is connected with the drastically dominant behavior
of the effective phantom field, which is impossible as was explained above. The impossibility of approaching the
Big Rip singularity can be argued in a more rigorous way as follows. Approaching the Big Rip, one has a growing
behavior of the scale factor $a(t)$ of the type $a(t) \sim (t_{BR}-t)^{-q}$, where $q >0$. Then the Hubble
parameter is
\begin{equation}
h(t) = \frac{q}{t_{BR} -t} \label{PT:Hubble3}
\end{equation}
and its time derivative
\[
\dot{h}(t) = \frac{q}{(t_{BR}-t)^2}.
\]
Then, according to equation~(\ref{PT:PDL}),
\begin{equation}
\frac12\dot{\xi}^2 - \frac12\dot{\phi}^2 = \frac{q}{3(t_{BR}-t)^2}\ . \label{PT:BR}
\end{equation}
Substituting equations~(\ref{PT:BR}), (\ref{PT:Hubble3}) into the Friedmann equation (\ref{PT:Friedmann1}), we come to
\[
\frac{q}{(t_{BR}-t)^2}=\frac{q}{3(t_{BR}-t)^2}+Ae^{\alpha\phi}-Be^{-\beta\xi}\ .
\]
In order for this to be satisfied and consistent, the potential of the scalar field $\phi$ should behave as
$1/(t_{BR}-t)^2$. Hence the field $\phi$ should be
\begin{equation}
\phi = \phi_0 -\frac{2}{\alpha} \ln (t_{BR} - t)\ , \label{PT:log}
\end{equation}
where $\phi_0$ is an arbitrary constant. Now substituting equations~(\ref{PT:Hubble3}) and (\ref{PT:log}) into the
Klein-Gordon equation for the scalar field $\phi$ (\ref{PT:KG}), the condition of the cancellation of the most
singular terms in this equation which are proportional to $1/(t_{BR}-t)^2$ reads
\begin{equation}
2 + 6q + A\alpha^2 \exp(\alpha{\phi_0}) = 0\ . \label{PT:cond}
\end{equation}
This condition cannot be satisfied because all the terms in the left-hand side of equation~(\ref{PT:cond}) are positive.
This contradiction demonstrates that it is impossible to reach the Big Rip.

\begin{figure}[htp]\centering
\begin{minipage}{.5\textwidth}\centering
\includegraphics[scale=1]{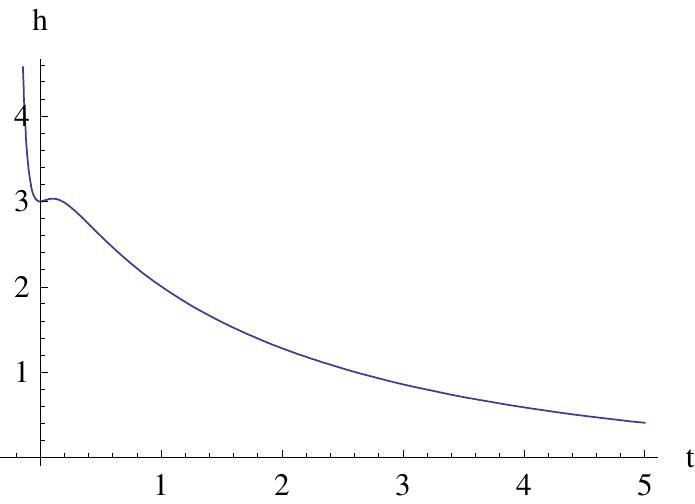}
\end{minipage}%
\begin{minipage}{.5\textwidth}\centering
\includegraphics[scale=1]{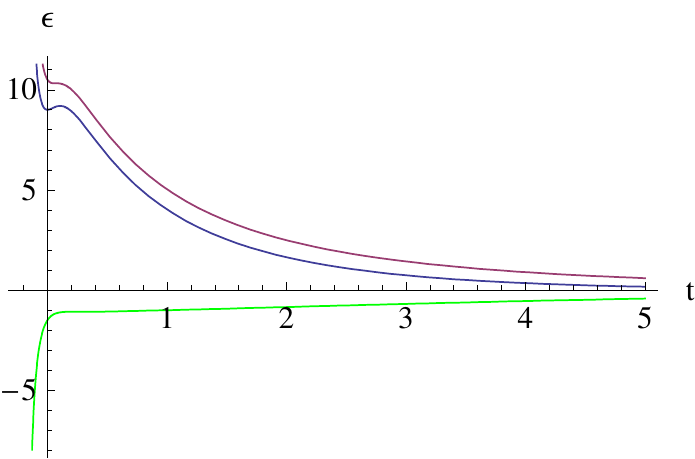}
\end{minipage}
\caption{(left) Plot of the Hubble parameter representing the cosmological evolution. The evolution starts from a
Big Bang-type singularity and goes through a transient phase of super-accelerated expansion (``phantom era''),
which lies between two crossings of the phantom divide line (when the derivative of $h$ crosses zero). Then the universe expands
infinitely. (right) Plots of the total energy density (blue), and of the energy density of the normal field
(purple) and of the phantom one (green).}\label{PT:fig1} 
\end{figure}

\begin{figure}[htp]\centering
\begin{minipage}{.5\textwidth}\centering
\includegraphics[scale=.8]{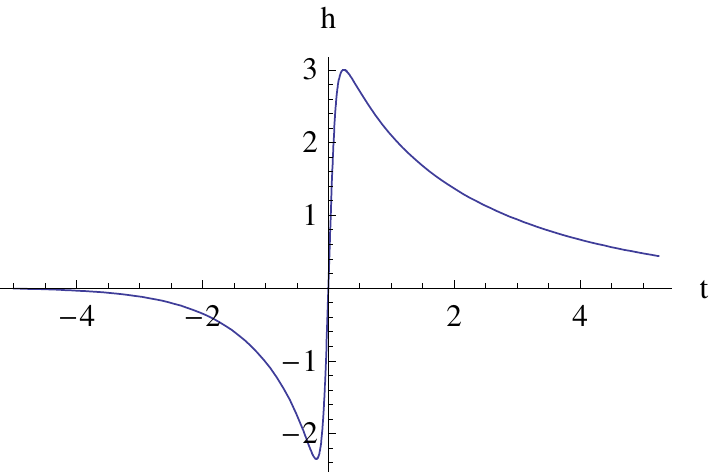}
\end{minipage}%
\begin{minipage}{.5\textwidth}\centering
\includegraphics[scale=.8]{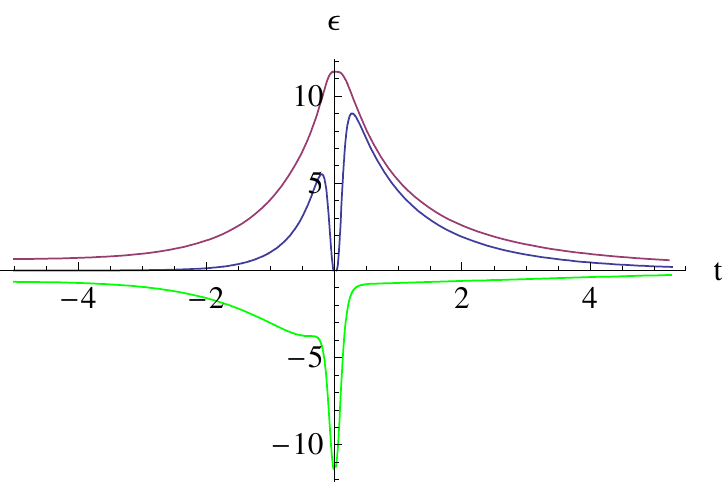}
\end{minipage}
\caption{(left) The evolution starts with a contraction in the infinitely remote past. At the first phantom divide line
crossing the contraction becomes super-decelerated and  turns in a super-accelerated expansion when $h$ crosses
zero. The second  crossing ends the ``phantom era''; the decelerated expansion continues till the universe
begins contracting. In a finite time a Big Crunch-type singularity is reached. (right) Plots of the total energy
density (blue), and of the energy density of the normal field (purple) and of the phantom one
(green).}\label{PT:fig2}
\end{figure}

\begin{figure}[htp]\centering
\begin{minipage}{.5\textwidth}\centering
\includegraphics[scale=.8]{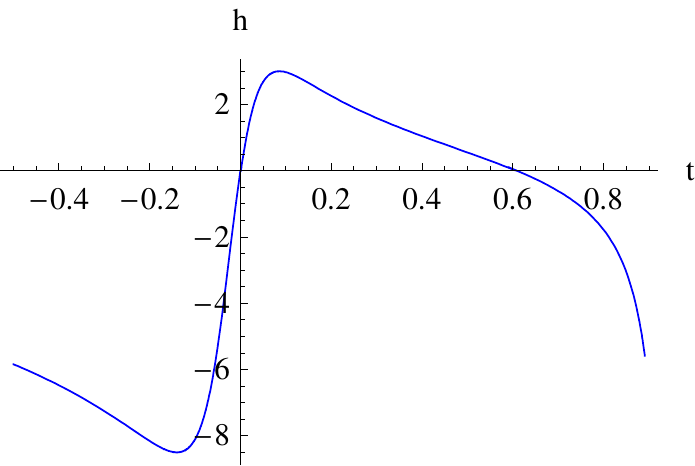}
\end{minipage}%
\begin{minipage}{.5\textwidth}\centering
\includegraphics[scale=.8]{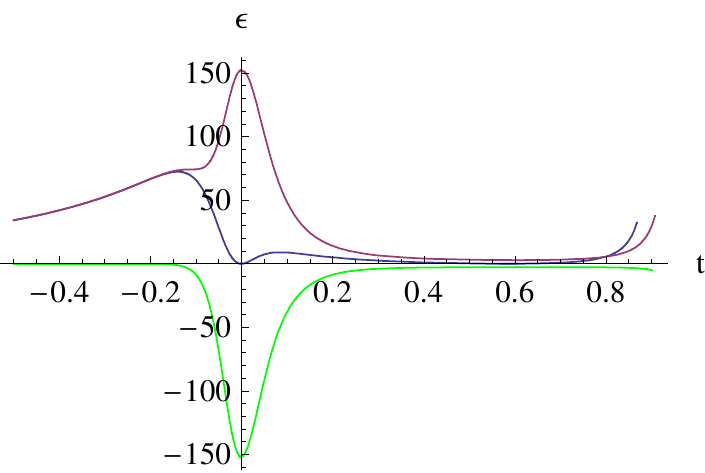}
\end{minipage}
\caption{(left) The cosmological evolution begins with a contraction in the infinitely remote past. With the
first phantom divide line crossing the contraction becomes super-decelerated until the universe stops ($h=0$) and starts
expanding. With the second crossing the ``phantom era'' ends and the expansion continues infinitely. (right)
Plots of the total energy density (blue), and of the energy density of the normal field (purple) and of the
phantom one (green).}\label{PT:fig3}
\end{figure}

\begin{figure}[htp]\centering
\begin{minipage}{.5\textwidth}\centering
\includegraphics[scale=.8]{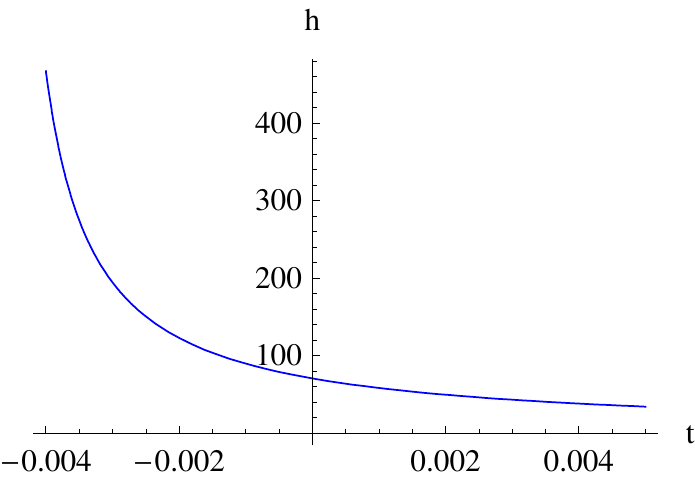}
\end{minipage}%
\begin{minipage}{.5\textwidth}\centering
\includegraphics[scale=.8]{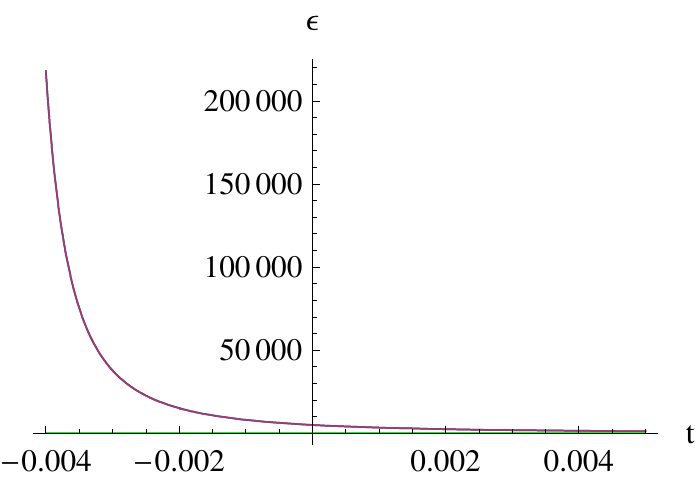}
\end{minipage}
\caption{(left) Evolution from a Big Bang-type singularity to an infinite expansion, without any crossing of the phantom divide line.
This evolution is thus guided by the ``normal'' field $\phi$. (right) Plots of the total energy density (blue),
and of the energy density of the normal field (purple) and of the phantom one (green). Notice that the energy
density of the phantom field (green) is very close to zero, thus the total energy density is mainly due to the
standard field.}\label{PT:fig4}
\end{figure}

Now I would like to describe briefly some examples of cosmologies contained in our model, deduced by numerical analysis of the system of equations of motion.\\
In figure~\ref{PT:fig1} a double crossing of the phantom divide line is present. The evolution starts from a Big Bang-type singularity and
goes through a transient phase of super-accelerated expansion (``phantom era''), which lies between two crossings
of the phantom divide line. Then the universe undergoes an endless expansion. In the right plot I present the time evolution of the
total energy density and of its partial contributions due to the two fields, given by the equations
\begin{eqnarray*}
&&\varepsilon_\phi = \frac{1}{2}\dot\phi^2+Ae^{\alpha\phi}\ ,\\
&&\varepsilon_\xi = -\frac{1}{2}\dot\xi^2-Be^{\beta\xi}\ ,\\
&&\varepsilon =\varepsilon_\phi+\varepsilon_\xi\ ,
\end{eqnarray*}
which clarify the roles of the two fields in driving the cosmological evolution.\\
The evolution presented in figure~\ref{PT:fig2} starts with a contraction in the infinitely remote past. Then the contraction becomes super-decelerated and  turns later in a super-accelerated expansion . With the second phantom divide line crossing the ``phantom era'' ends; the decelerated expansion continues till the universe begins contracting. After a finite time a Big Crunch-type singularity is encountered. From the right plot we can clearly see that the ``phantom era'' is indeed characterized by a bump in the (negative) energy density of the phantom field.\\
In figure~\ref{PT:fig3} the cosmological evolution again begins with a contraction in the infinitely remote past. Then the universe crosses the phantom line: the contraction becomes super-decelerated until the universe stops  and starts expanding. Then the "phantom era" ends and the expansion is endless.\\
In figure~\ref{PT:fig4} the evolution from a Big Bang-type singularity to an eternal expansion is shown. The phantom
phase is absent. Indeed the phantom energy density is almost zero everywhere.

\section{Conclusions}\label{sec:PT4}
As was already said many times,  the data are compatible with the presence of the phantom energy, which, in turn, can
be in a most natural way realized by the phantom scalar field with a negative kinetic term. However, such a field
suffers from the instability problem, which makes it vulnerable. Inspired by the development of $PT$~symmetric
Quantum Theory we introduced the $PT$~symmetric two-field cosmological model where both the
kinetic terms are positive, but the potential of one of the fields is complex. We studied a classical background
solution of two Klein-Gordon equations together with the Friedmann equation, when one of this fields (normal) is
real while the other is purely imaginary. The scale factor in this case is real and positive just like the energy
density and the pressure. The background dynamics of the universe is determined by two effective fields -- one
normal and one phantom -- while the lagrangian of the linear perturbations has the correct sign of the mass term.
Thus, so to speak, the quantum normal theory is compatible with the classical phantom dynamics and the problem of
instability is absent.

As a byproduct of the structure of the model, the phantom dominance era is transient, the number of the phantom
divide line crossings is even and the Big Rip singularity is avoided.
\chapter{Cosmological singularities with finite non-zero radius}\label{sec:soft}

{\it The present chapter is intended to present the results of \cite{mio_soft}. This is kind of \mbox{off-topic}, with respect to the preceding work, but not at all by far. Indeed the idea is to study the reconstruction of a \emph{single} scalar field model, able to reproduce a specific dynamical evolution, which seemed to us particularly interesting, since it presents some kind of mild version of the Big Bang singularity. Dynamical analysis is performed on the model and its phase space is divided into classes of qualitative different cosmic evolutions. Here there is no crossing of the phantom divide line, nor phantom fields.}

\section{Introduction and review of paper \texorpdfstring{\cite{mio_soft}}{on mild singularities}}
%

We have repeatedly said that the discover of cosmic acceleration~\cite{Riess,Perlmutter} has stimulated the research of new cosmological models. This development of model-designing art has revealed  cosmological evolutions
possessing various types of singularities, sometimes very different from the traditional Big Bang and 
Big Crunch. The most popular between them is, perhaps, the Big Rip cosmological singularity 
\cite{Rip1,Rip2} arising in super-accelerating models driven by some kind of phantom matter.
Other types of singularities are sudden singularities~\cite{sudden1,sudden2,sudden3}, Big Brake~\cite{we-tach}, and so on~\cite{other1}-\cite{other4}.
Here we would like to study the singularities which are close to the known Big Bang, Big Crunch and Big Rip singularities,
but arising at finite values of the cosmological factor (different from zero and infinity as well). 
Similar singularities were recently considered in \cite{freeze1,freeze2,freeze3}.

In the paper under consideration we construct potentials which can drive the cosmological evolution towards (or from) such 
singularities. Combining qualitative and numerical methods we study the set of possible cosmological histories 
in the suggested models to show that the presence of such singularities in a cosmological model under consideration 
depends essentially on initial conditions and that the same model can accommodate qualitatively different 
cosmological scenarios.  

The structure of what follows is: in section~\ref{sec:s1} we construct some potentials corresponding to evolutions 
with ``mild'' singularities. In the third section~\ref{sec:s2} we analyze their dynamics. The conclusion~\ref{sec:s3} is devoted to an interpretation 
of the obtained results. 

\section{Construction of scalar field potentials}\label{sec:s1}

Let me repeat here very briefly the idea of the reconstruction of potentials for cosmological models (cfr. section~\ref{sec:21}).

We shall consider flat Friedmann models with the metric~\eqref{FRWline}
The Hubble parameter $h(t) \equiv \dot a/a$ 
satisfies the Friedmann equation \eqref{fri3} and also its `manipulation'~\eqref{fri}.
%

If the matter is represented by a spatially homogeneous minimally coupled scalar field, then
the energy density and the pressure are given by the formul\ae\ \eqref{2:energy} and \eqref{2:pressure}, respectively; while equations \eqref{2:poten} and \eqref{2:phi-dot} give the expression of the potential and of $\dot\phi$ as functions of time.
Integrating equation (\ref{2:phi-dot})
one can find the scalar field as a function of time. Inverting this dependence we can obtain the time parameter as
a function of $\phi$ and substituting the corresponding formula into equation (\ref{2:poten}) one arrives to the uniquely
reconstructed potential $V(\phi)$. It is necessary to stress that this potential reproduces a given cosmological evolution
only for some special choice of initial conditions on the scalar field and its time derivative.
  
It is known that the power-law cosmological evolution is given by the Hubble parameter $h(t) \sim \frac{1}{t}$. 
We shall look for a ``softer'' version of the cosmological evolution given by the law 
\begin{equation}
h(t) = \frac{S}{t^{\alpha}}\ ,
\label{s:soft}
\end{equation}
where $S$ is a positive constant and $0 < \alpha < 1$. At $t = 0$ a singularity is present, but it is different from the 
traditional Big Bang singularity. Indeed, integrating we obtain
\begin{equation}
\ln \frac{a(t)}{a(0)} = \frac{S}{1-\alpha} t^{1-\alpha}.
\label{s:soft1}
\end{equation}
If $t > 0$ the right-hand side of equation~(\ref{s:soft1}) is finite and hence one cannot have $a(0) = 0$ in the left-hand side of this equation, because it would  imply a contradiction, making $\ln\frac{a(t)}{a(0)}$ divergent. Hence $a(0) > 0$, while 
\begin{equation}
\dot{a} = a(0)\frac{S}{t^{\alpha}}\exp\left(\frac{S}{1-\alpha}t^{1-\alpha}\right) \xrightarrow[t\rightarrow 0]{}\infty\ .
\label{s:soft2}
\end{equation}
This type of singularity can be called ``mild'' Bing Bang singularity because the cosmological radius is finite (and non-zero) while 
its time derivative, the Hubble variable and the scalar curvature are singular. 
It is interesting to note that when $t \rightarrow \infty$ both $a(t)$ and $\dot{a}(t)$ tend to infinity, but 
they do not encounter any cosmological singularity because the Hubble variable and its derivatives tend to zero.

Let us reconstruct the potential of the scalar field model, producing the cosmological evolution (\ref{s:soft}) using the 
technique described above. 
Equation~(\ref{2:phi-dot}) gives 
\begin{equation}
\dot\phi=\pm\sqrt{\frac{2}{3}\alpha S}\;t^{-\frac{\alpha+1}{2}}\ .
\label{s:phi-dot1}
\end{equation}
We shall choose the positive sign, without loosing generality.
Integrating, we get
\begin{equation}
\phi(t)=\sqrt{\frac{2}{3}\alpha S}\;\frac{2t^{\frac{1-\alpha}{2}}}{1-\alpha}\ ,
\label{s:phi}
\end{equation}
up to an arbitrary constant.
Inverting the last relation we find 
\begin{equation}
t(\phi)=\left(\left(\frac{3}{2\alpha S}\right)^{1/2}\frac{1-\alpha}{2}\phi\right)^{\frac{2}{1-\alpha}}.
\end{equation}
Hence, using equation~(\ref{2:poten}) we obtain
\begin{equation}
V(\phi)=\frac{S^2}{\left(\sqrt{\frac{3}{2\alpha S}}\frac{1-\alpha}{2}\phi\right)^{\frac{4\alpha}{1-\alpha}}}-\frac{\alpha S}{3\left(\sqrt{\frac{3}{2\alpha S}}\frac{1-\alpha}{2}\phi\right)^{\frac{2(\alpha+1)}{1-\alpha}}}\ ,
\end{equation}
This potential provides the cosmological evolution (\ref{s:soft}) if initial conditions compatible with equations~(\ref{s:phi-dot1}) and 
(\ref{s:phi}) are chosen. Naturally, there are also other cosmological evolutions, generated by other initial conditions, which 
will be studied in the next section.

\section{The dynamics of the cosmological model with \texorpdfstring{$\alpha = \frac12$}{alfa=1/2}}\label{sec:s2}

In order to achieve some simplification of calculations we shall consider a particular model, namely the one with the choice 
$\alpha = \frac12$.  In this case 
\begin{equation}
a(t)=a(0)e^{2S\sqrt{t}}\ ,
\label{s:a-new}
\end{equation}
and 
\begin{equation}
V(\phi)=\frac{16S^4}{(\sqrt{3}\phi/2)^4}-\frac{32S^4}{3(\sqrt{3}\phi/2)^6}\ .
\label{s:poten-new}
\end{equation}
The Klein-Gordon equation reads
\begin{equation}
\ddot\phi+3\dot\phi\ {\rm sign}(h)\sqrt{\frac{1}{2}\dot\phi^2+\frac{16S^4}
{(\sqrt{3}\phi/2)^4}-\frac{32S^4}{3(\sqrt{3}\phi/2)^6}}
-\frac{32\sqrt{3}S^4}{(\sqrt{3}\phi/2)^5}+\frac{32\sqrt{3}S^4}{(\sqrt{3}\phi/2)^7}=0\ .
\label{s:Klein-Gordon}
\end{equation}
This equation is equivalent to the dynamical system  

\begin{equation}
\left\{
\begin{array}{ll}
\dot\phi=&x,\\
\dot x=&-3x\ {\rm sign}(h)\sqrt{\frac{x^2}{2}+\frac{16S^4}{(\sqrt{3}\phi/2)^4}-\frac{32S^4}{3(\sqrt{3}\phi/2)^6}} \\
&+\frac{32\sqrt{3}S^4}{(\sqrt{3}\phi/2)^5}-\frac{32\sqrt{3}S^4}{(\sqrt{3}\phi/2)^7}.
\label{s:dyn1}
\end{array}\right.
\end{equation}
The qualitative analysis of dynamical systems in cosmology was presented in detail in~\cite{Bel-Khal}.

First of all, let us notice that the system has two critical points:
\mbox{$\phi = \pm \frac{2}{\sqrt{3}}, x = 0$.}
We consider the linearized system around the point with the positive value of $\phi$:
\begin{equation}\left\{
\begin{array}{ll}
\dot{\varphi}=& x,\\
\dot x =& -3x\ {\rm sign}(h)\frac{4}{\sqrt{3}}S^2+32\cdot 3 S^4\varphi,
\end{array}\right.
\label{s:dyn2}
\end{equation}
where $\varphi\equiv \phi-2/\sqrt{3}$.

The Lyapunov indices for this system are (for $h > 0$)
\begin{equation}
\lambda_1=-8\sqrt{3}S^2\ ,
\label{s:lambda1}
\end{equation}
\begin{equation}
\lambda_2=4\sqrt{3}S^2.
\label{s:lambda2}
\end{equation}
For negative $h$, corresponding to the cosmological contraction, the signs of $\lambda_1$ and $\lambda_2$ are changed. 
The eigenvalues are real and have opposite signs, hence both the critical points are saddles. The universe being in one of these 
two saddle points means that it undergoes a de~Sitter expansion or contraction, according to the sign of $h$, with the 
value of $h = h_0$ given by 
\begin{equation}
h_0 = \frac{4S^2}{\sqrt{3}}\ .
\label{s:deSitter}
\end{equation}
For each saddle point there are four separatrices
which separate four classes of trajectories in the phase plane $x,\phi$ corresponding to four 
types of cosmological evolutions.

\begin{figure}
\centering
\includegraphics[scale=1]{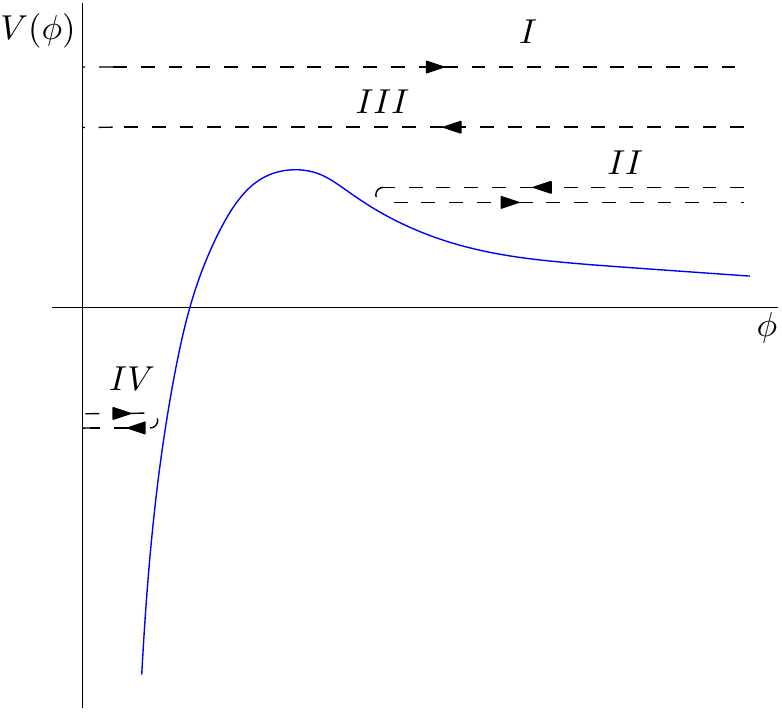}
\caption{Plot of the potential $V(\phi)$ as given by equation~(\ref{s:poten-new}). Here only the positive $\phi$-axis is shown, which is enough to understand the behavior thanks to the parity of the function. We have also delineated the four different dynamical behaviors of the scalar field, clearer to understand taking into account the phase portrait (see figure~\ref{s:fig2}).}\label{s:fig1}
\end{figure}

\begin{figure}
\centering
\includegraphics[scale=1]{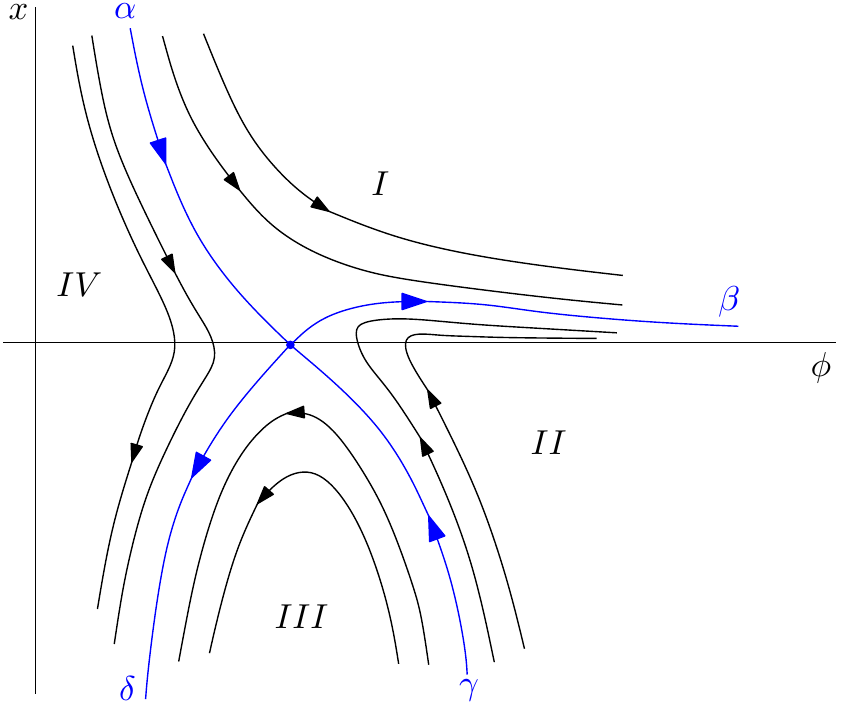}
\caption{Positive $\phi$-axis phase portrait for the dynamical system (\ref{s:dyn1}) with $h\geq 0$. The four separatrices of the saddle point ($\alpha,\beta,\gamma,\delta$) individuate four regions (\emph{I,II,III,IV}) with different behaviors of the trajectories, as explained in the text.}\label{s:fig2}
\end{figure}

In order to simplify the study of the dynamics let us note that the potential is an even function of the scalar field $\phi$ and that the saddle points are also symmetrical with respect to the $x$ axis. Thus, it is sufficient to consider only one of this saddle points. We shall carry out our qualitative analysis taking into 
account both figure~\ref{s:fig1}, giving the form of the potential, and  figure~\ref{s:fig2}, representing the phase portrait in the plane $\phi,x$.

First let us consider trajectories which begin at the moment $t = 0$, when the initial value of the scalar field is infinite, its 
potential is equal to zero and the time derivative of the scalar field is infinite and negative. In terms of the figure~\ref{s:fig1} it means that we consider the motion of the point beginning at the far right on the slope of the potential hill and moving towards the left (i.e. towards the top of the hill) with an infinite initial velocity.  Such a motion for $h > 0$ describes a universe born from the 
standard Big Bang singularity. Further details of this evolution depend on the asymptotic ratio between absolute values of 
$\dot\phi$ and $\phi$ at $t \rightarrow 0$. If this ratio is smaller than some critical value then the scalar field does not 
reach  the top of the hill and at some moment it begins to roll down back to the right. During this process of rolling down the scalar field 
increases, the potential is decreasing and the velocity $\dot{\phi}$ becomes positive and increasing. However the universe expansion works as a friction and at some moment its influence becomes dominant causing an asymptotic damping to zero of the velocity. The universe expands infinitely with $h(t) \rightarrow 0$. In the phase portrait (figure~\ref{s:fig2}) such trajectories 
populate the region $II$. This region is limited by the separatrices $\beta$ and $\gamma$. The first one corresponds to the positive (for $h > 0$) eigenvalue $\lambda_2$, while $\gamma$ corresponds to $\lambda_1$.   
If the ratio $\dot\phi/\phi$ has the critical value, then the scalar field reaches asymptotically the top of the hill of the potential, meaning that the universe becomes asymptotically de~Sitter: in the phase portrait it 
is nothing but the curve $\gamma$.

When the ratio introduced above is larger than the critical one we encounter a different regime. In this case the scalar field 
passes with non-vanishing velocity the top of the hill and begins to roll down in the abyss on the left. The velocity is growing, 
but the potential becomes negative and at some moment the total energy density of the scalar field vanishes together with the 
Hubble parameter $h$: this means that the universe starts contracting. This contraction provides the growing of the absolute 
value of the velocity of the scalar field and the kinetic term again becomes larger then the potential one. Moreover, both the terms in the Klein-Gordon equation increase the absolute value of $\dot{\phi}$. One can easily show that the regime in which the time derivative $\dot{\phi}$ becomes equal to $-\infty$ at some \emph{finite} value of $\phi$ is impossible, because it implies a contradiction 
between the asymptotic behavior of different terms in equation~(\ref{s:Klein-Gordon}). Thus, the universe tends to the singularity 
squeezing to the state with the value of $\phi$ equal to zero and an infinite time derivative $\dot{\phi}$. To understand which 
kind of  singularity the universe encounters, we need  some detail about the behavior of the scalar field. 
Let us suppose that, approaching the singularity at some moment $t_0$, the scalar field behaves as
\begin{equation}
\phi(t) = \phi_0(t_0-t)^{\mu}\ ,
\label{s:appr}
\end{equation}
where $0 < \mu < 1$. Then the first and second time derivative are 
\begin{equation}
\dot{\phi}(t) = -\mu\phi_0(t_0-t)^{\mu-1},\ \ \ddot{\phi}(t) = \mu(\mu-1)\phi_0(t_0-t)^{\mu-2}.
\label{s:appr1}
\end{equation}
The potential behaves as 
\begin{equation}
V = -\frac{2048 S^4}{81\phi_0^6(t_0-t)^{6\mu}}\ .
\label{s:appr-pot} 
\end{equation}
To have the Hubble variable well defined we require that the kinetic term is larger than the absolute value of the  negative potential term 
(\ref{s:appr-pot}), i.e. 
$
2\mu-2 \leq -6\mu,
$
or $\mu \leq \frac14$. Now two opposite cases may hold: (i) the friction term in the Klein-Gordon equation could dominate the potential term or (ii) the opposite situation. For (i) to be the right case,
one should require $2\mu -2 < -7\mu$ or $\mu < \frac29$. In this situation the asymptotic behavior of the second time derivative 
of $\phi$ should be equal to that of the friction term, or, in other words $\mu-2 = 2\mu-2$, that is $\mu = 0$, which 
obviously is not relevant. Thus we have to consider the range $\frac29 < \mu \leq \frac14$. In this case the potential term 
should be equal to the second time derivative of $\phi$, which implies:
\begin{equation}
\mu = \frac14
\label{s.mu}
\end{equation}
and 
\begin{equation}
\phi_0 = 4\sqrt{\frac{S}{3}}\ .
\label{s:phi0}
\end{equation}
Substituting the values of $\mu$ and $\phi_0$ into the expression for $h(t)$,
we obtain 
\begin{equation}
h(t) = -\frac{S}{\sqrt{t_0-t}}\ .
\label{s:apprh}
\end{equation}
Thus, we see that the singularity we are approaching is of the ``mild'' Big Crunch type.

\begin{figure}
\centering
\includegraphics[scale=1]{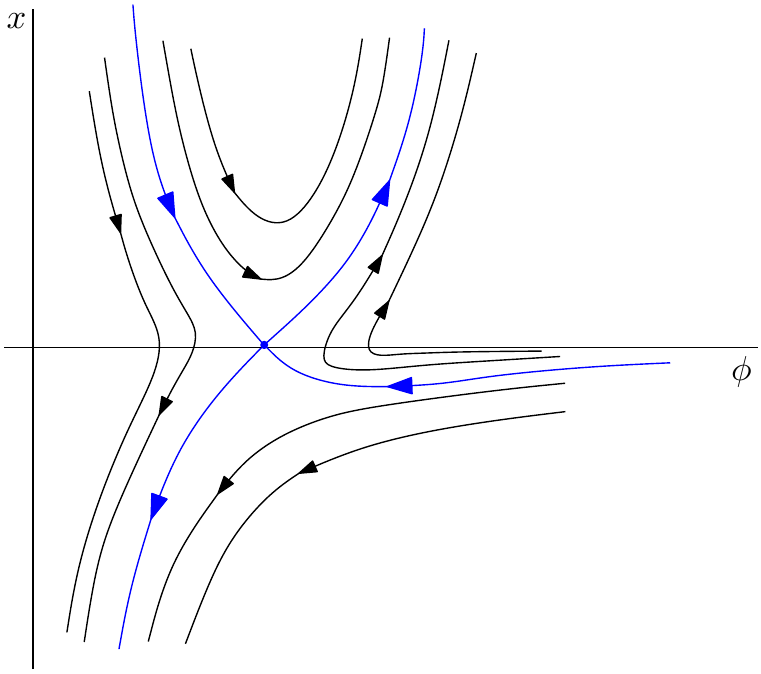}
\caption{Phase portrait for the dynamical system under discussion (equation~(\ref{s:dyn1})) with $h\leq 0$, i.e. describing evolutions of the universe characterized by contraction. As for the $h$-positive case, we can see the four separatrices of the saddle point and the corresponding four regions of different dynamical behaviors of the trajectories.}\label{s:fig3}
\end{figure}

In the phase portrait (figure~\ref{s:fig2}) these trajectories occupy the region $III$ limited by the separatrices $\gamma$ and $\delta$. The cosmological 
evolutions run from the Big Bang singularity to the mild Big Crunch one. However, the figure~\ref{s:fig2} is not sufficient to describe the 
complete behavior of the universe under consideration, because at some moment the Hubble variable changes sign and we 
should turn to the figure~\ref{s:fig3},  giving the phase portrait for the contracting universe $h < 0$.  Note that increasing the velocity 
$\dot{\phi}$ with which the scalar field overcomes the top of the hill, implies increasing the moment of time when the point 
of maximal expansion of the universe is reached. The limiting case in which this moment tends to infinity corresponds to the 
separatrix $\delta$. 

The third regime begins from the mild Big Bang singularity, when the scalar field is equal to zero and its time derivative is 
infinite and positive. In figure~\ref{s:fig1} that situation is represented by the point climbing from the abyss to the top of the hill. If the velocity term 
is not high enough, at some moment the field stops climbing and rolls back down. During this fall the Hubble variable changes
sign and the universe ends its evolution in the mild Big Crunch singularity. The corresponding 
trajectories belong to the region $IV$ of our phase plane, bounded by the separatrices $\delta$ and $\alpha$. The universe 
has its finite  life time between the mild Big Bang and the mild Big Crunch singularities. The situation when the scalar field arrives exactly to the 
top of the hill and stops corresponds to the separatrix $\alpha$. 

The fourth set of cosmological trajectories is generated by the scalar field climbing from the abyss and overcoming 
the top of the hill with the subsequent infinite expansion: the scalar field is infinitely growing and the Hubble parameter 
tends to zero. These trajectories occupy the region $I$ and our original cosmological evolution (\ref{s:soft}) belongs to this 
family.     Naturally there are other four classes of cosmological evolutions, which can be easily obtained by inverting the time direction.

\section{Conclusions}\label{sec:s3}
Let me sum up the results. Wishing to describe a cosmological evolution beginning from the singularity characterized by a finite and non-zero
initial (or final) cosmological radius and an infinite value of the scalar curvature due to the infinite value of the Hubble parameter, we have 
constructed a scalar field potential, providing such an evolution. Then, using the methods of qualitative analysis of the 
differential equations, we have shown that the proposed model accommodates four different classes of cosmological evolutions,
depending on initial conditions. Numerical simulations have confirmed our predictions. 

The main results of this work are: (i) the realization 
of a concrete cosmological model with a scalar field, where finite cosmological radius singularities are present and (ii) 
the complete description of all the possible evolutions of this model depending on the initial conditions. It is important to remark that, given 
the fixed scalar field potential, one has different types of evolutions encountering different kinds of singularities.     

Let us note, that there are some studies \cite{sudden1,sudden2,sudden3}, \cite{other1}-\cite{other4}, \cite{freeze1,freeze2,freeze3} devoted to the general analysis of various kinds 
of new cosmological singularities. In this work we were not looking for an exhaustive classification of different possible 
cosmological models possessing  some kind of singularities, but rather we wanted to study in a complete way a particular cosmological model, having some interesting properties.

\part{Loop Quantum Gravity and Spinfoam models}\label{part:2}
 \chapter*{Introduction\markboth{Introduction}{Introduction}} \addcontentsline{toc}{chapter}{Introduction}\label{sec:lqg_intro}
 
The title of the present thesis talks about the ``relation between Geometry and Matter''. Just to be clear from the very beginning: I will not talk about the coupling of matter with gravity in the Loop Quantum Gravity framework. Even though some work has been done in this respect\footnote{See e.g. \cite{Thiemann_matter1,Thiemann_matter2,Ma_matter} for matter coupling in the canonical LQG approach, and \cite{SF_matter1}-\cite{SF_matter5} for the spinfoam approach.} ,  this is not quite the sense we had in mind when we thought about that title. Let me spend some words about it.

Gravity, or the gravitational field, is indeed a rather peculiar object in physics and -- I dare say -- in Nature. This is true even at the classical level. This has become apparent when Einstein discovered the particular ``position'' of the gravitational field with respect to all the other (matter) fields. Indeed the gravitational field \emph{is} a field like all the others, with its dynamical equations and so on. But it is the background on which all the other fields live as well. All the matter fields are defined using the metric $g$ in their equations, they live and ``play'' on a certain metric manifold -- the ``game rules'' -- which ``is'' the gravitational field itself. Moreover, there is interplay between them, since Einstein's equations tell us that the game rules depend on how the matter plays and viceversa, in a recursive/non-linear interplay. This duality of being both a dynamical field and the background spacetime on which the fields live is at the heart of the peculiar nature of Gravity.

This is a classical picture. When we quantize the matter fields in the Quantum Field Theory framework, we somewhat forget this relation, and assume a fixed flat background, without gravity. We can try at best to do Quantum Field Theory on curved background, with all its  subtleties, still relying on that critical and peculiar relation I was talking about; and, most importantly, we still need a background spacetime metric on top of which our matter fields live. 

But now a question urges: what if we quantize the gravitational field? We can argue that this quantization is doomed to revolutionize once again the way in which we think about matter fields living on a spacetime, just as General Relativity revolutionized the concept of fixed spacetime. Indeed  Loop Quantum Gravity gives a picture of space, and of spacetime, which is purely combinatorial and relational, where the very concept of ``event'' looses its meaning: totally different from the smooth metric manifold to which we are accustomed. The loss of a background is crucial in the way we think matter fields.

In this precise sense I think that the study of Quantum Gravity is fundamental to understand the ``relation between Geometry and Matter''.

The following is a brief review of the main aspects of Loop Quantum Gravity and of spinfoam theory, the latter being an attempt to understand the quantum dynamics of the gravitational field. After this review I will present the research work I have done together with Eugenio Bianchi and Carlo Rovelli, during my period of research in the Marseille Quantum Gravity group.  It is about a rather technical aspect of the spinfoam approach to quantum gravitational dynamics.

\section*{Non-perturbative Quantum Gravity: some good reasons to consider this way\markright{Non-perturbative Quantum Gravity: some good reasons to consider this way}}\addcontentsline{toc}{section}{Non-perturbative Quantum Gravity: some good reasons to consider this way}

Loop Quantum Gravity is a form of canonical quantization of General Relativity. We'd better say an attempt in this direction, since the target is still far from being reached. However, there are quite amazing and interesting results that should not be under-estimated. 
I will briefly review the standard approach to the canonical quantization of gravity and then pass to a review of the fundamental concepts of LQG. But first let us stress the two main ingredients that has permitted LQG to become what it is: 
\begin{itemize}
\item background independence,
\item focus on connection (rather than metric).
\end{itemize}
A few words on each of these concepts are due, before diving into the theory. 
Quantization of gravity means quantization of the gravitational field, the metric tensor field $\bm{g}(x)=g_{\mu\nu}(x)\d x^\mu\otimes\d x^\nu$, i.e. the geometry of spacetime. One should naively expect that such a quantization would bring to some form of ``quantized geometry'', meaning a quantization of distances, volumes and such. This should happen at a very small distance scale, of course. Dimensionally speaking, we know that this scale should be the Planck length, which is of the order of $10^{-33}$~cm. Moreover, Quantum Mechanics has taught us that linear superpositions of (quantum) states must be taken into account, in order to give a proper (correct) picture of nature. These two (quite speculative) reasonings are here used just to introduce the concept of background independence. Background independence is a technical way of saying nothing more than this: if one seeks for a complete quantization of the metric, one should quantize the metric itself and not its small perturbations around a fixed background geometry. In fact, the latter is what is done in the perturbative (or background-dependent) approach to Quantum Gravity: to assume the following splitting
\be
g_{\mu\nu}(x)=g^0_{\mu\nu}(x)+h_{\mu\nu}(x)\ ,
\label{back}
\ee 
where $g^0_{\mu\nu}$ is treated as a classical field, a background metric, and $h_{\mu\nu}$ are its quantum fluctuations, i.e. it is what one tries to quantize. This is obviously a legitimate procedure if one wants to analyze the behavior of a (self-interacting) spin-2 particle on a (generally curved) $g^0_{\mu\nu}$ background. The problem is the following: this splitting breaks down at high energies. Namely this kind of theory has been proved to be UV-unrenormalizable \cite{Goroff:1985sz}. Background-dependence supporters believe that this unrenormalizability is a hint of a more fundamental theory still to be found. This is a chance, indeed. Background-independence supporters instead believe that this is just a hint of a non-proper way to face the problem, namely to believe that the splitting \eqref{back} should be valid at all energy scales. Indeed in the regime we are interested in (namely the Planck scale) it is reasonable to argue that the fluctuations won't be small; thus it would be a nonsense (read `wrong') to assume that the geometry (along with the causal structure) is determined by $g^0_{\mu\nu}$ alone. We do not know what is the fabric of spacetime at the Planck scale, but -- as argued above -- it is likely to be some quantum superposition, some grain-like structure, no more capable (in general) to be described by a smooth field + small fluctuations. In this respect background independence is both more conservative -- because after all makes use of the ``old'' covariant approach -- and more radical -- since it starts from the belief that `pure' quantum spacetime should be deeply different from that of QFT.

Let us come to the second issue: the focus on connection. In GR one usually takes the metric as the ``main character'' of the play. It is seen as the fundamental field of the theory. In LQG instead, the metric tensor is more like a secondary player. Actually it is no news at all that GR can be recast in the language of differential forms and tetrad field (we will soon see in what sense), and the language of tetrads -- as opposed to that of metric tensor -- leads naturally to put the focus on connections. At a classical level, everything is equivalent\footnote{Actually there is  one  (to my knowledge) difference: the tetrad formulation is ``an extension'' of standard GR. We shall see in section~\ref{sec:tetrad} in what sense.}, but for what concerns quantization, the choice of variables will prove to be crucial, and this is one of the key points of the LQG approach. 


\chapter{Canonical Quantum Gravity: from ADM formalism to Ashtekar variables}

{\it The present chapter is dedicated to review the basics of canonical quantization of gravity. Everything is well-known and established, so I will skip tedious  calculations and try to present only the key passages of this subject, especially the ones which turn to be important in the LQG approach\footnote{I refer the reader to the good textbook by Baez and Muniain~\cite{Baez_book} for a thorough and somewhat LQG-oriented presentation of this subject.}.}

\section{Hamiltonian formulation of General Relativity}
The starting point is the well known Einstein-Hilbert action:
\be
S[g_{\mu\nu}]=\frac{1}{16\pi G}\int\d ^4x\sqrt{-g}R\ .
\label{EH}
\ee
In the following I shall always set $16\pi G=1$ for simplicity, and I will recover the constants when to stress the importance of physical scales.

In order to perform the canonical (i.e. hamiltonian) analysis, one needs to know which are the conjugated variables to the metric and then perform the Legendre transform. Indeed the hamiltonian formulation looses the manifest covariance, separating time from space -- so to say -- in order to find the conjugate momenta. 
To this aim, one assumes that it is possible to foliate the spacetime manifold ($\mathcal{M}$ from now on) into spacelike three dimensional hypersurfaces. This amounts to say that $\mathcal{M}$ is diffeomorphic to $\Sigma\times\mathbb{R}$. This assumptions is actually not that restrictive: a theorem by Geroch \cite{Geroch:1970uw,Bernal:2003jb} proves that it has to be so if  $\mathcal{M}$ is globally hyperbolic, namely if it has no causally disconnected regions.
Obviously we are far from saying that we are defining an absolute time coordinate! One can choose each timelike direction as ``time''. This amounts to say that there will be an infinity of diffeomorphisms $\phi :\mathcal{M}\rightarrow\Sigma\times\mathbb{R}$, each inducing a time coordinate $\tau=\phi_* t$  on $\mathcal{M}$ (namely the pullback of the $t$ coordinate in $\mathbb{R}$). This splitting procedure is known as the Arnowitt-Deser-Misner decomposition~\cite{ADM}.

\begin{figure}\centering
\centering
\includegraphics{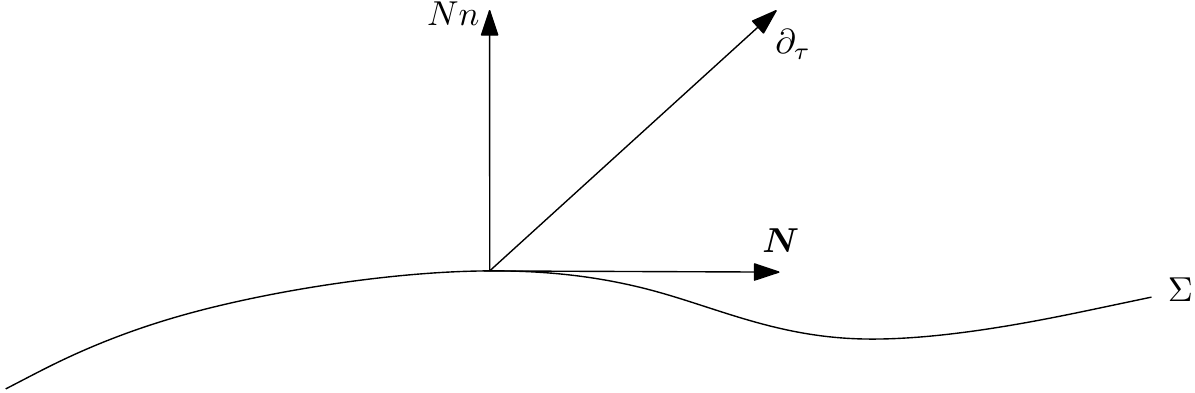}
\caption{Splitting of $\pp_\tau$ into normal and tangential components with respect to $\Sigma$.}
\label{fig:lapse_shift}
\end{figure}

Now we decompose the timelike (coordinate) vector field $\pp _\tau$ into its components tangential and normal to $\Sigma$ (figure~\ref{fig:lapse_shift}):
\be
\pp_\tau =Nn+\vec{N}\ ;
\ee
the \emph{shift vector} (field) $\vec{N}$ belongs to the tangent bundle $T\Sigma$ of $\Sigma$;  $n$ is the unit normal to $\Sigma$ (i.e. $\bm{g}(n,v)=0\ \forall v\in T\Sigma\ ,\ \bm{g}(n,n)=-1$) and its coefficient $N$ is called the \emph{lapse function}.
Using the following identities\footnote{From now on latin indices from the beginning of the alphabet ($a,b,c$\ldots) will denote spacelike indices ranging from 1 to 3.} 
\be
g_{00}=\bm{g}(\pp_\tau,\pp_\tau)=-N^2+g_{ab}N^aN^b\ ;\quad g_{0a}N^a=\bm{g}(\pp_\tau,\vec{N})=N_aN^a\rightarrow g_{0a}=N_a\ ,
\ee
the expression of the spacetime line element in terms of these variables is:
\be
\d s^2=g_{\mu\nu}\d x^\mu\d x^\nu=-(N^2-N_aN^a)\d t^2+2N_a\d t\d x^a+g_{ab}\d x^a\d x^b\ .
\ee
Notice that the induced $\bm{g}$-metric on $\Sigma$ is given by
\be
q_{\mu\nu}=g_{\mu\nu}-n_\mu n_\nu\ ,
\ee
(which has, correctly, $q_{00}=0$) and thus we will talk of $q_{ab}$ as the metric on $\Sigma$.
The object $q^\mu_\nu=\delta^\mu_\nu-n^\mu n_\nu$ is indeed the projector on $\Sigma$
\begin{equation*}
q^\mu_\nu v^\nu=v^\mu-\bm{g}(n,v)n^\mu=\left\{\begin{aligned} 	0;\quad v \text{ normal to }\Sigma\ ,\\
								v;\quad v \text{ tangent to }\Sigma\ .
					      \end{aligned}\right.
\end{equation*}

We want to use these variables $q_{ab},N_a,N$, to rewrite the Einstein-Hilbert action \eqref{EH}. Namely, defining the \emph{extrinsic curvature}
\be
K_{ab}=\mathcal{L}_n q_{ab}=\f12 N^{-1}(\dot{q}_{ab}- ^3\nabla_aN_b- ^3\nabla_b N_a)\ ,
\ee
where $\mathcal{L}_n$ stands for the Lie derivative along the $n$ direction (thus the extrinsic curvature contains informations on how $\Sigma$ is embedded in $\mathcal{M}$) and $^3\nabla$ is the covariant derivative compatible with $q_{ab}$, it is a matter of algebra to find
\be
L=R\sqrt{-g}=q^{1/2}NR=q^{1/2}N(^3R+\Tr(K^2)-(\Tr K)^2)\ .
\ee
The momenta conjugate to the 3-metric are
\be
p^{ab}=\f{\pp L}{\dot q_{ab}}=q^{1/2}(K^{ab}-\Tr K q^{ab})\ ,
\ee
and there are no momenta for $N$ and $N^a$ (they are identically zero).
We are now able to perform the Legendre transform and write down the action \eqref{EH} as
\begin{multline}
S[\pi^{ab},q_{ab},N_a,N]=\int_{\mathbb{R}\times\Sigma}\d t\ \d ^3x\ \left[\pi^{ab}\dot q_{ab}+2N_b\nabla^{(3)}_a(q^{-1/2}\pi^{ab})\right.\\
\left.+N\left(q^{1/2}[R^{(3)}-q^{-1}\pi_{ab}\pi^{ab}+\tfrac12 q^{-1}\pi^2]\right)\right]\ ,
\label{EH1}
\end{multline}
where $^3R$ is the scalar curvature of $q_{ab}$ and $\pi=\pi^{ab}q_{ab}$.
Notice that the absence of momenta for the lapse function and the shift vector means they carry no dynamics, namely that their equation of motion are the following  \emph{constraint} equations:
\be
V^b(q_{ab},\pi^{ab})=-2\nabla^{(3)}_a(q^{-1/2}\pi^{ab})=0
\label{vec}
\ee
and
\be
S(q_{ab},\pi^{ab})=-q^{1/2}[R^{(3)}-q^{-1}\pi_{ab}\pi^{ab}+\tfrac12 q^{-1}\pi^2]=0\ ,
\label{sca}
\ee
usually referred to as the \emph{vector constraint} and the \emph{scalar constraint}, respectively.
With this symbol convention the action \eqref{EH1} becomes simply
\be
S[\pi^{ab},q_{ab},N_a,N]=\int_{\mathbb{R}\times\Sigma}\d t\ \d ^3x\ \left[\pi^{ab}\dot q_{ab}-N_bV^{b}-NS\right]\ .
\label{EH2}
\ee
We can also readily identify the hamiltonian density
\be
\mathcal{H}(\pi^{ab},q_{ab},N_a,N)=N_bV^{b}(\pi^{ab},q_{ab})+NS(\pi^{ab},q_{ab})\ .
\label{H}
\ee
Notice that the hamiltonian \eqref{H} is a combination of constraints, i.e. it vanishes identically on solutions, which is a standard feature of generally covariant systems.

The symplectic structure is very easily found to be
\be
\{p^{ab}(x),q_{cd}(y)\}=\delta^a_{(c}\delta^b_{d)}\delta(x-y)
\label{ADM_poisson}
\ee
and zero otherwise.

Imagine we want to (canonically) quantize this system. Firstly we should find a Hilbert space in which to represent the algebra \eqref{ADM_poisson}. Notice that our configuration space is the space of all the 3-metric on the space manifold $\Sigma$. We shall call this space $\text{Met}(\Sigma)$. Thus what we are looking for is something like $L^2(\text{Met}(\Sigma),\mu)$ with respect to a suitable measure $\mu$ defined on $\text{Met}(\Sigma)$. This is the first great problem of our program of quantization. It is extremely difficult to define such a measure! Indeed this space in not even a vector space, since Einstein equations are non-linear. 

Moreover, there are other difficulties. Namely We have to impose the constraints \eqref{vec}, \eqref{sca}. They're expression in terms of configuration variables and momenta is extremely unfit to quantization, they are non-polynomial and contain nasty factors like $\sqrt{q}$ creating creates all sorts of ambiguities.

By the way, these kind of difficulties are much the same as the ones of Yang-Mills QFT. In that case, however, the perturbative approach (namely to `linearize' the equations of motion) proves to be successful, i.e. the Yang-Mills theory is renormalizable.

\section{The tetrad/triad formalism}\label{sec:tetrad}

We have some more work to do in order to have a suitable form to quantize. In this section we shall talk a bit about the introduction and the importance of the tetrad/triad formalism. 
Let's for a moment forget the ADM spacetime splitting and work covariantly in four dimensions.
The tetrad field, or vierbein field, or frame field is a rule that assigns to each point of spacetime an orthonormal local inertial frame.
Let's do it carefully, since this is often a misleading concept.
You recall that on each spacetime point it is possible to choose a coordinate system in which the Levi-Civita connection is zero (but not its derivatives!). This set of coordinates is usually called Riemann normal coordinate system, or ``free fall'' coordinates. Indeed they are just the mathematical expression of the equivalence principle: this system embodies the possibilities of (locally) ``eliminating'' the gravitational field by ``falling'' in it.
Let us call $\xi^I$, $I=0,1,2,3$ these coordinates.  Obviously we can't define in general a single flat coordinate system on the whole manifold\footnote{Or, better, these coordinates will be flat only in a small region of spacetime.}; if so the manifold would be flat. 
Their key feature is that they are orthonormal, that is
\be
g_{IJ}=\eta_{IJ}\ ,
\ee 
\emph{locally}, i.e. near a specific point: this expression is not true on all the manifold. If we move to a ``distant'' event, it breaks down, in general. However we can always \emph{change coordinates} to \emph{another} $\xi^I$ system, in which this flatness is true in that point.
The transformation matrix between these free fall coordinates-field and a generic coordinate system, is what we call a tetrad\footnote{Actually these are the components of a co-tetrad, since it is a 1-form; it is however common to call it tetrad as well.}, namely
\be
e_\mu^{I}(x)=\f{\pp\xi^{I}}{\pp x^\mu}(x)\ .
\ee
The important thing is that on each point we can put a tetrad $e^\mu_I(x)$ which is thus a tetrad \emph{field}.
There is a fundamental relation between the tetrad and the metric:
\be
\d s^2=\eta_{IJ}\d\xi^I\d\xi^J=\eta_{IJ}e^I_\mu(x)e_\nu^J(x)\d x^\mu\d x^\nu=g_{\mu\nu}\d x^\mu\d x^\nu\ ,
\ee
which means
\be
\quad g_{\mu\nu}(x)=e^I_\mu(x) e^J_\nu(x)\eta_{IJ}\ .
\label{g_e}
\ee
Thus one can always ``reconstruct'' the metric from the tetrad. By the way, the tetrad carries some redundancies: indeed the metric has 10 independent components, while the tetrad 16. This is due to the fact that we can always Lorentz-rotate a free fall system to get another \emph{equivalent} free fall system: 
\be
e^I_\mu(x)\rightarrow \tl{e}^I_\mu(x)=\Lambda^{I}_{J}(x)e^J_\mu(x)\ .
\label{lorentz_gauge}
\ee
This is true whatever Lorentz transformation we attach at each particular point, i.e. it is a Lorentz \emph{gauge} transformation. 
Indeed the Lorentz group SO(3,1) is a 6-parameter group, which matches our counting of independent components.

This is the intuitive physical picture. Actually not so many words are necessary to formally introduce the tetrads. Tetrads are simply a basis of orthonormal vector fields. I.e. 
they are four vector fields $e_I$ such that $\bm{g}(e_I,e_J)=\bm{\eta}_{IJ}$. Every vector can be expanded in a coordinate basis $\pp_\mu$ thus
\be
e_I=e^\mu_I\pp_\mu\ ,
\ee
while, in general, the tetrad are not a coordinate basis.

There is however a high brow (but useful) way of saying these same things, in the language of bundles. We are indeed picking a local isomorphism (sometimes called a \emph{trivialization})
\be
e:\mathcal{M}\times\mathbb{R}^n\rightarrow T\mathcal{M}\ ,
\ee
which permits us to describe (again, locally) the tangent bundle $T\mathcal{M}$ as the trivial bundle $\mathcal{M}\times\mathbb{R}^n$ (obviously for us $n=4$, but this is a rather general construction)\footnote{Notice that the existence of this local isomorphism is part of the definition of a differentiable manifold, namely to locally look like $\mathbb{R}^n$.}. 

This trivialization is what high brow people call a frame field.
Notice that, seen from this perspective, everything sounds quite natural: a section of the trivial bundle is just a function on $\mathcal{M}$ with value in $\mathbb{R}^n$, and we can take a basis of these sections $\xi_I$. Each section of the trivial bundle can thus be expanded as $s=s^I\xi_I$~\footnote{Notice that here we are in the trivial bundle, which is not the tangent bundle, they are only locally isomorphic (this is the same as saying that flat coordinates can be defined only locally). This justifies our use of different kind of indices, capital latin letters, which are sometimes called flat indices.}.
If we take $e(\xi_I)$ we obtain a basis of sections of the tangent bundle, which can thus be expanded in a coordinate basis
\be
e(\xi_I)=e^\mu_I\pp_\mu\ .
\ee 
The trivial bundle carries a metric structure, the Minkowski metric. Thus we can rise and lower the flat indices with $\eta_{IJ}$. This is a quite powerful tool, but it is useful only if we require
\be
\bm{g}(v,w)=\bm{\eta}(e(v),e(w))\ ,
\ee
namely that the local isomorphism sends $\bm{\eta}$ to $\bm{g}$. The condition for this to be fulfilled is simply 
\be
\bm{g}(e(\xi_I),e(\xi_J))=\eta_{IJ}\ 
\ee
that is, we want \emph{orthonormal frame fields} in order to work with the flat metric in the trivial bundle and then simply ``translate'' the results in $T\mathcal{M}$ via $e$.

To get the expression of equation \eqref{g_e} between the metric $\bm{g}$ and the trivial metric, we need the inverse frame field $e^{-1}$ (the co-tetrad, which is usually called tetrad as well, as argued above). Indeed
\be
g_{\mu\nu} =\bm{\eta}(e^{-1}(\pp_\mu),e^{-1}(\pp_\nu))=\bm{\eta}(e^{I}_\mu\xi_I,e^J_\nu\xi_J)=\eta_{IJ}e^J_\nu e^{I}_\mu\ .
\ee

\begin{rem}
An important remark is to be done now. We have a natural step forward to do here: recognize that, along with the trivial vector bundle, we get -- by virtue of the invariance under \eqref{lorentz_gauge} -- a principal $SO(3,1)$-bundle, sometimes called the \emph{frame bundle}. We can think of it as if we were attaching a copy of $SO(3,1)$ to each point in spacetime; a section of this bundle (namely a choice of a specific tetrad among the equivalent ones) is just a gauge choice.
\end{rem}

Now I want to briefly show (or hint) that the whole General Relativity, and in particular the Einstein-Hilbert action \eqref{EH}, can be seen as a theory of a tetrad field $e_\mu^I(x)$ and connection. First of all, we have seen that there are two bundles in the play: a vector bundle $T\mathcal{M}$ and a principal bundle with $SO(3,1)$-fiber. We have a well defined connection in $T\mathcal{M}$, the Levi-Civita connection $\Gamma$, defined as the only torsionless connection compatible with the metric $\bm{g}$: $\nabla\bm{g}=0$, with $\nabla=\pp+\Gamma$ the covariant derivative associated with the parallel transport defined by $\Gamma$.
We can define a connection on  the frame bundle as well, we shall call it $\omega$. It will define a covariant differentiation which will ``act on'' the flat indices, i.e. it defines a parallel transport on the frame bundle.
When we have objects with `a leg in $T\mathcal{M}$ and one in the frame bundle' we have the following ``total'' covariant derivative.
\be
\mathcal{D}_\mu v^I_\nu=\pp_\mu v^I_\nu-\Gamma^\alpha_{\mu\nu} v^I_\alpha+\omega^{I}_{\mu J} v_\nu^J\ .
\ee
Now, the requirement is that this covariant derivative has to be compatible with the tetrad (in order to have scalar products invariant under parallel transport both in the tangent and in the frame bundle), i.e.
\be
\mathcal{D}_\mu e^I_\nu=0\ .
\ee
This requirement links together the Levi-Civita connection and the frame bundle connection, namely
\be
\omega^I_{\mu J}=e^I_\nu\nabla_\mu e^\nu_J\ ,
\label{spin_connection}
\ee
thus implying that the frame bundle connection $\omega$ is not arbitrary and contains informations on the Levi-Civita connection as well. Such a connection is called the \emph{spin connection}, and we shall indicate it as $\omega(e)$ when we want to stress that it is tetrad-compatible (namely that equation \eqref{spin_connection} holds).
One can intuitively think that the Levi-Civita connection is the gauge potential due to the diffeomorphism gauge freedom of general relativity, while the $\omega$ connection is the gauge potential of $SO(3,1)$ gauge freedom. 

The curvature is defined as usual
\be
F^{IJ}=\d\omega^{IJ}+\omega^I_K\wedge\omega^{KJ}\ ,
\ee
and it's of course a 2-form.
If we use the spin connection ``definition'' \eqref{spin_connection} we have easily
\be
F^{IJ}_{\mu\nu}(\omega(e))=e^{I}_\alpha e^{J}_\beta R^{\alpha\beta}_{\ \ \mu\nu}\ ,
\ee
where $R$ is the Riemann tensor constructed out of the metric defined by $e$ (through the usual relation \eqref{g_e}).
This is a central result for our goal: the Riemann tensor is nothing but the curvature of the spin connection.
With this machinery at hand, it is now a matter of calculations to show that the following action
\be
S[e,\omega]=\frac12 \epsilon_{IJKL}\int_\mathcal{M}e^I\wedge e^J\wedge F^{KL}(\omega)
\label{we}
\ee
has the same equations of motions of the Einstein-Hilbert action \eqref{EH}.
Notice that we have explicitly underlined that the action \eqref{we} is to be considered with both $e$ and $\omega$ as independent variables. Many of the readers have certainly recognized the first-order or Palatini formulation of GR. The variation with respect to the connection simply implies that $\omega$ \emph{is} the spin connection on-shell. The vanishing variation with respect to the tetrad gives Einstein equations. 
We could have done the same thing with a second order (standard) formulation, in which the action is the same as \eqref{we}, but we take the curvature to be the one of the spin-connection \emph{ab initio}: again, the equation of motions for $e$ are Einstein field equations. However, it was our goal to stress that Einstein gravity can be seen as a theory of `tetrads and connections as independent variables'.

\begin{rem}
 Notice that the action \eqref{we} is actually an extension of the Einstein-Hilbert gravity, in fact \eqref{we} is well defined also in the degenerate case where the determinant of the metric is 0, i.e. when the metric is not invertible. In this case the tetrad/triad formulation does not crash and still has something to say.  
\end{rem}

Let us now go back to our ADM splitting. We introduce the frame field formalism, but we shall need it only on the spatial manifold $\Sigma$. Thus we call it a \emph{triad} field. The analogous of equation \eqref{g_e} shall be:
\be
q_{ab}=e^i_ae^j_b\delta_{ij}\ ,
\label{e}
\ee
where now we use $i$, $j$\ldots to label the internal 3-space.
We introduce also the the densitized triads
\be
E^a_i=\frac12 \epsilon^{abc}\epsilon_{ijk}e^j_be^k_c\ ,
\label{E}
\ee
and
\be
K^i_a=\frac{1}{\sqrt{\det (E)}}K_{ab}E^b_j\delta^{ij}\ .
\label{K}
\ee
One can check that
\be
\pi^{ab}\dot q_{ab}=2E^a_i\dot K^i_a\ .
\ee
Thus we now have, for the action \eqref{EH2} 
\begin{multline}
S[E_j^a,K^j_a,N_a,N,N^j]=\int_{\mathbb{R}\times\Sigma}\d t\ \d ^3x\ \left[E_i^a\dot K^i_a\right.\\
\left.-N_bV^b(E_j^a,K^j_a)-NS(E^a_j,K^j_a)-N^iG_i(E^a_j,K^j_a)\right]\ .
\end{multline}
where
\be
G_i(E_j^a,K^j_a)=\epsilon_{ijk}E^{aj}K^k_a\ .
\ee

Some remarks: 
\begin{itemize}
 \item
there are here \emph{three} constraints the \emph{vector constraint} $V^b$ imposing spatial diffeomorphism-invariance, the \emph{scalar constraint} $S$ imposing time diff-invariance and the \emph{Gauss} constraint $G_i$, which is due to the redundancy in using tetrads instead of metrics: namely we have a $SO(3)$ gauge freedom (all orthonormal frames in an euclidean 3d manifold are such up to rotations), which is the spatial analog of the Lorentz gauge freedom analyzed above. The $G_i$ constraint is there in order to impose physical states to be $SO(3)$-invariant. 
\item  This action is written in terms of densitized triads $E$ and their conjugate momenta, the $K$'s defined in \eqref{K}. The latter are, essentially, a redefinition fo the extrinsic curvature. Thus here the connection is still in a second plane: we shall need another change of variables in order to have it clearly in the play. 
\end{itemize}

\section{The Ashtekar-Barbero variables}

First of all, let us note that the spin connection $\omega_a^{ij}$ (based on $\Sigma$)  is a 1-form with value in the Lie algebra $\mathfrak{so}(3)$\footnote{This is always the case: a connection on a principal $G$-bundle is always a 1-form with values in $\mathfrak{g}$.}
Thus, we can expand $\omega$ on a basis of the algebra, for example on the standard Pauli matrices $\tau_i=\sigma_i/2$
\be
\omega^{ij}_a=\omega^k_a(\tau_k)^{ij}\ ,
\ee
which (trivially) allows us to speak of $\omega^i_a$, with a single internal index $i$.

There is now one key change of variables that can be done now. We define a new object as follows
\be
A^i_a=\omega^i_a+\gamma K^i_a\ ,
\label{AB}
\ee
where $\gamma$ is a real arbitrary parameter, called the \emph{Immirzi parameter}.
It is easy to see that $A$ is still a connection. Indeed $K^i$ transforms as a vector under local $SO(3)$ transformations, while $\omega$ transforms as a connection, namely 
\be
\omega^i\xrightarrow[]{g} g(x)\omega^i(x)g^{-1}(x)+g\d g^{-1}(x)\ ,\quad  K^i\xrightarrow[]{g} g(x)K^i(x)g^{-1}(x)\ ,
\ee
thus $A^i$ transforms as a connection as well.
This is known as the \emph{Ashtekar-Barbero connection}\footnote{For more details on the Ashtekare-Barbero formulation of GR you can see~\cite{Ashtekar:2004eh}. For the original Ashtekar's contribution see~\cite{Ashtekar:1987gu}.}. The remarkable fact about this variable, is that its conjugate momentum is just the densitized triad, i.e.
\be
\{A^i_a(x),E^b_j(y)\}=\gamma\delta^i_j\delta^b_a\delta(x-y)\ ,\quad\{E^a_i(x),E^b_j(y)\}=\{A^i_a(x),A^j_b(y)\}=0\ ,
\label{pb}
\ee

The Einstein-Hilbert action becomes
\begin{multline}
S[E^a_j,A^j_a,N_a,N,N^i]=\int_{\mathbb{R}\times\Sigma}\d t\ \d ^3x\ \left[E_i^a\dot A^i_a\right.\\
\left.-N_bV^b(E_j^a,A^j_a)-NS(E^a_j,A^j_a)-N^iG_i(E^a_j,A^j_a)\right]\ .
\label{EH3}
\end{multline}
with constraints
\begin{subequations}\label{constraints_AB}
\begin{align}
&V_b(E_j^a,A^j_a)=E^a_jF_{ab}-(1+\gamma^2)K^i_aG_i\ ,\\
&S(E^a_j,A^j_a)=\f{E^a_iE^b_j}{\sqrt{\det(E)}}\left(\epsilon^{ij}_{\ \ k}F^k_{ab}-2(1+\gamma^2)K^i_{[a}H^j_{b]}\right)\ ,\\
&G_i(E^a_j,A^j_a)=D_aE^a_i=\pp_aE^a_i+\epsilon_{ijk}A^j_aE^{ak}\ ,
\end{align}
\end{subequations}
where $F_{ab}$ is the curvature of the Ashtekar-Barbero connection and $D_a$ is its covariant derivative.

We have reached our goal: a formulation of gravity in terms of connection and its conjugate momentum, the triad. It is nothing but a $SO(3)$ gauge theory, plus constraints that implement space diffeomorphism invariance and time diffeomorphism invariance. In a compact sentence we can say we have reduced Einstein gravity to a \emph{background independent $SO(3)$ Yang-Mills theory}.

\section{Smearing of the algebra}\label{smearing}
In order to give a geometric interpretation to the actual variables $(E^a_j,A^i_b)$ we shall now smear them appropriately. Let us start from the densitized triad $E^a_j(x)$. This object is a 2-form, thus we shall smear it on a surface as follows
\be
E_i(S)=\int_SE_i^a(x(\sigma))n_a\d ^2\sigma\ ,
\ee
where $n_a=\epsilon_{abc}\f{\pp x^b}{\pp\sigma_1}\f{\pp x^c}{\pp\sigma^2}$ is the normal to the $S$ surface. The object here defined, $E_i(S)$ is clearly the \emph{flux} of the triad field across the surface, and is generally referred to simply as `the flux'.

The Ashtekar-Barbero connection \eqref{AB} is a one form, thus it is natural to smear it along a one-dimensional path. Take a  path $\gamma$ with parametrization $x^a(s):[0,1]\rightarrow\Sigma$ and take $A_a=A^i_a\tau_i\in SU(2)$, with $\tau_i$ generators of $SU(2)$.
The we can take the integral
\be
\int_\gamma A=\int_0^1A^i_a(x(s))\tau_i\f{\d x^a(s)}{\d s}\d s\ .
\ee
As one can see the triad field is identified on \emph{surfaces} and its conjugate momentum, the connection, on \emph{paths}. This is rather natural on a three dimensional manifold $\Sigma$.

It is useful for future developments,  to define the \emph{holonomy} of the connection along the path $\gamma$
\be
h_\gamma[A]=\mathcal{P}\exp\left(\int_\gamma A\right)\ ,
\ee
where the $\mathcal{P}$ symbol stands for path ordering.\\[14pt]

For future reference it is worth to introduce the so-called \emph{Holst action} for GR. We have seen that Einstein-Hilbert action~\eqref{EH} is equivalent to~\eqref{we}, which is written in terms of tetrad fields and connections, in a first order formalism.
We can add to \eqref{we} a purely topological sector, which classically does not give any contribution to the dynamics
\be
S[e,\omega]=\frac{1}{2}\epsilon_{IJKL}\int e^I\w e^J\w F^{KL}(\omega)+\f{1}{\gamma}\int e_I\w e_J\w F^{IJ}(\omega)\ .
\label{Holst}
\ee
The presence of the Immirzi parameter $\gamma$ is of course arbitrary at this level. I shall not go into details, however it can be proved that this action~\eqref{Holst} is a 4d manner to introduce the Ashtekar connection. Namely, introducing that topological sector amounts to shift the connection variable with the prescription~\eqref{AB}. See e.g.~\cite{Ashtekar:2004eh,Rovelli_book,Dona:2010hm}.

\chapter{Loop Quantum Gravity}

\section{The program}
The program of Loop Quantum Gravity is organized as follows:
\begin{enumerate}
\item[(i)] One starts with the canonical quantization of the fundamental variables and choosing a representation reproducing the commutator algebra. The program of Loop Quantum Gravity starts with choosing the connection as the configuration space  (position-like) variable. Next the \emph{kinematical} Hilbert space $\mathcal{H}_\text{kin}$ will be defined as the space of square integrable functions on the space of all the possible connections $\psi(A)$, with respect to some appropriate measure defined on the space, known as the \emph{Ashtekar-Lewandowski measure}. Notice that this Hilbert space is called kinematical since it is not the actual Hilbert space of the theory, because we still have to deal with the constraints, which -- quite obviously -- restrict this space to one of its subspaces, the \emph{physical} Hilbert space $\mathcal{H}_\text{phys}$.  
\item[(ii)] the following step is to  deal with the constraints. The Gauss and Diffeomorphism constraints have a natural unitary action on $\mathcal{H}_\text{kin}$, thus their quantization is straightforward. Quite loosely the subspace thus obtained is often indicated as $\mathcal{H}_\text{kin}$ as well. Sometimes, to stress the actual distinction, one can find $\mathcal{H}^D_\text{kin}\subseteq \mathcal{H}^G_\text{kin}\subseteq \mathcal{H}_\text{kin}$ where the sup-scripts stand for Diff constrained and Gauss constrained. Thus one obtains the space of solutions to six of the nine constraint equations.
\item[(iii)] the problem and still an open issue of Loop Quantum Gravity, is of course to find the space of solution of the scalar constraint, which actually drive the time evolution of the system, i.e. the dynamics. Thus the physical Hilbert space $\mathcal{H}_\text{phys}$ is still to be found, and also a proper scalar product between physical states, i.e. quantum transition amplitudes, which are the real clue of a every quantum theory.
\end{enumerate}
We will now briefly sketch each of these three steps, i.e. the definition of the kinematical Hilbert space with the Ashtekar-Lewandowski measure, the implementation of the Gauss and diffeomorphism constraints, and we shall discuss a bit the problems inherent to the quantization of the scalar constraint.

I encourage the interested reader to have a look to~\cite{Ashtekar:2004eh,Rovelli_book,Perez:2004hj,Dona:2010hm} for very good reviews of the basics of LQG and to~\cite{new look,Rovelli:2011eq} for more modern approaches to this subject.

\section{The kinematical Hilbert space \texorpdfstring{$\mathcal{H}_\text{kin}$}{}}\label{sec:kin}
Firstly we define the \emph{space of cylindrical functions based on a graph $\gamma$} denoted $\text{Cyl}_\gamma$\footnote{I strongly recommend the reading of~\cite{Baez:1993id} for a very clear and suggestive presentation of this subject.}.

A graph $\gamma$ is a collection of simple paths which meet at most at their endpoints. We call \emph{links} $l$ all these paths and denote with $L$ their total number. We say that$\gamma$ is in $\Sigma$ when all the links (as paths) belongs to $\Sigma$ themselves.
Given a graph $\gamma\subset\Sigma$ and a smooth function $f:SU(2)^{L}\rightarrow\mathbb{C}$, we define a cylindrical function $\psi_{\gamma,f}\in\text{Cyl}_\gamma$ as
\be
\psi_{\gamma,f}[A]= f(h_{l_1}[A],h_{l_2}[A],\ldots h_{l_{L}}[A])\ .
\label{cyl}
\ee
You have to think these paths as the ones of section~\ref{smearing}, i.e. as the natural one-dimensional objects embedded in the spatial hypersurface $\Sigma$ upon which the connection $A$ is integrated.

Next we define the \emph{space of cylindrical function on $\Sigma$} as
\be
\text{Cyl}=\bigcup_{\substack{\gamma\subset\Sigma}}\ \text{Cyl}_\gamma\ ,
\ee
where the union means over all the graphs in $\Sigma$. 
Now for the measure.
Let's define the \emph{Ashtekar-Lewandowski measure}~\cite{Ashtekar:1994mh}:
\be
\mu_{AL}(\psi_{\gamma,f})=\int\prod_{l\subset\gamma}\d h_l\ f(h_{l_1},h_{l_2},\ldots h_{l_{L}})\ ,
\ee
where $\d h$ is the normalized Haar measure over $SU(2)$. Recall that the Haar  measure is  normalized to 1, thus $\mu_{AL}(1)=1$.

With this structure one can define the inner product in the space of $Cyl$:
\be
\bk{\psi_{\gamma,f}}{\psi_{\gamma',g}}=\mu_{AL}(\psi_{\gamma,f}^*\psi_{\gamma',g})=\int\prod_{l\subset\gamma\cup\gamma'}\d h_l\ f^*(h_{l1}\ldots h_{l_{L}})\ g(h_{l1}\ldots h_{l_{L}})\ .
\label{inner}
\ee
Now we are able to define the kinematical Hilbert space $\mathcal{H}_\text{kin}$ as the Cauchy-completion of the space of cylindrical function in the Ashtekar-Lewandowski measure. That this is indeed a Hilbert space is not obvious, and it was proved by Ashtekar and Lewandowski~\cite{Ashtekar:1994mh}.

Having defined properly the Hilbert space, we can now move to find a basis. We shall use the Peter-Weyl theorem, which can be seen as a form of generalized Fourier expansion. Namely, every function $f\in L^2[SU(2),\d\mu_{\text{Haar}}]$ can be written as follows
\be
f(g)=\sum_j\sqrt{2j+1}f^{mm'}_j D^j_{mm'}(g)\ ,
\label{PW}
\ee
where the $D$'s are nothing but the Wigner $SU(2)$-representation matrices, $j$ is the spin label and run from 0 to infinity in half integer steps, and the $f$ (Fourier) coefficients are given by
\be
f^{mm'}_j=\sqrt{2j+1}\int_{SU(2)}\d\mu_{\text{Haar}}f(g)D^j_{mm'}(g)\ .
\ee
The completeness relation reads
\be
\delta(gh^{-1})=\sum_j(2j+1)\Tr(D^j(gh^{-1}))\ .
\ee
We can easily apply this to any cylindrical function
\be
\psi_{\gamma,f}[A]=\prod_{i=1}^{L}\sqrt{2j_i+1}\sum_{j_1\ldots j_{L}}f^{m_1\ldots m_{L},n_1\ldots n_{L}}_{j_1\ldots j_{L}}D^{j_1}_{m_1n_1}(h_{l_1}[A])\ldots D^{j_{L}}_{m_{L}n_{L}}(h_{l_{L}}[A])
\ee
where the Fourier coefficient are obtained by taking the inner product \eqref{inner} of the $\psi$ function with the tensor product of the Wigner matrices.
Thus we have found a complete orthonormal basis of $\mathcal{H}_\text{kin}$, namely -- calling $\phi^j_{mm'}=\sqrt{2j+1}D^j_{mm'}$ the normalized Wigner matrix -- the functions
\be
\prod_{i=1}^{L}\phi^{j_i}_{m_im'_i}\ ,
\label{basis}
\ee
with the remark of taking all the possible graphs in $\Sigma$ and all the values of spins associated to the links of the graph.

\section{The Gauss constraint. \texorpdfstring{$\mathcal{H}^G_\text{kin}$}{}}\label{sec:G}
Now we want to restrict the Hilbert space on its gauge invariant subspace, i.e. the subspace in which the solutions of the Gauss constraint live. What is to do, is simply to take only the states of $\mathcal{H}_\text{kin}$ which are $SU(2)$ invariant. For this purpose, let us try to understand how $SU(2)$ gauge transformations act on the basis \eqref{basis}.

It is indeed very easy to infer the result of a (finite) $SU(2)$ gauge transformation on a general cylindrical function from the behavior of the holonomy:
\be
h_l[A]\xrightarrow{g} g(x(0))h_l[A]g^{-1}(x(1))\ ,
\ee
where $x(s)$ is the parametrized path.
Usually, given a path/link $l$, one calls $s(l)$ the \emph{source} of the path, i.e. the $x(0)$ point, and $t(l)$ the \emph{target} $x(1)$, thus writing the previous equation as: $h_l[A]\xrightarrow{g} g_{s(l)}h_l[A]g_{t(l)}^{-1}$.
Then, the action on a general graph is
\be
\psi_{\gamma,\ f}[A]=f(h_{l_1},\ldots h_{l_L})\xrightarrow{g} f(g_{s(l_1)}h_{l_1}[A]g^{-1}_{t(l_1)},\ldots g_{s(l_L)}h_{l_L}[A]g^{-1}_{t(l_L)})\ ,
\label{gauge_state}
\ee
and we would like to find a basis of the space of cylindrical functions with such a property.
Let us denote with $U(g)$ the operator that acts the transformation \eqref{gauge_state}, i.e. -- writing it for the basis \eqref{basis} --
\be
U(g)\phi_{mm'}^j(h_l)=\phi^j_{mm'}(g_{s(l)}h_lg_{t(l)}^{-1})\ ,
\ee
or, more generally,
\be
U(g)\prod_{i=1}^{L}\phi_{m_im_i'}^{j_i}(h_{l_i})=\prod_{i=1}^{L}\phi^{j_i}_{m_im_i'}(g_{s(l_i)}h_{l_i}g_{t(l_i)}^{-1})\ .
\ee
Notice that the action of a gauge transformation is \emph{on the nodes} only: We can focus on the request of invariance on a single node, and then extend trivially the results to every node of a graph.
In order to make things as clear as possible and also useful for future reference, let us take a 4-valent node $n_0$. Let us write down in a smart way its generic state
\be
\psi_{\gamma ,\ f}[A]=\sum_{j_1\ldots j_4}\left(\phi^{j_1}_{m_1m'_1}(h_{l_1}[A])\ldots \phi^{j_4}_{m_4m'_4}(h_{l_4}[A])\right)R_{j_1\ldots j_4}^{m_1m'_1\ldots m_4m'_4}\ ,
\ee
where $R$ is all the rest of the factorization. $l_1$ to $l_4$ are the four links converging at $n_0$.
What we are going to do is to render invariant this small piece of graph, that is the one sourrounding this node. The idea is to group average, i.e. to take the action of a gauge transformation in $n_0$ and to integrate over all the possible such transformations:
\be
\phi^{j_1}_{m_1m'_1}(h_{l_1}[A])\ldots \phi^{j_4}_{m_4m'_4}(h_{l_4}[A])\rightarrow\int_{SU(2)}\d g\ \phi^{j_1}_{m_1m'_1}(gh_{l_1}[A])\ldots \phi^{j_4}_{m_4m'_4}(gh_{l_4}[A])
\ee
where we have assumed that $n_0$ is the source for all the four links, but it would be the same otherwise. It is clear that the result is invariant under the action of $U$ on the node $n_0$, since the Haar measure is such!
Now, we can simply do this on each node of a graph and obtain the desired invariant basis. But let us investigate a bit more such a projection on the single node $n_0$:
since the $\phi$ functions are essentially Wigner matrices it is obvious that
\be
\phi^j(gh)=D^j(g)\phi^j(h)\ ,
\ee
and then the projected state looks like
\be
\int_{SU(2)}\d g\ \left(D^{j_1}_{m_1m'_1}(g)\ldots D^{j_4}_{m_4m'_4}(g)\right)\phi^{j_1}_{m_1m'_1}(h_{l_1}[A])\ldots \phi^{j_4}_{m_4m'_4}(h_{l_4}[A])\ .
\ee
Let us give a name to this operator 
\be
P^{n_0}_{m_1m'_1\ldots m_4m'_4}=\int_{SU(2)}\d g\ \left(D^{j_1}_{m_1m'_1}(g)\ldots D^{j_4}_{m_4m'_4}(g)\right)\ .
\ee
This is indeed \emph{a projection operator from the tensor product of the four representation spaces to its $SU(2)$ invariant subspace}, where the projection property $P^{n_0}P^{n_0}=P^{n_0}$ is due to the Haar measure invariance, as it is easy to check. We can write it as
\be
P^{n_0}:V^{j_1}\otimes V^{j_2}\otimes V^{j_3}\otimes V^{j_4}\rightarrow \text{Inv}[V^{j_1}\otimes V^{j_2}\otimes V^{j_3}\otimes V^{j_4}]\ ,
\ee
We can pick an orthonormal basis in $\text{Inv}[V^{j_1}\otimes V^{j_2}\otimes V^{j_3}\otimes V^{j_4}]$ -- let us call it $\ket{\i}$ -- and decompose $P^{n_0}$ as
\be
P^{n_0}=\sum_{\i}\ket{\i}\bra{\i}\ .
\label{Pint}
\ee
Obviously the dimension of the invariant subspace depends on the spins that contribute to the node.
The basis vectors $\ket{\i}$, i.e. an orthonormal basis of the invariant subspace of a tensor product of vector spaces, are usually known as \emph{intertwiner operators} or simply \emph{intertwiners}.

\begin{calc}
We can do a very simple calculation just to explain how things works. Take a 3-valent node with spins $j_1,j_2j_3$. we have to deal with $V^{j_1}\otimes V^{j_2}\otimes V^{j_3}$, which is often written in the sloppy way $j_1\otimes j_2\otimes j_3$. Then we decompose it into the sum of irreducible representations, namely
\be
j_1\otimes j_2\otimes j_3=(|j_1-j_2|\oplus\ldots j_1+j_2)\otimes j_3\ ,
\ee
and so on, which is of course the well-known Clebsh-Gordon condition.
Take three values, for example $j_1=1/2,j_2=1,j_3=3/2$, then the decomposition reads
\be
\f12\otimes 1\otimes\f32=(\f12\oplus\f32)\otimes\f32=1\oplus2\oplus0\oplus1\oplus2\oplus3\ .
\ee
The invariant subspace is, by definition, the $0$ representation space. The dimension of $\text{Inv}[j_1\otimes j_2\otimes j_3]$ is given by the multiplicity of the $0$ representation in the decomposition. It is not hard to see that for a 3-valent node there will be at most a one dimensional invariant subspace, and this is the case when the `selection rule'
\be
j_3\in \{|j_1-j_2|\ldots j_1+j_2\}\ 
\label{sel_rule}
\ee
holds. This means that for every 3-valent node that satisfy the selection rule \eqref{sel_rule}, the projector into the invariant subspace is unique up to normalization. Actually it is just the (normalized) Wigner $3j$-symbol~\cite{wigner}
\be
\bk{\alpha_1,\alpha_2,\alpha_3}{\i}=\i^{j_1,j_2,j_3}_{\alpha_1,\alpha_2,\alpha_3}\sim\begin{pmatrix}j_1 & j_2 &j_3\\ \alpha_1 & \alpha_2 & \alpha_3\end{pmatrix}\ .
\ee
Every node with valence more than 3 can be decomposed into 3-valent contractions. For instance, a 4-valent intertwiner can be written as
\be
v^{\alpha_1,\alpha_2,\alpha_3,\alpha_4}_i=\i_i^{\alpha_1,\alpha_2,\beta}\i_{i,\beta}^{\ \ \alpha3,\alpha_4}\ ,
\ee
the label $i$ is a spin representation, it ranges over the Clebsh-Gordon-allowed representations joining the four spins. Graphically is everything quite self-explanatory: you can see the 4-valent intertwiner as
\begin{equation*}
\mathgraphics{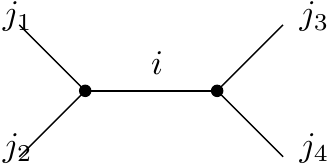}
\end{equation*}
where $i$ is sometimes said to be a virtual link, since it is a spin, but it is  spans all the allowed spins. Of course one can join the four different links in more than a way. In this regard the following equality holds
\be
\mathgraphics{fig/4vint-eps-converted-to}=\sum_j(2j+1)\begin{pmatrix} j_2 & j_1 & j\\j_3 & j_4& i\end{pmatrix}\mathgraphics{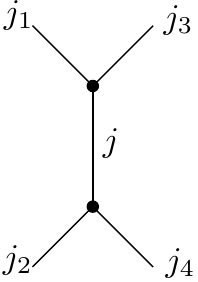}
\label{recoupling}
\ee

I refer to the appendix of~\cite{Rovelli_book} for more details on $SU(2)$ recoupling theory.
\end{calc}

In general, for a $v$-valent node $n$, one can define with the same procedure of group averaging, a projection operator $P^n$, given by
\be
P^n_{m_1\ldots m_v,n_1\ldots n_v}=\sum_{\alpha_v}\i_{m_1\ldots m_v}^{\alpha_v}\i_{n_1\ldots n_v}^{\alpha_v\ *}\ ,
\ee
where $\alpha_v$ denotes the elements of the basis.
This is clearly a generalization of equation \eqref{Pint}.		

Thus, we have come to a result: an orthonormal basis of $\mathcal{H}_\text{kin}^\mathcal{G}$ is given by the following
\be
s_{\Gamma,\{i\},\{j\}}[A]=\bigotimes_{l}\phi^{j_l}(h_l[A])\cdot\bigotimes_n \i_n\ ,
\label{spin_net}
\ee
where $l$ runs over the links and $n$ overs the nodes of the graph $\Gamma$. The $\cdot$ stands for the contraction between the Wigner matrices and the intertwiner operators.
These elements are manifestly $SU(2)$ invariant, thanks to the contraction with the invariant intertwiner operators, and they are know as \emph{spin network states}\footnote{See \cite{Baez:1994hx,Baez:1995md} for interesting details.}.

For future reference, and also to understand better the relation between the graph and the intertwiners defined on it, let us analyze a bit the situation of a 4-valent node.

\section{The vector/diffeomorphism constraint. \texorpdfstring{$\mathcal{H}^D_\text{kin}$}{}}\label{sec:diffs}

We shall now deal with space diffeomorphisms. We proceed just as for $\mathcal{H}^G_\text{kin}$, i.e. we first see how diffeomorphisms act on the vectors of $\mathcal{H}^G_\text{kin}$, then we try to ``average'' over all the possible actions, obtaining a diff-invariant subspace. Here, however, one must be more careful, since diff-invariant functions surely won't be inside $\mathcal{H}^G_\text{kin}$, because the orbits of the action of diffeomorphisms are not compact. It is just as if you want to constrain wave functions defined on a cylinder $\psi(\theta,x)$, $\psi\in L^2(S^1\times\mathbb{R})$, to $\hat{p}_\theta\psi=0$ and $\hat{p}_x\psi=0$.
This constraints say simply that $\psi$ cannot depend neither on $x$ nor on $\theta$, it is a constant. But while the integration on $\theta$ gives no problem, since $S^1$ is compact, the integration over $x$ is not bounded, and the  constrained states do not belong to $L^2$.
However it is always possible to choose a suitable dense subset $\Phi\subset L^2$ of test functions and define the constrained states as distributions on this space. Then one gets the Gelfand triple (see, e.g. \cite{Gelfand}) $\Phi\subset L^2\subset\Phi^*$, where $\Phi^*$ is the `extension' of $L^2$ we are looking for. 

If $\phi$ is a diffeomorphism of $\Sigma$, then its action on a cylindrical function $\psi_{\gamma,f}$ \eqref{cyl} is obvious
\be
U_{D}[\phi]\psi_{\gamma,f}[A]=\psi_{\phi^{-1}\gamma,f}[A]\ .
\label{Udiff}
\ee
Then the vector constraint can be rephrased as
\be
U_{D}[\phi]\psi=\psi\ ,
\ee
where, as  we have just said, the solutions can be found only in distributional sense, namely $\text{Cyl}\subset\mathcal{H}_\text{kin}\subset\text{Cyl}^*$, with $\psi\in\text{Cyl}^*$ the space of linear functionals on $\text{Cyl}$.
Now it is not worth to go on in technicalities: one simply average the distributions on $\text{Cyl}$ over all possible diffeomorphisms, thus obtaining only diff-invariant states. A look to \eqref{Udiff} suggests that the resulting space is formed by the equivalence classes of graphs under spatial diffeomorphisms. These graphs are, mathematically speaking, \emph{knots}\footnote{A knot is an embedding of a circle in a 3d space, up to isotopies. See, e.g. \cite{Baez_book} for details.}, and the spin network states after this procedure are also known as \emph{s-knots}. They are a basis of the space $\mathcal{H}_\text{kin}^{D}$.

Summarizing/simplifying: We have first obtained the kinematical Hilbert space from a suitable definition of a measure on $\text{Cyl}$. Then we have seen that imposing the Gauss constraint amounts to insert intertwiner operators at each node of the basis functions. Finally we have shown that the diff constraints simply say that we have to take the equivalence classes of the base graphs (knots).

\section{Quantization of the algebra and geometric operators}\label{sec:geom}

\begin{figure}\centering
 \includegraphics[scale=1]{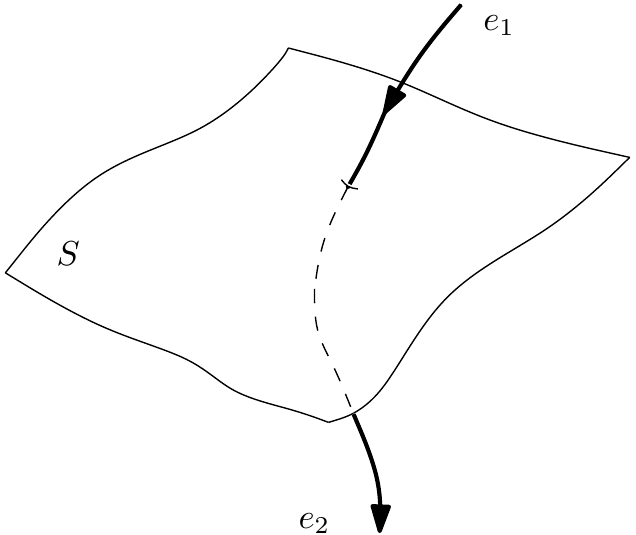}
\caption{The flux surface $S$ `punctured' by an holonomy $e=e_1\cup e_2$.}
\label{fig:puncture}
\end{figure}

We have been able to interpret and ``solve'' the Gauss constraint without actually quantize its corresponding operator in equation \eqref{constraints_AB}. Here we shall face the problem of quantizing the triad operator $E^a_i$ and the connection operator, or, better, their smeared version as introduced in section \ref{smearing}.
It is rather simple: we have defined the Hilbert space $\mathcal{H}_\text{kin}$ as (an opportune definition of) $L^2(\mathcal{A})$, thus on then spin-network basis \eqref{spin_net}, the connection operator shall act by multiplication, i.e. -- considering for simplicity the fundamental representation $h_e=D^{1/2}(h_e)$ --
\be
\hat{h}_\gamma[A] h_e[A]=h_\gamma[A] h_e[A]\ .
\ee
The flux shall act by derivation
\be
\hat{E}_i(S) h_e[A]=-i\hbar\gamma\int_S\d^2\sigma\f{\delta h_e[A]}{\delta A^i_a(x(\sigma))}=\pm i\hbar\gamma h_{e_1}[A]\tau_ih_{e_2}[A]\ .
\label{flux_quant}
\ee
Some remarks: the flux should be seen as a surface (namely the $S$ surface) (see figure~\ref{fig:puncture}); the holonomy (i.e. the smeared connection) is represented by the path $e$. If the surface $S$ and the path $e$ does non cross each other then the action above is identically zero. If they do cross each other, then the path is ``cut'' in two sub-paths $e_1$ and $e_2$ and the flux operator inserts an $SU(2)$ generator among them.
The sign in front depends on the relative orientation of $S$ and $e$.
Notice that recovering all the fundamental constants one gets an overall Planck length squared $l^2_p$ for $\hat E_i(S)$ (replacing the $\hbar$),
\be
l_p=\sqrt{\f{G\hbar}{c^3}}\simeq 10^{-33}{\rm cm}\ ;
\ee
this will be important to understand the scales we are working with.

Now that we have understood how the fundamental variables of our theory work, once quantized as operator living in our Hilbert space, we can turn to really interesting questions about the structure that emerges from this framework of canonical Quantum Gravity.

Actually it is quite simple to see that the area of a surface $S$ embedded in our space manifold $\Sigma$ can be written in a amazingly simple way in terms of triad variables, namely:
\be
A(S)=\int_S\d\sigma^1\d\sigma^2\sqrt{E^a_iE^b_j\delta^{ij}n_an_b}\ ,
\label{area}
\ee
but in our quantum theory the triad is an \emph{operator} that acts on $\mathcal{H}_\text{kin}^\mathcal{G}$, what shall become the area of a surface?

First of all let us understand how the product of two triads acts on holonomies. With \eqref{flux_quant} is easy to see that 
\be
\hat{E}_i(S)\hat{E}_j(S)h_e[A]=-l_p^4\gamma^2h_{e_1}[A]\tau_i\tau_jh_{e_2}[A]\ .
\ee
When the two fluxes are contracted one obtains the Casimir operator of the representation. In this case we have simply holonomies, i.e. fundamental representation, thus $C^2=\tau_i\tau^i=-\tfrac{3}{4}\id_2$.
Notice that the Casimir commutes with all the group elements: this fundamental feature allows us to ``recompose'' the path $e$, namely
\be
\hat{E}_i(S)\hat{E}^i(S)h_e[A]=-l_p^4\gamma^2 C^2h_{e}[A]\ .
\ee
This means that the holonomy is an eigenstate of the square of fluxes! In a generic spin-$j$ representation one gets
\be
\hat{E}_i(S)\hat{E}^i(S)D^j(h_e[A])=l_p^4\gamma^2 j(j+1) D^j(h_{e}[A])\ ,
\ee
since the Casimir reads $C^2_j=-j(j+1)\id_{2j+1}$.

\begin{figure}\centering
 \includegraphics[scale=1]{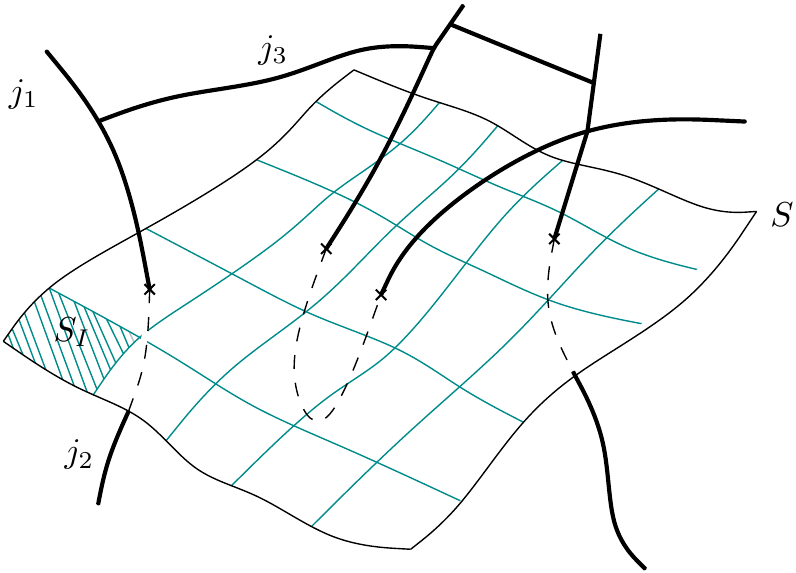}
\caption{A surface $S$ intersected by a generic spin network. You can see that refining again the regularization does not change the result, since each sub-cell has already at most one puncture.}
\label{fig:area}
\end{figure}

Now let us turn back to the area \eqref{area}. We have to write it in a way to include the smeared version of the triad, the flux, in order to be able to act with it on a general state vector.
Thus we have to pick up a regularization of the area \eqref{area}: We decompose the surface $S$ into $N$ 2-cells, and write the integral as the limit of the Riemann sum
\be
A(S)=\lim_{N\rightarrow\infty}A_N(S)\ ,
\ee 
with
\be
A_N(S)=\sum_{I=1}^N=\sqrt{E_i(S_I)E^i(S_I)}\ .
\ee
Summarizing: This amounts to divide $S$ in $N$ 2-cells, take the area of each of them $A_N(S)$, and let $N$ go to infinity. It is just a way to regularize the integral over $S$. Next we have to deal with $A_N(S)$.
We shall see what kind of action the operator  $\sqrt{\hat{E}_i(S_I)\hat{E}^i(S_I)}$ does on a generic state vector $\psi_\Gamma$.
The trick is to take the regularization sufficiently fine so that each $S_I$ intersects once and only once the embedded graph $\Gamma$. If so, we already know how this operator acts, namely
\be
\sqrt{\hat{E}_i(S_I)\hat{E}_j(S_I)}\ \psi_\Gamma=\gamma l_p^2\sqrt{j_p(j_p+1)}\ \psi_\Gamma\ ,
\ee
where $p$ denotes the single link of $\Gamma$ that puncture that specific $S_I$. Refining again the decomposition will have no consequences, since at most we have no intersection at all for some cells, but that gives a zero contribute to the area (see figure~\ref{fig:area}). Thus, at the end, 
\be
\hat{A}(S)\ \psi_\Gamma=\sum_{p\in S\cup \Gamma}\gamma l_p^2\sqrt{j_p(j_p+1)}\ \psi_\Gamma\ .
\ee

The great clues of this formula are: 
\begin{itemize}
\item the area operator is \emph{discrete},
\item its eigenfunctions are just the spin network states,
\item a part from the Immirzi parameter (which is a free parameter of LQG, even if there are some proposals to fix its value through the Black Hole entropy calculation~\cite{Domagala:2004jt,Meissner:2004ju}) the area eigenvalues are of the scale of the Planck length squared, which is precisely what intuitively we expected.
\end{itemize}
We shall see in a moment what a powerful intuitive picture these two key points provide. First, let us deal with the volume operator. We have to say that this issue contains some technicalities, for instance there are at present two distinct mathematically well defined volume operators. The distinctions come out in the regularization process. We shall not discuss it here, we merely present the results that are in agreement for both the versions, which -- incidentally -- are the truly interesting part of the story.
The volume operator has the same two remarkable features of the area operator, namely it has discrete spectrum and it is diagonal in the spin network basis, and is of order $l_p^3$.

Now we are ready to argue the picture that quantization is suggesting us: \emph{space geometry is discrete at the Planck scale}. Each spin network is a polymer-like excitation of space: volume excitation being dual to the nodes of a spin network and area excitation dual to the links. The area excitation is proportional to the spin quantum number, attached to every link of a spin network, while volume ones are determined by the intertwiner space of the node to which it is dual. This interpretation has also the power of being fully background independent, space geometry is determined in a pure combinatorial way by the spin network state, which thus can be seen as a \emph{quantum geometry} of space. Something which is obvious, but it is worth some words, is that this is not a built-in discretization, as it for instance in the lattice formulation of Quantum Gravity: quantization is telling us that space has a discrete nature, whose building blocks are of the order of the Planck scale.

Of course one can rise many objections to all the structure we have built so far, but it cannot be denied that this picture is extremely suggestive and powerful, and this is at the heart of Loop Quantum Gravity itself.

\section{The scalar constraint}\label{sec:scalar_constraint}
One step remains to be analyzed in order to complete the program of (canonical) Loop Quantum Gravity, i.e. to deal with the scalar constraint \eqref{constraints_AB}. Let us recall its expression, in a smeared form
\begin{align}
S(N)&=\int_{\Sigma}\d ^3x\ N \f{E^a_iE^b_j}{\sqrt{\det(E)}}\left(\epsilon^{ij}_{\ \ k}F^k_{ab}-2(1+\gamma^2)K^i_{[a}H^j_{b]}\right)\\
&=S^E(N)-2(1+\gamma^2)T(N)\ .
\label{ham}
\end{align}
The second line is a usual shorthand notation to separate the so-called euclidean contribution
\be
S^E(N)=\int_{\Sigma}\d ^3x\ N \f{E^a_iE^b_j}{\sqrt{\det(E)}}\epsilon^{ij}_{\ \ k}F^k_{ab}
\ee
from the rest
\be
T(N)=\int_{\Sigma}\d ^3x\ N \f{E^a_iE^b_j}{\sqrt{\det(E)}}\ K^i_{[a}H^j_{b]}\ .
\ee
The non-linearity if this expression is really awkward, in order to be able to quantize it properly. Notice that this is just the same problem one had before the loop representation. And it is no mystery that this problem remains the real big problem of canonical Quantum Gravity.

However, the rich structure we have built so far, surely has made possible many important steps towards a better comprehension of the quantum scalar constraint and (particularly) of its action on the kinematical state space. We shall not deal here with the many complicated open issues in this respect~\cite{Thiemann:1996aw,Thiemann_book} but we want at least to explain what has and what has not been achieved by now.

\begin{figure}\centering
 \includegraphics[scale=1]{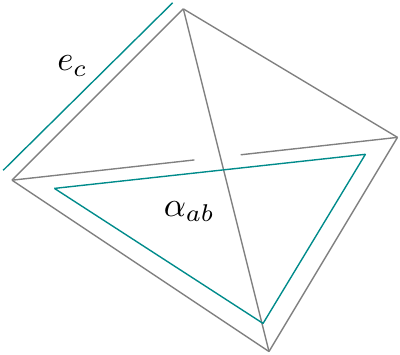}
\caption{One tetrahedron of the regularization cell-decomposition. One face-loop and one edge-path are shown.}
\label{fig:scalar}
\end{figure}

One important improvement has been put forward by Thiemann~\cite{Thiemann:2000bw}. He observed that, if one introduces
\be
\tl{K}=\int_\Sigma K^i_aE^a_i\ ,
\ee
then the following identities hold
\begin{align}
&K^i_a=\gamma^{-1}(A^i_a-\Gamma^i_a)=\gamma^{-1}\{A^i_a,\tl{K}\}\ ,\\
&\tl{K}=\f{\{S^E(1),V\}}{\gamma^{3/2}}\ ,\\ 
&\f{E^a_iE^b_j}{\sqrt{\det(E)}}\epsilon^{ijk}\epsilon_{abc}=\f{4}{\gamma}\{A^k_a,V\}\ ,
\end{align}
with $V=\int\sqrt{\det(E)}$.
This is actually very useful, since \eqref{ham} becomes
\be
S^E(N)=\int\d ^3x\ N\epsilon^{abc}\delta_{ij}F^i_{ab}\{A^j_c,V\}\ ,
\ee
and
\be
T(E)=\int\d ^3x\ \f{N}{\gamma^3}\epsilon^{abc}\epsilon_{ijk}\{A^i_a,\{S^E(1),V\}\}\{A^j_b,\{S^E(1),V\}\}\{A^k_c,V\}\ .
\ee
Now the trick proceeds by rewriting both the connection and the curvature in terms of holonomies and so come to an expression that involves only the volume and the holonomies. This expression can be transformed into an operator, since we already know how $\hat{h}$ and $\hat{V}$ act on a generic state of our Hilbert space. 
We do not list here all the passages, be sufficient to say that thanks to the well known
\be
h_{e_a}[A]\simeq 1 + \varepsilon A^i_a\tau_i+O(\varepsilon^2)\ ,
\ee
-- where $e_a$ is a path along the $x^a$ direction -- one can express both $A$ and $F$ in terms of $h$.
The integral in \eqref{ham} must be regularized as for the area and volume operator, thus the space must be decomposed into a cellular decomposition, a triangulation, i.e. a collection of tetrahedra bound together. We give here the regularized expression for the euclidean part of \eqref{ham}:
\begin{multline}
S^E(N)=\lim_{\varepsilon\rightarrow 0}\sum_IN_I\varepsilon^3\epsilon^abc\Tr(F_ab(A)\{A_c,V\})=\\
\lim_{\varepsilon\rightarrow 0}\sum_I N_I\epsilon^{abc}\left[(h_{\alpha_{ab}^I}[A]-h^{-1}_{\alpha^I_{ab}}[A])h^{-1}_{e^I_c}[A]\{h^{-1}_{e^I_c}[A],V\}\right]\ ,
\label{SE}
\end{multline}
$\varepsilon^3$ is the volume of a cell, $\alpha_{ab}$ is an infinitesimal loop in the $ab$-plane, around the face of the $I$-th cell; while $e_a^I$ is an infinitesimal path along the $a$-direction, along an edge of the $I$-th cell (see figure~\ref{fig:scalar}). Notice that the dependence on the cell scale disappears in terms of holonomies.

\begin{figure}\centering
 \be
\hat H\quad \mathgraphics[scale=.8]{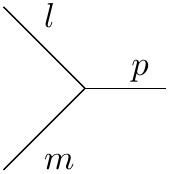}\quad\sim\quad \mathgraphics[scale=.8]{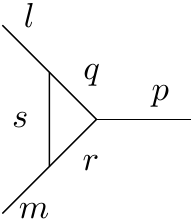}\ .
\ee
\caption{Typical action of the hamiltonian operator on a node of spin network. In the framework we have sketched $s=1/2$, but we shall see later that links with generic spin label could be consistent as well, and this is one of the (many) ambiguities in the quantization of the scalar constraint.}
\label{fig:Haction}
\end{figure}

Now we can formally `quantize' this operator \eqref{SE}  by putting an hat over the volume and the holonomies and see how it acts on a spin network:
\be
\hat S^E(N)=\lim_{\varepsilon\rightarrow 0}\sum_IN_I\epsilon^{abc}\left[(\hat h_{\alpha_{ab}^I}[A]-\hat h^{-1}_{\alpha^I_{ab}}[A])\hat h^{-1}_{e^I_c}[A]\{\hat h^{-1}_{e^I_c}[A],\hat V\}\right]\ .
\label{qSE}
\ee
This operator is well-defined and it is very easy  to see how it acts on $\mathcal{H}_\text{kin}^D$: it has the property of the volume operator to act only on nodes of spin networks, while the holonomies in \eqref{qSE} modifies the spin network by creating new links in the fundamental representation (spin 1/2) (see figure~\ref{fig:Haction}).

(We have dealt only with the euclidean part, but a similar analysis can be done for the $T$ part of \eqref{ham} as well, and the following results still hold.) 
This way of acting is at the real clue of Loop Quantum Gravity with respect to the quantization of the scalar constraint. For example, take all the spin network states with no nodes: they are just \emph{Wilson loops}, i.e. the trace of the holonomy around a closed path. All the Wilson loops, irrespectively from the spin, are solutions of the quantum scalar constraint equation! Actually this simple but important fact is what started the interest in Loop Quantum Gravity: it provides a set of explicit solutions of the quantum theory of gravity.
Moreover, equation \eqref{qSE} tells us that the action on nodes is rather peculiar: it creates special links, sometimes called \emph{exceptional links}. They are special in the sense that the new nodes they carry are exactly of zero volume, thus being `invisible' to a further action of the hamiltonian operator.
This allow the following picture: a generic solution of the quantum scalar constraint is labeled by graphs with nodes of the following kind
\be
\mathgraphics{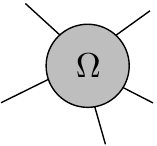}=\alpha\mathgraphics{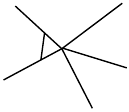}+\ldots +\beta\mathgraphics{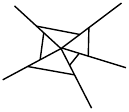}+\ldots +\gamma\mathgraphics{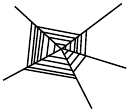}+\ldots\ ,
\ee
also known as \emph{dressed nodes}, i.e. with infinite superpositions of exceptional edges. The
 $\Omega$ label stands for the collection of weights of the superposition.

Up till now the good points. Now let us spend some words for the big problems still to be solved:
\begin{itemize}
 \item the action of the hamiltonian operator that we have described above is said to be \emph{ultra-local}~\cite{Smolin:1996fz}. This feature has raised some concerns whether this kind of theory is at all capable of reproducing general relativity in the classical limit~\cite{Smolin:1996fz}.
\item there is a large degree of ambiguity in the definition of the quantum scalar constraint! One of such ambiguities arises from using the holonomies in the fundamental representation: we could have used any representation, thus the action of the scalar operator would have been different, creating links with arbitrary spin. This doesn't affect the ultra-locality issue, but certainly we would end up with an infinite set of dynamical theories, all different. Another ambiguity is in the regularization scheme: the scalar quantum operator is regularization-dependent!
\item
there is still a mathematical problem in analyzing thoroughly the limit in \eqref{qSE} (and the analogous one for the $T$ part of \eqref{ham}), the limit must exist and should give a well defined operator acting on the space of s-knots. 
\end{itemize}

These kind of ambiguities have stimulated both the research into a better comprehension of the scalar constraint itself, for example the Master constraint approach~\cite{Thiemann:2005zg}\footnote{See also the book by Thiemann~\cite{Thiemann_book} for a more self-contained and thorough presentation of the entire scalar constraint problem.}, and new paths of research, the most important of which is the spinfoam formalism, to which we dedicate the next sections of this chapter.

\section{Concluding remarks}

I have tried in this chapter to review the very basics of LQG. Of course many are the arguments I had to omit not to go beyond the goals I have in mind. Here I show a schematic list of the most important issues I have not treated, but that certainly deserve attention, in order to understand the global importance of such theory:
\begin{itemize}
 \item First of all the scalar constraints definitely deserves more attention. I refer to \cite{Thiemann_book}.
\item \emph{Black Hole entropy}. The idea is to quantize a sector of the theory containing
an isolated horizon and then to count the number of physical states compatible with a given macroscopic area of the horizon. See \cite{Ashtekar:2004cn} for detailed calculations and \cite{Rovelli_book} for discussions.
\item \emph{Loop Quantum Cosmology}. This is the cosmological sector of LQG, which has rapidly developed since its birth in 2000.
There are indeed quite interesting and stimulating results, particularly concerning the initial singularity of GR, i.e. precisely where we expected Quantum Gravity to tell something. I refer to the review~\cite{Bojowald:2008zzb} and to the papers~\cite{Bojowald1}-\cite{Bojowald4}. Recently also spinfoam calculations of cosmological problems have been performed~\cite{SFcosm1,SFcosm2,SFcosm3}.

\end{itemize}

\chapter{The spinfoam approach to the dynamics}\label{sec:spinfoam}

We have up till now described what ``happens quantum mechanically'' \emph{on spatial hypersurfaces}, and we have argued the difficulties in dealing with the evolution of these quantum states in a timelike direction, difficulties connected with the hamiltonian operator.
However -- with in mind the goal of some kind of sum-over-histories formulation of the dynamics -- we can take a quantum 3-geometry, for simplicity a single spin network state (i.e. an eigenstate of area and volume) $\ket{s'}$ and imagine it evolving in a timelike direction. Imagine its ``world-sheet'': every node would sweep timelike paths, that we call \emph{edges}, every link would sweep \emph{faces} and there will be some spacetime points in which two or three edges meet, that is events in which, for instance, one edge split in two (or whatever) or two or more edges converge in one, etc\ldots We call these points \emph{vertices} and they are the points in which the original quantum state $\ket{s'}$ changes by means of the dynamics, i.e. by \emph{action of the hamiltonian operator}\footnote{thus we have already an idea about how two quantum states differing from a single vertex -- i.e. a  action on a node by the hamiltonian operator -- should be like, see section \ref{sec:scalar_constraint}.}.
Time evolution has thus produced a 2-complex with vertices, edges and faces. This is a dressed 2-complex, since the coloring of the spin network will induce a coloring in the 2-complex as well: irreducible representations to faces and intertwiners to edges. This dressed 2-complex is what is called a \emph{spinfoam}.

The spinfoam approach to (Loop) Quantum Gravity began with the work of Reiseberger and Rovelli~\cite{Reisenberger:1996pu,Rovelli:1998dx}, which first put out the idea of defining quantum amplitudes as sum over histories starting from the hamiltonian formulation (see next section for more details). Actually, work had already been done in this respect (see for example~\cite{Halliwell:1990qr}), but not in the LQG framework.
The general idea of a spinfoam derived from a path-integral discretization procedure (see section~\ref{sec:BF}), was instead proposed by Baez in the seminal paper~\cite{Baez:1997zt} (in which the name ``spin foam'' was first used) and in the lectures~\cite{Baez:1999sr}, which are still a very good `beginner's guide' to spinfoam models, perhaps one of the best in the literature. For other good, but a bit dated, reviews see for example~\cite{Rovelli_book,Perez:2004hj}; for more up-to-date (but a bit more technical) resumes see instead~\cite{new look,Rovelli:2011eq,Livine:2010zx}.

\begin{figure}\centering
 \includegraphics{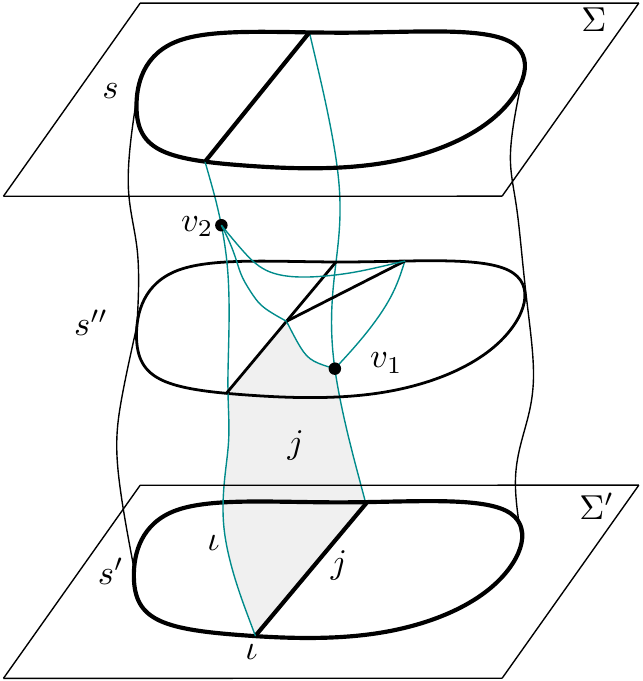}
\caption{Example of spinfoam: evolution from a spin network $s'$ to $s$ passing through an intermediate spin network through two vertices, i.e. actions of the hamiltonian operators. You can see the edges (evolution of nodes of $s'$), faces (evolution of links of $s'$ -- one of the faces has been colored as an example) and vertices.}\label{fig:sf}
\end{figure}

\section{Sum-over-histories from hamiltonian formulation}\label{sec:ham_spinfoam}

We can now make a rather heuristic reasoning trying to ``derive'' a path integral formulation from the hamiltonian formulation, just as one does in standard Quantum Mechanics. I want to stress again the heuristic approach of what follows, that is mainly taken by~\cite{Rovelli_book} (but you can find more detailed discussions in~\cite{Rovelli_prop}). In the next section we shall see a far more rigorous (and alternative) definition of the sum over histories approach to the Quantum Gravity dynamics. Rigorous proofs that the two derivations are actually the same thing have been found \emph{only for the three dimensional case}~\cite{Reisenberger:2001pk}.

The spirit of the spinfoam approach is to try a Feynman-like procedure in a gravitational \emph{and} background independent context.
In standard Quantum Mechanics the Feynman idea is somehow summarized in the following expression
\be
\bk{y,t'}{x,t}\sim\int_{\substack{q(t)=x\\q(t')=y}}\mathcal{D}q\ e^{iS[q]}\ .
\ee
On the left hand side we have the scalar product (transition amplitude) between two different position eigenstates at two different times. The same amplitude can be written in the Heisenberg picture as $\bek{y}{U(t,t')}{x}$, with $U(t,t')$ evolution operator form time $t$ to $t'$.
In a gravitational context one should then give sense to the following expression
\be
\int\mathcal{D}g_{\mu\nu}\ e^{iS_{EH}[g]}
\ee
or, to be more specific,
\be
A(g,g')\sim\int_{\substack{g_{|t=1}=g'\\g_{|t=0}=g}}\mathcal{D}g_{\mu\nu}\ e^{iS_{EH}[g]}\ ,
\label{F1}
\ee
which formally represents the transition amplitude from a space with metric $g$ to one with metric $g'$. The specific values of the time label are irrelevant for the diffeomorphism invariance holds.
There is one big trouble when approaching an integral like the one in \eqref{F1}: we do not know a non perturbative definition of the measure $\mathcal{D}g_{\mu\nu}$ and perturbatively we know that the theory is non-renormalizable (and moreover in this case we should break the background independence). 

In order to give a concrete significance to the expression \eqref{F1} one usually starts from the canonical formulation, which is just what is done for the actual \emph{definition} of path integrals in all the others fields. 
Let's for a moment resume what is the standard way to \emph{define} the \emph{sum-over-paths} procedure.
The idea is to take the time evolution operator $e^{-iHt}$ and to have it acting step by step (in each step you have simply the action of $H$ on a state, which you can calculate), then taking the limit of this time step going to zero. This gives a mathematically precise definition of the Feynman path integral representation of the propagator.

Unfortunately, we are not able to do the same thing in the Quantum Gravity context. The problem is again the hamiltonian operator: we do not know how to quantize it in a proper way, as was pointed out in~\ref{sec:scalar_constraint}. Thus, we cannot follow this route pretending to obtain new rigorous insights.
 
However, we can still derive some basic properties that the Quantum Gravity path integral should satisfy. We shall see that this will be enough to find at least a form for Quantum Gravity transition amplitudes. Then, in the following section, we shall see an alternative way to follow, somewhat dual to the derivation from the canonical formulation, which is called ``spinfoam'' approach.\\[14pt]

Consider the integral \eqref{F1}: between which states the amplitude should be computed? It should be computed between eigenstates of the three-geometry, i.e. with states with a definite 3d metric: but these are nothing but the spin-network states. Thus what we would like to calculate is actually something like
\be
A(s,s')=\bk{s}{s'}_\text{phys}=\bek{s}{A}{s'}_\text{kin}\ .
\label{transamp}
\ee 
Let us try to explain thoroughly this last expression. $\ket{s}$ is the usual spin-network state \eqref{spin_net}, or, better, the s-knot states, i.e. with basis graph an equivalence class of graphs under spatial diffs (see section \ref{sec:diffs}). This is also the reason for the subscript ``kin'', to stress that the spin networks belong to the kinematical Hilbert space $\mathcal{H}_\text{kin}$, but not to the physical Hilbert space. So $A$ (for `amplitude') represents the projector on the kernel of the hamiltonian operator $\hat H$, i.e. the projector onto the physical Hilbert space. If we assume for simplicity that the hamiltonian $\hat H$ has a non-negative spectrum, then we could (formally) write
\be
A=\lim_{t\rightarrow\infty}e^{-Ht}\ ,
\ee
indeed, if $\ket{n}$ is a basis that diagonalizes $H$ (with eigenvalues $E_n$), then
\be
A=\lim_{t\rightarrow\infty}\sum_n\ket{n}e^{-E_nt}\bra{n}=\sum_n\delta_{0,E_n}\ket{n}\bra{n}\ ,
\ee
namely $A$ projects onto the lowest-energy subspace, i.e. the kernel of $\hat H$, if we assume a non-negative spectrum.

Proceeding with these formal manipulations, we can also write
\be
A=\lim_{t\rightarrow\infty}\prod_x e^{-H(x)t}=\lim_{t\rightarrow\infty} e^{-\int\d ^3 x\ H(x)t}\ ,
\ee
hence
\be
A(s,s')=\lim_{t\rightarrow\infty}\bek{s}{e^{-\int\d ^3 x\ H(x)t}}{s'}_\text{kin}\ .
\ee
Quite loosely, if we want the propagator to be 4d diff invariant, then the limit is irrelevant, so
\be
A(s,s')=\bek{s}{e^{-\int_0^1\d t\ \int\d ^3 x\ H(x)t}}{s'}_\text{kin}\ .
\ee
Now we can split this expression by inserting identities in the form $\id =\sum_s\ket{s}\bra{s}$, obtaining something like
\begin{align}
A(s,s')=\lim_{N\rightarrow\infty}\sum_{s_1\ldots s_N}&\bek{s}{e^{-\int\d ^3 x\ H(x)\d t}}{s_N}_\text{kin}\bek{s_N}{e^{-\int\d ^3 x\ H(x)\d t}}{s_{N-1}}_\text{kin}\\
&\ldots\bek{s_1}{e^{-\int\d ^3 x\ H(x)\d t}}{s'}_\text{kin}\ .\nonumber
\end{align}
We can see then, that the transition amplitude between two spin-network stats, can be expressed as a sum, as follows
\be
A(s,s')=\sum_\sigma A(\sigma)\ ,
\ee
i.e. as a sum over histories $\sigma$ of spin network. A history $\sigma$ is a discrete sequence of spin-network
\be
\sigma=(s,s_N,\ldots,s_1,s')\ .
\ee
Moreover the amplitude is just a product of the amplitudes between the single steps in the history
\be
A(\sigma)=\prod_v A_v(\sigma)\ ,
\label{amp_prod}
\ee
where we have labeled with $v$ each of the above said single steps (this terminology will be clarified in a moment).
Now, to take a further step, let us recall that the hamiltonian operator acts only on the \emph{nodes} of the spin network. Thus the single amplitude $A_v$ in \eqref{amp_prod} is non-vanishing only between spin-network that differ at a node by the action of $H$.

A history of spin-networks $\sigma$ is what is called a \emph{spinfoam}, and it is exactly the ``world-sheet'' of the spin network, as pointed out at the very beginning of this section: \emph{it is the time evolution of a spin network state}.

As a conclusion of this brief passage I want to recall again two things: 1. the heuristic value of the former passages and 2. the fact that a \emph{rigorous derivation} of a sum-over-histories formula for Quantum Gravity transition amplitude can be given in the three dimensional case~\cite{Reisenberger:2001pk} (and moreover it matches the result of the following section).

\section{Path integral discretization: \emph{BF} theory}\label{sec:BF}

Now we would like to introduce the spinfoams formalism in a different (and somewhat clearer) way. We shall try to ``discretize'' the path-integral itself. 
Or, better, we shall discretize spacetime with triangulations \emph{à la Regge}~\cite{regge1,regge2}, and try to read out how the path integral of general relativity can be adapted on this triangulation.

This is an approach somewhat dual to the hamiltonian derivation of the path integral: there one take the physical inner product defined in terms of an evolution operator (the projector into the hamiltonian kernel) and ``decompose'' it by inserting identity resolutions, \emph{de facto} giving a discrete definition of the path integral. The present approach is the other way around, i.e. one defines the amplitude as the path integral of (an appropriate form of) the General Relativity action, and discretize this integral trying to give it a rigorous meaning. For a good introduction to this approach see~\cite{Baez:1999sr}.

\begin{rem}
I point out from the very beginning that the discretization we are imposing to define the path integral is not at all as the one of, say, relativity on the lattice. Indeed, there one uses the lattice just as a \emph{regularization} procedure, to be removed at the end by an appropriate continuous limit. Here instead, quantization tells us that the fundamental theory \emph{is} discrete! So we are justified in our discretization procedure, and no continuous limit is to be done. Another problem will be the triangulation dependence of our final results, but this is a completely different issue.   
\end{rem}
We shall start with $BF$~theory, which is actually a trivial theory, but it is up till now the only theory in which the spinfoam approach (=discretization of path integral) can be completely gone through.

To set up a general $BF$~theory we need a \emph{principal bundle} -- let's call it $P$ --  with base space the spacetime manifold $\mathcal{M}$, fiber a gauge Lie group $G$. The basics fields in the theory are a connection $A$ on $P$ and an $\text{ad}(P)$-valued $(n-2)$-form $B$ on $\mathcal{M}$.
Here $\text{ad}(P)$ is the \emph{associated vector bundle} to $P$ via the adjoint action of the group $G$ on its Lie algebra. However, we can forget for the time being all these technicalities and just see how things work in calculations. 
The lagrangian of $BF$~theory is the defined as follows
\be
\mathcal{L}=\Tr (B\wedge F)\ ,
\label{BFlagrangian}
\ee
where, as usual, $F$ is the curvature of the connection $A$.
Now it easy to see that this theory is (at least classically) trivial, indeed the equations of motion state that
\be
F=0\ ;\quad \d _A B=0\ .
\ee
The first says that the manifold is flat, there are no local degrees of freedom. Indeed the second says that the parallel transport rule is trivial.
Now, take the simplest path integral you can imagine for this theory, i.e. the partition function 
\begin{align}\label{bf_path}
\mathcal{Z}(\mathcal{M})&=\int\mathcal{D}A\mathcal{D}Be^{i\int_\mathcal{M}\Tr (B\wedge F)}\\ 
&=\int\mathcal{D}A\delta{F}\ \nonumber ,
\end{align}
where we have integrated out the $B$ field, using its equation of motion $F=0$.
This expression has, by itself, no rigorous meaning. 
To give a sense to the above formal expression the next step -- as stated at the beginning of this section -- is the triangulation of the manifold $\mathcal{M}$ and the definition of a discretized version of \eqref{bf_path}.

Let's recall that a triangulation of a (sufficiently smooth) manifold is obtained by a discretization by means of simplices, precisely $n$-simplices.
Recall that in the sum-over-histories framework, we argued that spacetime should be seen as the \emph{dual} to what we called a spinfoam, i.e. dual to spin networks world- sheets (just as 3-volumes are dual to nodes of a spin network on space-hypersurfaces). This picture will somehow guide us in the discretization of the path integral, indeed we shall define the discrete versions of $B$'s and $A$'s on the dual of the triangulation.

We call $\Delta$ such a triangulation. Now we take the so called \emph{dual $2$-skeleton} of this triangulation. It is built in this way: put a \emph{vertex} at the center of each $n$-simplex; an \emph{edge} intersecting each $(n-1)$-simplex and one (polygonal) \emph{face} intersecting each $(n-2)$-simplex. We could go on, defining dual volumes intersecting $(n-3)$-simplices, and so on till defining a dual $n$-complex to each point (0-simplex) of $\Delta$, thus building the dual triangulation $\Delta^*$. But we shall need only its two dimensional subset, the skeleton.

\begin{center}
 
\begin{tabular}{c c}
 $\Delta$ & 2-skeleton $\subset\Delta^*$\\
\hline
n-simplex & vertex $v$\\
(n-1)-simplex & edge $e$\\
(n-2)-simplex & face $f$\\
\end{tabular}

\end{center}

For example, if $n=3$, then the triangulation is made up by tetrahedra glued along  faces. For each tetrahedron we have one vertex of the dual skeleton, 4 edges (one for each triangle) and 6 faces (one for each edge of the tetrahedron).

\begin{center}
 
\begin{tabular}{c c}
 $\Delta$ & 2-skeleton $\subset\Delta^*$\\
\hline
tetrehedra & vertex $v$\\
triangle & edge $e$\\
edge (of a triangle) & face $f$\\
\end{tabular}

\end{center}

While, for $n=4$ -- the interesting case -- we have 

\begin{center}
\begin{tabular}{c c}
 $\Delta$ & 2-skeleton $\subset\Delta^*$\\
\hline
 4-simplex & vertex $v$\\
tetrehedra & edge $e$\\
triangle & face $f$\\
\end{tabular}

\end{center}

Now we want to define our $BF$ field theory on this dual skeleton, i.e. the points on the manifold will be ``replaced'' by the vertices, edges and faces of the dual skeleton.

A connection is a one form (precisely a $\mathfrak{g}$-valued one form), thus it is natural to associate it to a $n-1$ dimensional object in spacetime. And this is just what we have called \emph{edge} of the dual skeleton\footnote{In this sense, the discretization defined on the dual of $\Delta$ seems a rather natural choice.}. Thus our connection is a prescription to associate a group element to each edge of the skeleton (think of it as the holonomy of $A$ along that edge). If the connection has to be flat (as is the case in $BF$~theory) then we should want that
\be
g_{e_{1f}}\ldots g_{e_{Nf}}=1\ ,
\ee
where: $g$ stands for a group element; $e_f$ are the edges that surround the face $f$, and there are $N$ of them (i.e. we have called $N$ the number of edges of the face $f$\footnote{Notice that the dual faces can have an arbitrary number of edges, it depends on how the triangulation is done and on the manifold $\mathcal{M}$.}.
This formula thus states that the holonomy around each face is the identity and this is the flatness of the manifold.
The $B$ field is a $\mathfrak{g}$-valued 2-form on $\mathcal{M}$ and we discretize it as a map assigning to each $(n-2)$-simplex -- i.e. to each face $f$ of the skeleton -- a $\mathfrak{g}$ element $B^{IJ}_f$.

Equation \eqref{bf_path} in its discrete version, as just prescribed, becomes
\be
\mathcal{Z}_\Delta(\mathcal{M})=\int\prod_{e\in\mathcal{E}}\d g_e\ \prod_{f\in\mathcal{F}}\d B_f e^{i\Tr[B_fU_f]}=     \int\prod_{e\in\mathcal{E}}\d g_e\ \prod_{f\in\mathcal{F}}\delta (g_{e_1f}\ldots g_{e_Nf})\ ;
\label{discrete_path}
\ee
where  with $\mathcal{E}$ and $\mathcal{F}$ are denoted the set of all the edges and of all the faces of the 2-skeleton, respectively. I have put a subscript to stress that this discrete definition is, in general, triangulation -dependent.In these last equalities we have performed the discretized version the integral over $B$. A remark on the expression $\Tr[B_fU_f]$: $U_f=g_{e_1f}\ldots g_{e_Nf}$ is the (discrete) holonomy around the $f$ face; using the identity $U_f\simeq\id+F_f$ where $F_f\in\mathfrak{g}$ we get the exponent of \eqref{discrete_path}.

The importance of this last formula \eqref{discrete_path} is that it has a precise and definite meaning, and it is no more a purely formal prescription. It is actually the first spinfoam path integral we encounter in this thesis. Here it is written in terms of \emph{group variables}. Now we shall see how to evaluate the integrals and pass to the \emph{spin/intertwiner representation}, more suitable for an intuitive grasp and for a direct link wit spin networks, but completely equivalent.

Let us use the following decomposition of the group delta function (Peter-Weyl decomposition)
\be
\delta (g)=\sum_{\rho\in\text{Irrep}(G)}\text{dim}(\rho)\Tr(\rho(g))\ ,
\ee 
i.e. a sum over the all irreducible representations of the group of the character of the representation weighted by its dimension.
Thus we have
\be
\mathcal{Z}_\Delta(\mathcal{M})=\sum_{\rho: \mathcal{F}\rightarrow\text{Irrep}(G)}\int\prod_{e\in\mathcal{E}}\d g_e\ \prod_{f\in\mathcal{F}}\text{dim}(\rho_f)\Tr(\rho_f(g_{e_1f}\ldots g_{e_Nf}))\ ;
\label{bf}
\ee
where the sum is over all the manners to associate to the faces of the 2-skeleton irreducible representation of the group (operation usually called ``coloring'').
We can go further. It is crucial the following identity:
\be
\int\d g\ \rho_1(g)\otimes\rho_2(g)=\frac{\i\i*}{\text{dim}(\rho_1)}\ \delta_{12}
\label{id_simple}
\ee
if $\rho_1\simeq\rho_2*$ and 0 otherwise. 
Maybe it is useful to write this identity graphically like this
\be
\int\d g\mathgraphics{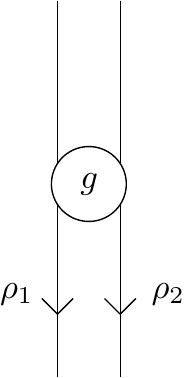}=\frac{\i\i*}{\text{dim}(\rho_1)}\mathgraphics{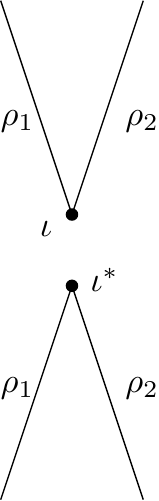}\ .
\ee
A generalization of the above formula is the following:
\be
\int\d\ g =\mathgraphics{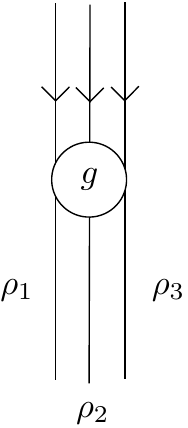}=\sum_\i\mathgraphics{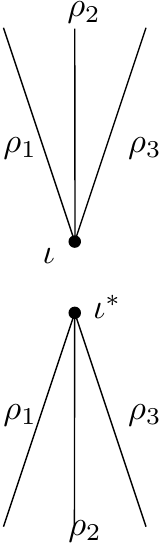}\ ,
\label{formula3}
\ee
and so on, with arbitrary number of representations involved. It is important to catch the meaning of these last formul\ae. Actually it is quite simple: the left hand side is a group averaging of the product of a certain number of representations, thus it belongs to the invariant subspace of the tensor product of those representations. The right hand side is obviously a projector onto the invariant subspace: the two sides are the same thing.

\begin{figure}\centering
\includegraphics{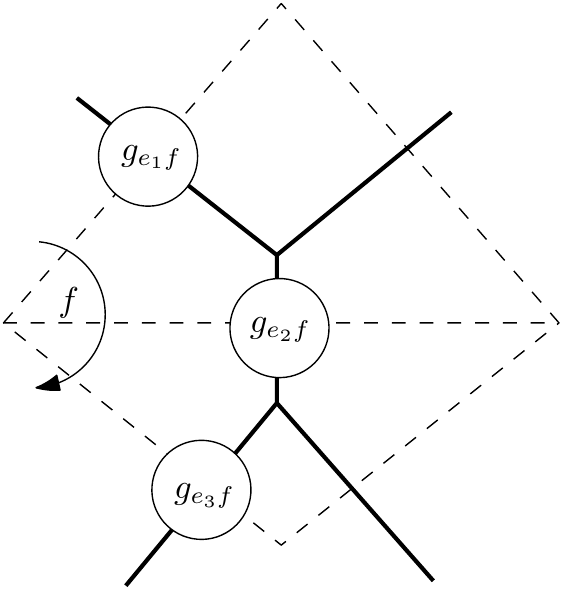}
\caption{(Part of the) triangulation of the manifold (dashed lines) and its dual 2-skeleton (thick lines). One face $f$ has been labeled with its group elements $g_f$'s on the edges.}
\label{fig:2d}
\end{figure}

Let's go back to our $BF$~theory. Let's make the things as simple as possible first: $n=2$, our spacetime is a surface. Thus it is triangulated by triangles and the dual 2-skeleton is made by polygons. A look to figure~\ref{fig:2d} will certainly clarify what's going on. It is easy to see that each group element is shared by two faces (which is the same thing of saying that each edge is common to two and only two faces). In our path integral formula \eqref{bf} we thus see that each integration actually concerns the product of two representations, precisely the two faces that share the edge of that group element. Thus we can use formula \eqref{id_simple} which implies that the integral is non-vanishing only if all the representations (i.e. on all the faces) are equal.
The integral can then be evaluated and the final result is
\be
\mathcal{Z}_\Delta^{(2)}(\mathcal{M})=\sum_{\rho\in\text{Irrep}(G)}\text{dim}(\rho)^{\chi(\mathcal{M})}\ ,
\label{BF2d}
\ee
where $\chi(\mathcal{M})=|\mathcal{V}|-|\mathcal{E}|+|\mathcal{F}|$ is the Euler characteristic of the manifold, and it is a topological invariant quantity. For a manifold of genus $g$ it equals $2-2g$. Thus our partition function for a 2d $BF$~theory converges only for surfaces with genus greater than 1 (thus it diverges for sphere and torus). See section~\ref{sec:divergences} for some discussions about divergences in spinfoam models.

\begin{figure}\centering
\includegraphics{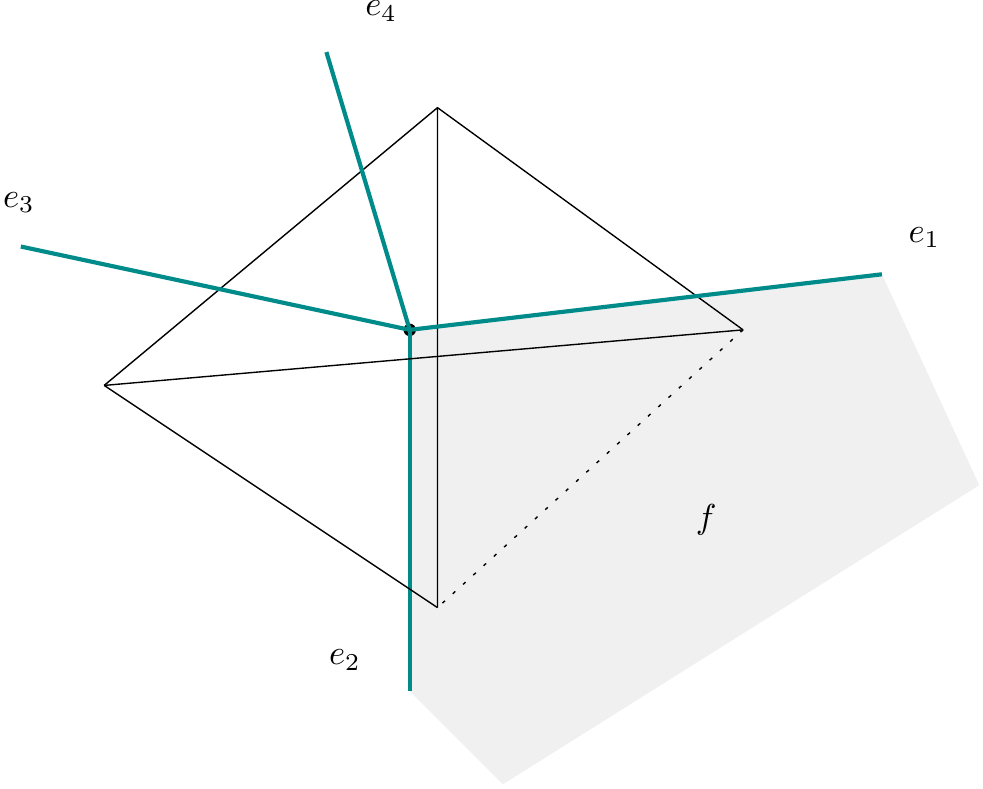}
\caption{A tetrahedron of a 3d triangulation is shown. The thick lines are the four edges, one for each triangle. One (typical) face has been colored, the one corresponding to the dotted line.}\label{fig:3triang}
\end{figure}

\begin{figure}\centering
\begin{minipage}{.4\textwidth}
\includegraphics[scale=.8]{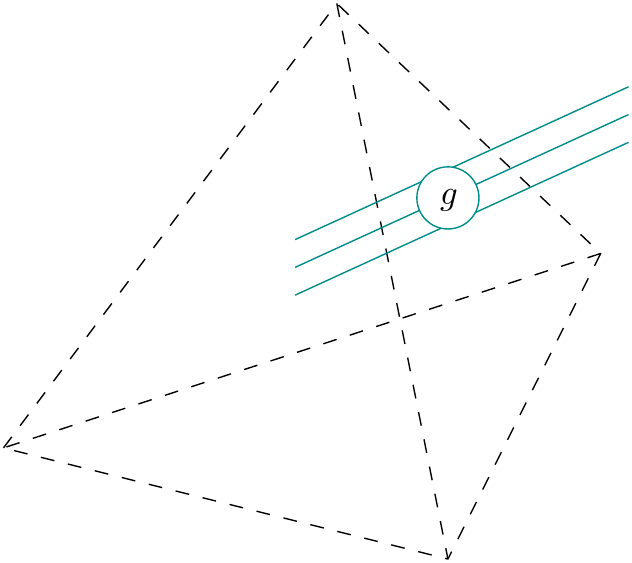}
\end{minipage}%
\quad\begin{minipage}{.4\textwidth}
\includegraphics[scale=.8]{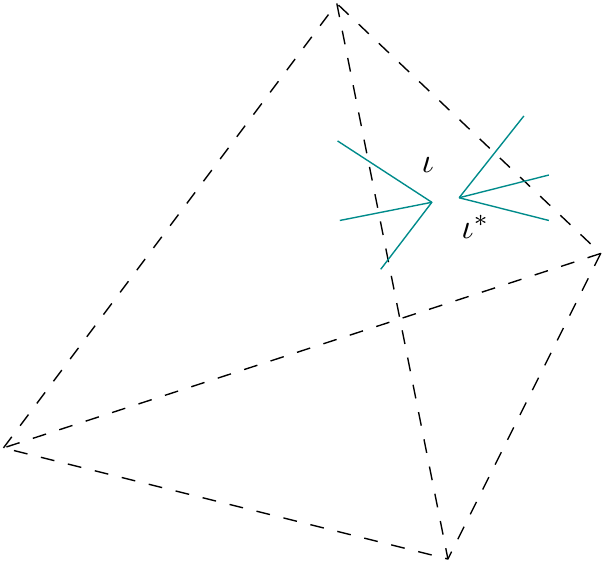}
\end{minipage}
\caption{(on the left) A tetrahedron of a triangulation of a 3d manifold. The three thick lines represent actually a single edge, the one dual to the triangle punctured, which is shared by three faces (one per each side of the triangle). On the right figure, the integration over the $g$ group element has been evaluated by means of equation \eqref{formula3}.}\label{fig:3d}
\end{figure}

Let's step up to dimension 3. In this case the triangulation of $\mathcal{M}$ is made up by tetrahedra. Again, it is worthwhile to have a look to figures~\ref{fig:3triang} and \ref{fig:3d}. We have shown an edge of the dual skeleton, which is shared by three faces. Thus the typical integral will be now of the kind of the left hand side of equation \eqref{formula3}, and thus the integrations split the edge in the sum of intertwiners of the kind shown in figure~\ref{fig:3d}. Now, it is not difficult to see that the intertwiners of a single tetrahedron combine to form a tetrahedron as well (a dual one) labeled by the representations of its 6 edges and by the intertwiners on its vertices.
Thus we have an explicit formula for the partition function, i.e.
\be
\mathcal{Z}_\Delta^{(3)}(\mathcal{M})=\sum_\rho\sum_\i\prod_{f\in\mathcal{F}}\text{dim}(\rho_f)\prod_{v\in\mathcal{V}}\quad\mathgraphics{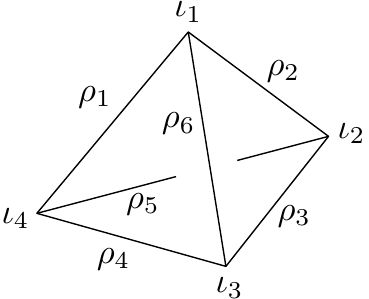}\ .
\label{BF3d}
\ee
The first sum is over $\rho:\mathcal{F}\rightarrow\text{Irrep}(G)$, i.e. a labeling of faces with $G$-representations, while the second on labeling of edges with intertwiners. 
The graphic ``atom'' in this formula is just the contraction of four 3-valent intertwiners, and depends obviously on 6 spins. If $G=SU(2)$, it is nothing but the Wigner $6j$-symbol. You can equivalently see this atom as a spin network (with 6 links and 4 nodes) in a tetrahedron-like pattern, with all the six group elements put on the identity, which is often called \emph{evaluation} of a spin network (recall what  a spin network is precisely, see equation~\eqref{spin_net}).

The trick is the same in 4d, \emph{mutatis mutandis}. In this case one has 4 faces that share an edge, thus we shall use the formula~\eqref{formula3} with 4 representations. In 4d the triangulation is made by a 4-simplex, and the integrations over the groups gives the contraction of intertwiners in a 4-simplex pattern. The final formula is
\be
\mathcal{Z}_\Delta^{(4)}(\mathcal{M})=\sum_\rho\sum_\i\prod_{f\in\mathcal{F}}\text{dim}(\rho_f)\prod_{v\in\mathcal{V}}\quad\mathgraphics{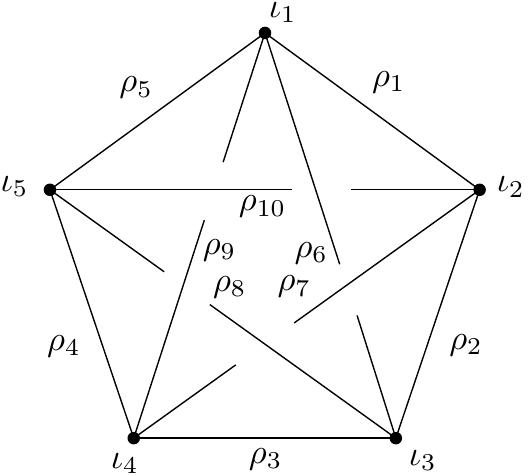}\ .
\label{BF4d}
\ee
The atom in this formula is the contraction of five 4-valent intertwiners in a 4-simplex like pattern. Incidentally, notice that the atom is precisely a 4-simplex (its projection on 2 dimensions).

Notice that one can always split a 4-valent intertwiner into two 3-valent ones\footnote{See section~\ref{sec:G} and in particular equation~\eqref{recoupling}.}, thus one would end with a trivalent spin-network with 15 spins. If the gauge group is $SU(2)$, this is called the `15j-symbol'.

\begin{rem}
Take a 3d model of gravity, i.e. 3d Riemannian General Relativity. Ponzano and Regge, in 1968, showed \cite{PonzanoRegge} that the $SU(2)$-$BF$ path integral (see equation \eqref{BF3d}) has the correct semi-classical limit, in the following sense: There is a discretized action (called the \emph{Regge action}), which is a sum over all the tetrahedra with which we have triangulated the 3d manifold of
\be
S_{R}=\sum_e l_e\theta_e\ ,
\ee
i.e. the sum over the 6 edges of each tetrahedron of the product of the length of that edge and the dihedral angle (the angle between the two normals to the faces incident to that edge) -- that is an approximation of the integral of the Ricci curvature, i.e. an approximation of the Einstein-Hilbert action.
Now, the tetrahedral spin-network of equation \eqref{BF3d}, with gauge group $SU(2)$,  has the following asymptotics for \emph{large spins}
\be
\sqrt{\f{2}{3\pi V}}\cos\left(S_{R}+\f{\pi}{4}\right)\ ,
\ee
where $V$ is the volume of the tetrahedron and $l_e=j_e+1/2$.

This is wonderful, since for large spins -- which means for scales much grater than the Planck length -- one exactly recovers something like $e^{iS}$ with the right General Relativity action!\footnote{Actually one gets just the real part of the exponential. This is a technicality, roughly speaking the reason is that given the lengths of the edges of a tetrahedron, we still can rotate and reflect the tetrahedron.} This means that, at least in this simple and physically un-interesting case, our discretization procedure is consistent, and -- reassured by this consistency -- we feel more confident with the four dimensional case as well.

The $BF$~theory 3d model is commonly known as the \emph{Ponzano-Regge} model.

\end{rem}

Up till now, we have dealt only with manifold \emph{without boundary}, i.e. -- to use an hamiltonian jargon -- we have dealt only with vacuum-to-vacuum transition amplitudes. In order to define something like $\bk{s}{s'}_\text{phys}$ (see \eqref{transamp}), we have to consider triangulation of manifold with boundary as well.
It is quite natural, but under this apparent simplicity many subtleties hide, so one must be careful. The simple part is: take a manifold without boundary and cut it in a spacelike direction. You get two manifold with spacelike boundary. Thinking in terms of the spinfoam, that is in terms of the skeleton of the triangulation, the boundary is made of links, results from the cutting of faces, and nodes, cut of edges. The reader has surely recognized that we are actually doing the inverse procedure with respect to world-sheet sweeping of a spinfoam from a spin-network (as in the hamiltonian-to-spinfoam approach \ref{sec:ham_spinfoam}) , i.e. spinfoam sectioning.

The variables of the cut edges and faces will be variables of the boundary -- be them group elements or spin/intertwiner (or whatever), depending on the representation we want to use -- not to be summed/integrated over. All the construction is exactly the same, with some variables fixed, the boundary variables. 

Now that we have the notion of spinfoam with boundary, we can take the following picture: take a spinfoam (with or without boundary) and imagine to surround each vertex $v$ with a little 3-sphere. You will get an ensamble of little bubbles -- sometimes called \emph{atoms} -- each representing a little spinfoam formed by a single vertex with a boundary $\psi_v$, which is a two dimensional graph with links and nodes representing the vertex. Look to formula \eqref{BF3d} or \eqref{BF4d}, this bubble is just the symbol on the far right, the vertex amplitude\footnote{Incidentally, this ``bubble picture'' is what suggested the name ``spinfoam'' in the early days of this theory.}. It is now clear (if it wasn't already) that a spinfoam amplitude is just a sum over labellings of product of amplitudes: the face amplitude (about which we will have much more to say) and, most important, the vertex amplitude.

This is quite often taken as a \emph{definition} of a spinfoam amplitude.

The subtleties about boundary states arise in the Quantum Gravity context, i.e. when one tries to define quantum transition amplitude for general relativity and not for a generic $G$-$BF$~theory. Indeed in that case, the boundary has to represent a state of $\mathcal{H}_\text{kin}$, the kinematical Hilbert space, i.e. $SU(2)$ spin network states.  This will be a key point.

\begin{figure}\centering
 \begin{minipage}{.5\textwidth}\centering
  \includegraphics[scale=1]{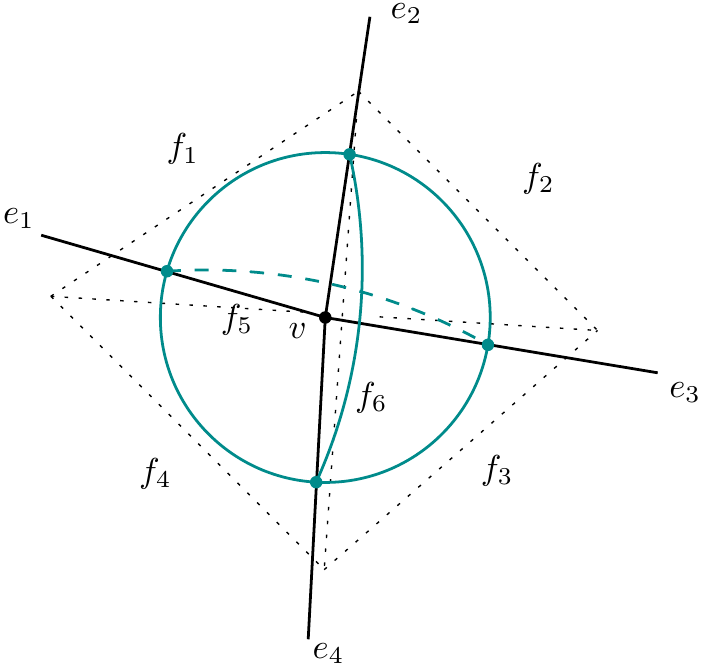}
 \end{minipage}%
\begin{minipage}{.5\textwidth}\centering
 \includegraphics[scale=1]{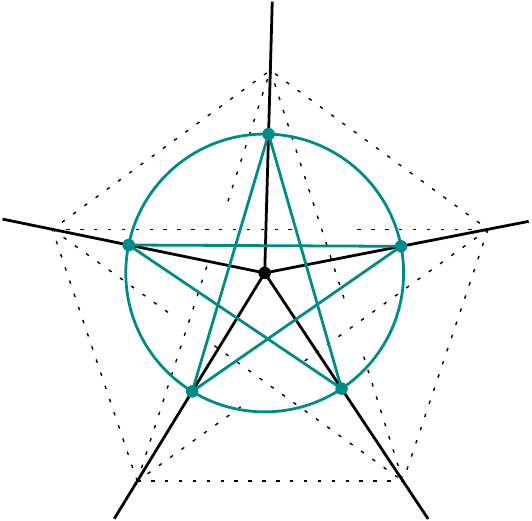}
\end{minipage}
\caption{(on the left) atom `bubble' of a 3d triangulation: dotted line for the $\Delta$-tetrahedron, thick lines for the edges of $\Delta^*$, colored lines for the ``cut'' tetrahedron. On the right the same graph for a 4d triangulation.}\label{fig:cut}
\end{figure}

Before going on to explain what happens in the (interesting) case of gravity, I want to focus on another little but useful issue in $BF$~theory.
I have said that in formul\ae~\eqref{BF3d}, \eqref{BF4d}, the term on the far right is actually the amplitude of a little spinfoam atom. What is like the amplitude of a spinfoam atom (i.e. the vertex amplitude) written in the terms of integral over the group (in the sense of equation \eqref{bf_path}?
To really understand this (and to really understand the vertex amplitude in general), one has to think at what happen when you cut away with a $(n-1)$-sphere from the dual skeleton of an $n$-triangulation. Let us focus on $n=3$ and $n=4$. Actually, it is far more easy to think a little about it and catch it by oneself, rather than long and cumbersome (and useless) explanations. I just give a few hints: take a tetrahedron, and draw the dual. There will be one vertex, four edges departing and 6 faces. Cutting this dual will give another (curved) tetrahedron graph (see figure \ref{fig:cut}), whose links are just the cut of the six faces of the spinfoam, and whose nodes are the cut of the edges. 
In the same manner, in a 4-simplex triangulation, when you cut away a $3-sphere$ around a vertex you get another 4-simplex whose edges and nodes are the cut of faces and edges of the spinfoam.

Thus, one can write the ampltiude simply as
\be
A_v(U_{ij})=\int\prod_i\d g_i\prod_{i<j}\delta(g_iU_{ij}g_j^{-1})=\int\prod_e\d g_e\prod_f\delta(g_{e_1}^fU_{l\equiv f}g_{e_2}^f)\ ,
\label{BFatom}
\ee
where $U_{ij}\ (i,j=1,\ldots ,5; i<j)$ or $U_l$ are the four(3d)/ten(4d) external variables attached to the links of the boundary graph, the notation is quite self-explanatory. The integrals are of course over $G$.
If one decompose the delta into representations and performs the group integrals -- just as we have done before for the partition function -- one ends precisely with
\be
A^{3d}_v(\rho_l,\i_n)=6j(\rho_l,\i_e)\ ,\quad A_v^{4d}(\rho_l,\i_n)=15j(\rho_l,\i_e)\ ,
\ee
i.e. the vertex amplitude in the spin/intertwiner representations (i.e. if we take as boundary variables irreducibles on links and intertwiners on nodes) is -- as it had to be -- the tetrahedron-like or 4-simplex-like intertwiner contraction.

\section{Spinfoam models for Quantum Gravity}

Let us now come to the interesting part of the story: the spinfoam `machinery' applied to General Relativity, rather than to the (trivial) $BF$~theory.
This is still a very open issue, thus everything that follows is in some degree work in progress.

The first fundamental step is to recognize that GR can be cast into ``$BF$~theory + constraints''. Thus the general guideline will be to treat the path-integral discretization \emph{à la} $BF$~theory, and then impose the constraints. It is not surprising that all the big troubles will be in that second step.
But let us do everything at its proper time. 

\subsection{The Barrett-Crane model}\label{sec:BC}

The first spinfoam model for quantum gravity is the Barrett-Crane model~\cite{Barrett:1997gw}.

Take an $SO(4)$-$BF$~theory in 4d. Its action can be written (see \eqref{BFlagrangian})
\be
S[B,\omega]=\int B_{IJ}\w F^{IJ}(\omega)\ ,
\label{BFaction}
\ee
where $F$ is the curvature of the connection $\omega$. If we replace
\be
B_{IJ}=\f12\epsilon_{IJKL}e^K\w e^L\ ,
\ee 
we get precisely Einstein-Hilbert action, in the form of equation \eqref{we}. Thus, we could formally write the action for General Relativity as
\be
S[B,\omega,\phi,\mu]=\int \left(B_{IJ}\w F^{IJ}(\omega)+\phi_{IJKL}B^{IJ}\w B^{KL}+\mu\epsilon^{IJKL}\phi_{IJKL}\right)\ .
\label{GRBFaction}
\ee
Let us check it: $\phi$ is a 0-form while $\mu$ is a 4-form, they both have no dynamics and thus, classically,  they just give constraints, namely
\be
\epsilon^{IJKL}\phi_{IJKL}=0\ ,
\label{mu}
\ee
i.e. $\phi_{IJKL}=-\phi_{JIKL}=-\phi_{IJLK}=\phi_{KLIJ}$, for the variation with respect to $\mu$, while for $\phi$, using \eqref{mu}
\be
\epsilon^{\mu\nu\rho\sigma}B^{IJ}_{\mu\nu}B^{KL}_{\rho\sigma}=e\ \epsilon^{IJKL}\ ,
\label{phi}
\ee
where $e=\tfrac{1}{4!}\epsilon_{IJKL}B^{IJ}_{\mu\nu}B^{KL}_{\rho\sigma}\epsilon^{\mu\nu\rho\sigma}$. This last expression \eqref{phi} contains 20 equations --   often called \emph{simplicity constraints} -- one for each independent component of $\phi$. They constrain 20 of the 36 independent components of the $B$ field. The solutions to \eqref{phi} are of two kinds
\be
B_{IJ}=\pm\f12\epsilon_{IJKL}e^K\w e^L\ ;\quad B^{IJ}=\pm e^I\w e^J\ ;
\ee
in terms of the 16 degrees of freedom of the tetrad field $e^I_a$. Only the first type of solution gives the GR action \eqref{we}\footnote{Up to an overall sign.}, while the second give a topological (trivial) action. This must be remembered when to quantize the action 

For the following, it is useful to rewrite the constraints \eqref{phi} as
\begin{subequations}\label{constr}
\begin{align}
 B^*_{\mu\nu}\cdot B_{\mu\nu}&=0\ ,\\\label{diag1}
 B^*_{\mu\nu}\cdot B_{\mu\sigma}&=0\ ,\\\label{offdiag1}
 B^*_{\mu\nu}\cdot B_{\sigma\tau}&=\pm 2\tl{e}\ ,
\end{align}
\end{subequations}
with $\mu\nu\sigma\tau$ all different, and with $e=\tfrac{\tl{e}}{4!}\epsilon_{\mu\nu\sigma\tau}\d x^{\mu}\w\d x^\nu\w\d x^\sigma\w\d x^\tau$.
The first two constraints in these form are also known as \emph{diagonal} and \emph{off-diagonal} simplicity constraints.

\begin{rem}
Notice that we have taken an $SO(4)$ connection -- or, better, its double cover $Spin(4)$ -- thus this is an \emph{euclidean} version of General Relativity. To obtain the actual GR one should take an $SL(2,\mathbb{C})$ (the double cover of $SO(3,1)$) $BF$~theory. The problem is the non compactness of $SL(2,\mathbb{C})$ that requires special care in handling the (divergent) integrals over the group. It is a general custom in Quantum Gravity to investigate the simpler euclidean case first. 
\end{rem}

Equation \eqref{GRBFaction} is our GR ``$BF$ + constraints'' action. Now the goal is to take the $BF$ spinfoam model and analyze what kind of consequences have the imposition of constraints and, firstly, to understand how to impose them. The intuitive idea is that the constraints should restrict in some (interesting) sense the spinfoam sum in \eqref{BF4d}. In particular the constraints must impose that on the boundary of the spinfoam one gets the kinematical Hilbert space of LQG, and so they must somehow restrict $SO(4)$ to $SU(2)$ on the boundary.
However, the imposition of the constraints is actually a rather cumbersome issue, to be dealt with very carefully. I shall sketch here briefly the key points. The interested reader is encouraged to read the review on Barrett-Crane model \cite{Perez:2003vx}. However, the issue of constraints imposition will be gone through more carefully in the case of the Engle, Pereira, Rovelli, Livine (EPRL)  model (see section~\ref{sec:EPRL} for details and references).

The idea is to take the following path integral
\be
\mathcal{Z}^{BC}(\mathcal{M})=\int\mathcal{D}A\mathcal{D}B\delta[B-(e\w e)^*]\exp\left(i\int_\mathcal{M} B\w F[A]\right)\ .
\label{constrainedpath}
\ee
Naively speaking, this means that one must restrict the integral to those configurations satisfying  the delta function in~\eqref{constrainedpath}. This turns out to be the case if one restricts the topological spinfoam sum \eqref{BF4d} -- with $G=Spin(4)$ -- to the so-called $Spin(4)$ \emph{simple representations}: recall that $Spin(4)=SU(2)\times SU(2)$, thus you can label $Spin(4)$ irreducible representations with couples of spins $(j^+,j^-)$: simple representations are the ones with $j^+=j^-$.

The spinfoam partition function results in
\be
\mathcal{Z}^{BC}_\Delta=\sum_{simple\ \rho}\prod_{f}\text{dim}(\rho_f)\prod_v\mathgraphics[scale=.8]{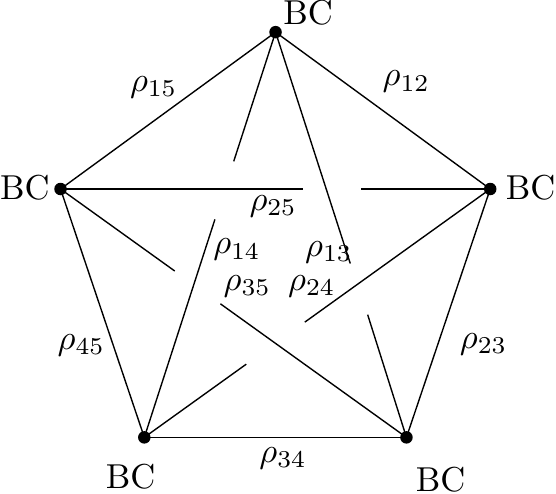}\ ,
\label{ZBC}
\ee
where, as just said, the sum is over simple $Spin(4)$ representations; BC is a fixed intertwiner operator between (and depending only on) the 4 representations converging in it (following the 4-simplex pattern). 
The vertex amplitude ultimately depends on 10 spins, and it is thus referred to as the $10j$-symbol. 
Of course, thanks to the `simplicity' of the representations, we can rewrite this partition function as a sum over $SU(2)$ irreducible, using the following property of the BC intertwiner
\be
\mathgraphics[scale=.7]{fig/10j-eps-converted-to}=\sum_{\i_1,\ldots ,\i_5}\mathgraphics[scale=.7]{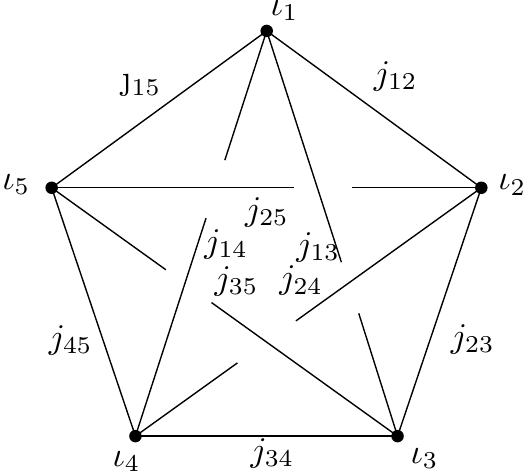}\ \mathgraphics[scale=.7]{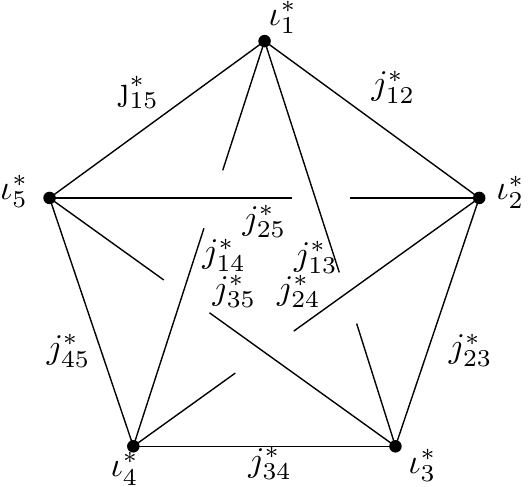}\ ,
\ee
namely
\be
\mathcal{Z}_\Delta^{BC}=\sum_{j}\prod_{f}(2j_f+1)^2\prod_v\sum_{\i_1,\ldots ,\i_5}\mathgraphics[scale=.7]{fig/BC4simplex-eps-converted-to}\ \mathgraphics[scale=.7]{fig/BC4simplex1-eps-converted-to}\ .
\label{ZBC1}
\ee
Notice that, according to \eqref{BF4d}, the face amplitude has been set equal to the dimension of the face representation. Here this representation is simple, so $\text{dim}(j\otimes j)=(2j+1)^2$.

However, this happens to be a subtle point, since we know (e.g. from QFT) that constraints change the integration measure of the path integral. Thus the use of the topological face amplitude is not justified at all in dealing with Quantum Gravity spinfoam models. This is a key point, regarding the choice of the face amplitude, and it will be treated more carefully in section~\ref{sec:face}.

We have sketched briefly the structure of the Barrett-Crane model. However it must be said that this spinfoam model, despite all its success in the few years after its first appearance~\cite{Barrett:1997gw}, has been proved to be unfit to Quantum Gravity~\cite{alesci}: the big problem is that the boundary states of the BC model are only a small subgroup of the spin network states. This is due to the way by which constraints are imposed.

This is the main reason for which much work has been done in order to ameliorate the model. There have been a few proposal starting from 2007. In the following section we present the model which is considered the present day best candidate to attempt doing calculations in Quantum Gravity.

\subsection{The EPRL model}\label{sec:EPRL}

Here I present the Engle, Pereira, Rovelli, Livine proposal for the vertex amplitude of spinfoam models. I refer the reader to the original papers~\cite{Engle:2007uq,Engle:2007qf,Pereira:2007nh,Engle:2007wy} for a thorough discussion.

I don't want to give here a rigorous (and cumbersome) derivation and justification of the model, I just want to stress the key points.
We start with a triangulation $\Delta$ with the usual association of group elements to (dual) edges etc\ldots (see section \ref{sec:BF}), but we use now the General Relativity action in its Holst form~\eqref{Holst}, which can be compactly written as
\be
S[e,\omega]=\int \left((e\w e)^*+\f{1}{\gamma}(e \w e)\right)\w F(\omega)\ .
\label{Holst1}
\ee
with $*$ the Hodge dual operator, namely
\be
F^*_{IJ}\equiv \frac12\epsilon_{IJKL}F^{KL}\ .
\ee
This action has the merit of including the Immirzi parameter $\gamma$ of Loop Quantum Gravity as a pre-factor of the topological sector, which has no consequences on the equations of motion (they remain the Einstein field equations) but allows the formulation in terms of Ashtekar-Barbero variables. 
The program is, as for the Barrett-Crane model, to write then action as `$BF$ + constraints' then write the $BF$ spinfoam sum and only at the end apply the constraints. The difference with respect to BC are in the starting action and (more important) in the way of imposing the constraints. Here we shall work directly in the Lorentz framework, thus with gauge group $SL(2,\mathbb{C})$ rather then $SO(4)$.

Holst action \eqref{Holst}, \eqref{Holst1} can be written as
\be
S[e,\omega]=\int \left(B+\f{1}{\gamma}B^*\right)\w F(\omega) + \text{constraints}.
\ee
where the constraints must impose, as in BC, $B=(e\w e)^*$.
The discrete variables shall be, as previously, $B_f\in \mathfrak{g}$ associated to triangles/dual faces and $g_e$ associated to tetrahedra/dual edges. We call $U_f$ the holonomy around the face $f$, i.e. the product of the group elements of the edges bounding $f$.

This time I shall give a more detailed derivation of the EPRL formula.

\begin{calc}
We begin by giving a discretized form of the simplicity constraints \eqref{constr}. 
We can easily discretize the constraints in the form \eqref{constr} as
\begin{subequations}\label{constr1}
\begin{align}
B^*_f\cdot B_f &=0\ ,\label{diag} \\
B^*_f\cdot B_{f'} &=0\ ,\label{offdiag}\\
B^*_f\cdot B_{f'} &=\pm 12 V\ ,\label{third}
\end{align}
\end{subequations}
where: $V$ is the volume of the simplex; in the second equations $f$ and $f'$ are faces sharing an edge (equivalently: they are associated to triangles living on the same tetrahedron) while in the third they are don't (equivalently: they are attached to triangles belonging to two distinct tetrahedra). Actually, the correct way to deduce the discrete $B_f$ variable from the 2-form $B_{\mu\nu}$ is the following\footnote{Actually the situation is a little more tricky, I refer to the good paper by Engle, Pereira and Rovelli \cite{Engle:2007qf} for a thorough explanation of this point.}
\be
B_f=\int_{f}B\ .
\ee
Instead of requiring the last constraint \eqref{third}, we take the following \emph{closure} constraint
\be
B_{f_1}+B_{f_2}+B_{f_3}+B_{f_4}=0\ ,
\label{clos}
\ee 
for the four faces sharing an edge. It is easy to check that diagonal+off diagonal+closure is an equivalent system of constraints.

\begin{rem}
 In the BC model the diagonal constraint implies the face representation to be simple, while the off diagonal one implies the uniqueness (and the specific form) of the BC intertwiner. Both constraints are imposed \emph{strongly}.
\end{rem}

We have seen that these constraints admit more solutions than GR, precisely a trivial topological sector. Now we exclude the trivial sector by imposing a slightly different form of the constraints, i.e. we require that for each tetrahedron exist a vector $n_I$ such that, for each triangle of the that tetrahedron, holds\footnote{Indeed the off diagonal simplicity constraints imply that the triangles of each tetrahedron lie on a common hypersurface. If they are satisfied, there will be a direction $n^I$ orthogonal to all the faces.}
\be
n_I(B^*_f)^{IJ}=0\ .
\ee
This constraint is intended to replace the off diagonal simplicity constraint \eqref{offdiag}. Geometrically the $n_I$ represents a vector normal to the tetrahedron/edge.

Having cleared the meaning of the discrete versions of the simplicity constraints \eqref{phi}, we rephrase them in a convenient way.
The conjugate momenta to the holonomies are
\be
J_f=B_f+\f{1}{\gamma}B_f^*\ .
\ee
Which, once inverted
\be
B_f=\left(\f{\gamma^2}{\gamma^2+1}\right)\left(J_f-\f{1}{\gamma}J_f^*\right)\ ,
\ee
thus we can reformulate the constraints as
\begin{align}\label{simp}
C_{ff}&=J_f^* J_f\left(1-\f{1}{\gamma^2}\right)+\f{2}{\gamma}J_fJ_f=0\ ,\\ \label{offsimp}
C_f^J&=n_I\left(J^{*\ IJ}+\f{1}{\gamma}J^{IJ}_f\right)=0\ .
\end{align}

Now we choose a specific $n_I$. A typical choice is $n_I=\delta^0_{I}$, that means that all the tetrahedra are spacelike. In other words, we are selecting a specific $SU(2)$ subgroup of the full $\sl$. Obviously this is a gauge choice, and must not influence physical results.
With this choice the constraint \eqref{offsimp} becomes
\be
C^j_f=\f12\epsilon^j_{\ kl}J^{kl}_f+\f{1}{\gamma}J^{0j}_f=L^j_f+\f{1}{\gamma}K^j_f=0\ ,
\label{offsimp1}
\ee
where $L$ are the generators of the $SU(2)$ subgroup of $SL(2,\mathbb{C})$ that leaves $n_I$ invariant; $K$ are the generators of boosts in the $n_I$ direction.

Let us now deal with the quantization of the constraints. In order to do that we must identify a Hilbert space in which we define operators. Taking a single vertex bubble, recall that the graph on its boundary $\gamma_v$ naturally defines the following boundary Hilbert space
\be
L^2(\sl^L)\ ,
\ee
where $L,N$ are the number of links and nodes of $\gamma_v$ (cut from the spinfoam by the 3-sphere). We impose the constraints as follows.

The closure constraint \eqref{clos} imposes $\sl$-invariance on this space, i.e. it implements the (usual) quotient
\be
L^2(\sl^L/\sl^N)\ .
\ee
The simplicity constraints \eqref{simp}, \eqref{offsimp1} are defined on faces, thus they act each on a single copy of the group, namely on $L^2(\sl)$.
The diagonal simplicity constraint \eqref{simp} on this space  reads
\be
C_1\left(1-\f{1}{\gamma^2}\right)+\f{2}{\gamma}C_1 ``=``0\ ,
\label{simp1}
\ee
with $C_1$ and $C_2$ the Casimir operator of $\mathfrak{g}$, i.e.
\be
C_1=J\cdot J=2(L^2-K^2);\quad C_2=J^*\cdot J=-4L\cdot K\ .
\ee
The quotations mark means that we have to decide how to impose this constraint. However notice that having expressed it in terms of Casimir operators eigenvalues, it commutes with all the other operators, thus we can impose it strongly, i.e. requiring it to annihilate physical states. 

The constraint \eqref{offsimp1}, on the other hand, is more subtle. The technique used in \cite{Engle:2007qf}, first proposed by Thiemann \cite{master} is to pack them into a \emph{master constraint}
\be
M_f=\sum_j (C^j)^2\ ``=``\ 0\ .
\ee
Classically it is of course equivalent to imposing the $C^j$ equal zero separately.
The power of this approach, is that $M_f$ is now a combination of Casimirs
\be
L^2\left(1+\f{1}{\gamma^2}\right)-\f{C_1}{2\gamma^2}-\f{C_2}{2\gamma}\ ``=``\ 0\ .
\label{offsimp2}
\ee 
Combining \eqref{simp1} and \eqref{offsimp2} we get the following (definitive) set of 2 constraints:
\begin{align}
&C_2\left(1-\f{1}{\gamma}\right)+\f{2}{\gamma}C_1 \ ``=``\ 0\ ,\\
&C_2-4\gamma L^2\ ``=``\ 0\ .
\end{align}
Having done this, we now simply have to see what consequences these constraints have on states of $L^2(\sl)$. It is easy to see that, labeling with $(p,k)$ ($p$ real and $k$ half-integer) the $\sl$-irreducible  representations, thus having the decomposition
\be
L^2(\sl)=\sum_{(p,k)}\mathcal{H}_{(p,k)}\otimes\mathcal{H}_{(p,k)}\ ,\quad\mathcal{H}_{(p,k)}=\bigoplus_{j'=k}^\infty j'\ ,
\ee
-- with $j'$ in the sum denoting the $SU(2)$ spin-$j'$ irreducible -- of $\sl$-irreducibles into $SU(2)$-irreducibles, the constraints impose, for each face,
\be
p_f=\gamma j_f\ ,\quad k_f=j_f\ ,
\ee
for some half-integer $j_f$, and they restrict the decomposition
 to the lowest spin, $j'=k=j$.

 You see that the constraints have selected a $SU(2)$ subgroup of $\sl$! Precisely they tell two things: 1. the permitted $\sl$ face labels are only the irreducibles of the type $(p_f=\gamma j_f,k_f=j_f)$ for some half-integer $j_f$\footnote{Notice that this also implies that the effective sum over $p$, which should be an integral, restricts to a sum.}; 2. given the face label $(\gamma j_f,j_f)$ they select out the spin-$j$ $SU(2)$ irreducible.
Let's give a precise definition of the embedding $L^2(SU(2))\rightarrow L^2(\sl)$:
\be
Y:j\rightarrow j\subset \mathcal{H}_{(\gamma j,j)}\subset L^2(\sl)\ ,
\ee
is the map that sends each $SU(2)$ spin $j$ irreducible in the lowest spin (i.e. -- of course -- the spin $j$) irreducible inside $(\gamma j,j)$.
Thus, the $Y$ map takes each state in $L^2(SU(2))$ to a state of $L^2(\sl)$.
In terms of Wigner matrices it is really simple
\be
D^j\xrightarrow{Y}YD^jY\dag=D^{(\gamma j,j)}\ .
\ee

This is very nice indeed, since we can define boundary variables to be $SU(2)$ states, and this is just what we expect to have, since on the boundary we want to put spin networks.

\begin{rem}
I have done all the calculation for $\sl$ since it is the ``reality''. However, as I had occasion to say, it is sometimes useful to see what happens in the simpler euclidean case, i.e. for $SO(4)$. I do not repeat the discussion -- which proceeds very similarly -- the result is the following: the simplicity constraints reduce the $SO(4)$ irreducibles $(j_f^1,j_f^2)$ to the ones given by $(\gamma_+ j_f,\gamma_- j_f)$, $j_f$ `running' over $SU(2)$ irreducibles and with~\cite{Engle:2007qf} 
\be
\gamma_\pm=\f{|1\pm\gamma|}{2}\ . 
\ee
Thus, in the euclidean $SO(4)$ case we have an embedding of $SU(2)$ in $SO(4)$, quite similarly to what happens in the lorentzian case.
\end{rem}

We are almost done. We just have to see the consequences of what I have just said in terms of concrete formul\ae.

Let us focus on a single vertex. The idea is to take the $BF$ vertex amplitude \eqref{BFatom} (or equivalently the 4-simplex spin network evaluation, i.e the graph on the far right of \eqref{BF4d}) and to restrict the boundary variables to satisfy the simplicity constraints, in the form we have just said:
\be
A_v^{EPRL}(U_l)=\int_{\sl^N}\prod_n\d g_n\prod_f P(U_l,g_{s(l)}g_{t(l)}^{-1}),
\label{EPRLdelta}
\ee
where we have no more a delta inside the face product, since now $U_l$ is an $SU(2)$ element, but a `generalized' delta, i.e.
\be
P(U,g)=\sum_j(2j+1)\Tr(D^j(U)Y\dag D^{(\gamma j,j)}(g) Y)\ ,
\ee
with $U\in SU(2)$, $g\in\sl$.
Performing the $SU(2)$ integrals, just as in the standard $BF$~theory case (\ref{sec:BF}), one can easily find the spin/intertwiner representation of the vertex amplitude
done in the $BF$~theory case.
\end{calc} 
One ends up with
\be
 A_v^{EPRL}(j_l,\i_n)=\sum_{k_n}\int\d p_n\left((k_
n/2)^2+p_n^2\right)\prod_n Y^{\i_n}_{\i_{(p_n,k_n)}}\times15j((\gamma j_l,j_l);\i_{(p_n,k_n)})\ ;
\label{EPRLatom}
\ee
where $Y^{\i_n}_{\i_{(p_e,k_e)}}=\bek{\i_{(p,k)}}{Y}{\i}$.
Putting together all the vertex amplitudes we get
\be
A_{\Delta}^{EPRL}(j_l,\i,n)=\sum_{j,\i}\prod_fA_f(j_f)\prod_vA^{EPRL}_v(j,\i;j_l,\i_n)\ ;
\ee
where the last parenthesis indicates that some simplices are cut by the boundary, thus some spins and intertwiners will be fixed (and not summed) to the boundary value, given in the left hand side\footnote{The partition function is obviously a particular case of this formula, i.e.
\be
\mathcal{Z}_\Delta^{EPRL}=\sum_{j,\i}\prod_fA_f(j_f)\prod_vA^{EPRL}_v(j,\i)\ .
\ee}

Notice again that we have no clue about the face amplitude: should it be the $SU(2)$ dimension of the representation attached to that face (as in a $SU(2)$-$BF$~theory)? Should it be the $\sl$/$SO(4)$ dimension (as in a $\sl$/$SO(4)$-$BF$~theory)? The imposing of constraint changes the measure of the path integral, and we have no control about that in this derivation.

\section{Spinfoam: a unified view}

After all this rather cumbersome (and quite chronological) model-making I would like to give a more coherent and compact view of spinfoam approach and of its relation with Loop Quantum Gravity.
As pointed out in \cite{new look} it is time to provide a ``top-to-bottom'' framework, in which set some properties we want these models to satisfy and deduce from them specific models. This has (at least) the merit of clarifying the mess of technicalities that plague -- in my opinion -- the spinfoam model-making approach.
\begin{rem}
 What follows is just an attempt to give a unified and more coherent framework of spinfoam approach, but \emph{it is not} a well established  and rigorous ``chapter'' of Loop Quantum Gravity. You should take what follows as a reasoning \emph{ex post} on how we can insert all the models into a single framework. For this section I mainly follow the paper \cite{new look}.
\end{rem}

First of all: what do we want from a spinfoam model? We want a way to compute \emph{transition amplitudes} 
\be
\bek{s}{A}{s'}\ ,
\ee
between Loop Quantum Gravity spin network states. $A$ represents the projection operator into the kernel of the hamiltonian operator (recall equation \eqref{transamp}). In a covariant 4d context this is rephrased: take a 4d manifold with boundary state $\psi$, we want to calculate
\be
\bk{A}{\psi}\ .
\label{amp}
\ee
The boundary state $\psi$ is typically formed by two spin networks (in terms of graphs $\Gamma_{\psi}=\Gamma_{\psi_1}\cup\Gamma_{\psi_2}$) the ``initial'' and ``final'' spin networks, but we can be as generic as we want. This last expression is the amplitude associated to the specific state $\psi$. The linear functional $A$ (or, if you want, $\bra{0}A$, but here we enter in the subtle (and fascinating) issue of vacuum in Quantum Gravity, maybe we will spend some words about it in the rest of the chapter)  is the heart of the spinfoam model, and must be thought of as an evolution operator.
What properties do we ask  \eqref{amp} to have?
\begin{itemize}
 \item \emph{Superposition}. Namely we want
\be
\bk{A}{\psi}=\sum_{\sigma}A(\sigma)\ ,
\ee
i.e. we want that the amplitude is expressed as a ``sum over histories''. This is what we argued with a rather heuristic touch in section \ref{sec:spinfoam}. Obviously the specific set over which to sum is at this level completely undetermined.

\item \emph{Locality}. This is a request on the single history-amplitude $A(\sigma)$. We require
\be
A(\sigma)\sim\prod_v A_v\ .
\ee
This amounts to ask that each history has an amplitude that is the product of elementary amplitudes, or vertex amplitudes (the ``atoms'' we have been talking about). 

\item \emph{Local Lorentz invariance}. Recall that classical general relativity, in tetrad formulation, has a local Lorentz invariance, namely a $\sl$ gauge invariance. However the boundary states know nothing of $\sl$, they are built up over $SU(2)$ gauge invariance. Thus there must be some embedding from $SU(2)$-gauge invariant states to $\sl$-invariant ones. This map, usually called simply $f$ is what actually determines the spinfoam amplitude. 
\end{itemize}
  
Now we have the key ingredients. We have to focus on a single vertex, surrounded by a kinematical state $\psi_I$. We \emph{define}
\be
A_v(I)=\bk{A_v}{\psi_I}=\bek{\id}{f}{\psi_I}\ ,
\ee
with 
\be
f=P_{\sl}\circ Y
\ee
the map that embeds the boundary state in $L^2(\sl^L)$ and then projects onto the  $\sl$-invariant on the nodes, i.e. onto $L^2(\sl^L/\sl^N)$. $\ket{\psi_I}$ stands for a state in the boundary Hilbert space of the vertex $v$, and $I$ denotes a set of quantum numbers (think of a spin network state).

Writing it a bit more explicitly
\be
\bk{A_v}{\psi_I}=\int_{\sl^N}\prod_n\d g_n \bra{\id}\mathcal{U}(g_n){Y}\ket{\psi_I}\ .
\ee
$\psi_I$ is a generic boundary state: it could be a spin network state $\ket{j_l,\i_n}$ giving $A_v(j_l,\i_n)$ or, in the holonomy(`position') representation, $\ket{U_l}$ giving $A_v(U_l)$. $\bra{\id}$ stands for the $\sl$ holonomy state where all the links are set to the identity (it corresponds to the \emph{evaluation} of the state on which it acts). $\mathcal{U}(g)$ is the action of $g\in\sl$ over the state $Y\ket{\psi_I}$.
In a suggestive way one could write
\be
\bra{A_v}=\int_{\sl^N}\prod_n\d g_n\bra{\id}\mathcal{U}(g_n)Y\ ,
\ee
as a functional over the kinematical Hilbert space of Loop Quantum Gravity.
Of course this definition gives the EPRL formula \eqref{EPRLatom}, when $\ket{\psi_I}=\ket{j_l,\i_n}$, and gives exactly \eqref{EPRLdelta} in terms of holonomies. 

Moreover, this framework is completely general, in the sense that it is able to reproduce every spinfoam model that  matches Loop Quantum Gravity kinematical Hilbert space on the boundary. Obviously various models can be reproduced by a choice of the $f$ function, i.e. of the $Y$ map, which is, ultimately, what determines the specific vertex amplitude, and -- indeed -- is itself completely determined by how we impose the constraints on the $BF$ model (recall that there are infinite $SU(2)$ irreducibles inside each $\sl$ irreducible). 

Once got the vertex amplitude one get the total amplitude by 
\be
\bk{A}{\psi_I}\simeq\sum_\sigma\prod_vA_v(I)\ ,
\ee
notice that this sum is two-fold: fix a triangulation and you  have to sum over the labeling of internal faces and edges; then you have to sum over all the possible triangulation in order to really catch all the possible ``paths'' bounded by $\psi_I$! This make apparent the problem in dealing with triangulation dependence and in how to weight different triangulations (the attentive reader was surely already aware of this problem since the very beginning of the discretization stuff). I will briefly review the topic (which is a totally open issue) in the following section. 

Moreover here there is no hint about something attached to the faces of each spinfoam. We know from the $BF$~theory derivation that something there must be, and that it very likely has to do with the dimension of the representations attached to the faces (see chapter~\ref{sec:face}).

\section{Spinfoams as a field theory: Group Field Theory}

The present section is a bit beyond the scope of this review of Loop Quantum Gravity and spinfoam models, particularly since to understand the research I did in this field (\ref{sec:face}) the argument of this section is not strictly necessary, nor -- I must admit -- closely related. 

However, I spent quite a long time in Marseille studying Group Field Theory, and I seriously think it has more than a few chances of becoming a bridge between Quantum Gravity in its spinfoam formulation and the prolific and well-established framework of Quantum Field Theory.

In a sense that will be clear -- I do hope -- from what follows, Group Field Theory is precisely a way of putting the spinfoam models in the framework of QFT, a very special kind of QFT.

GFT was first introduced by Boulatov in 1992 \cite{Boulatov}, and had, since then, a very promising development as a Quantum Gravity framework. 
The nowadays work in this topic is mainly to study the derivation of the spinfoam models (particularly the EPRL model) in this framework; to study quantum corrections and renormalization issues; control the sum over triangulations; all done with QFT-like tools. For a panoramic  of the present-day research see the following papers \cite{GFTnow,GFTnow1}, while for good reviews of GFT see \cite{GFTreviews1,GFTreviews2,GFTreviews3}.

The very basic idea is the following: find \emph{a field theory whose Feynman diagrams are spinfoams}.
The natural ancestor of GFT are the well known matrix models: their Feynman diagrams are ribbon-like, which can be seen as dual to a triangulated surface \cite{Brezin:1977sv,David:1984tx}.

A matrix model is a model whose action is something like
\be
S[M]=\f{N}{2}\Tr(M^2)+\f{N\lambda}{3!}\Tr(M^3)\ ,
\label{matact}
\ee
where $M$ is a $N\times N$ matrix. I have taken a potential term of order 3, but of course it is just an example.

Going directly to the quantum side of the story, let us try to calculate the partition function for this action
\be
\mathcal{Z}=\int\mathcal{D}M\ e^{-S[M]}=\int\mathcal{D}M\ e^{-\f{N}{2}\Tr M^2}\sum_n\f{1}{n!}\left(\f{N\lambda}{3!}\right)^n(\Tr M^3)^n\ .
\label{matpart}
\ee
What are the differences from standard QFT? There are (at least) two: 1. it is non local, since our `fields' $M_{ij}$ are here objects which depend on two `points'; 2. it is not a field theory, since the `space' variables are discrete, $i,j=1,\ldots ,N$.

We shall drop the non-field stuff in a moment; the true `news' is the dropping of locality, which might sound rather blasphemous, but recall that we are in a completely combinatorial framework and these fields do not pretend to represent causal propagating particles, but blocks of spacetime, as we shall see . Deriving the propagator and interaction vertex of the theory is straightforward
\begin{align}
G_{ijkl}&=\delta_{ij}\delta_{kl}&\mathgraphics{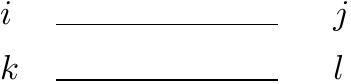}\ ,\\[10pt]
V_{ijklmn}&=\delta_{jm}\delta_{kl}\delta_{mn}&\mathgraphics{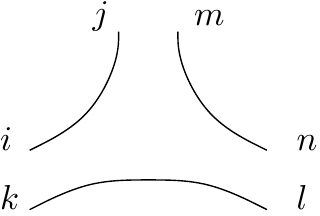}\ .
\end{align}
Now imagine to glue interaction vertices through propagators to obtain Feynman diagrams, you will get -- as hinted above -- ribbon-like structures. If we take the dual of each of these graphs you obtain precisely a triangulation of a 2-manifold. In this sense each propagator represents a side of a triangle, while the vertex represent the triangle itself, namely
\begin{equation*}
G\sim\mathgraphics{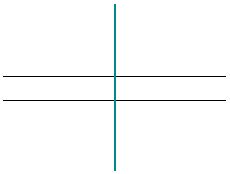}\ \ ,\qquad V\sim\mathgraphics{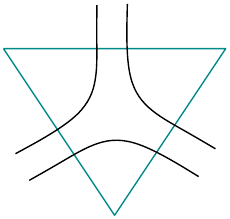}\ \ .
\end{equation*}
We can rephrase this by saying that each propagator is an edge and each vertex a face of the dual 2-skeleton of a triangulation.
This gives also a powerful tool for calculation explicitly the partition function \eqref{matpart}, indeed it has been proven that it is given by
\be
\mathcal{Z}=\sum_{g}w_g(\lambda)N^{2-2g}\ ,
\ee
where $g$ is the genus of the triangulated surface and $w$ is a weight depending on the diagram symmetry factor.  
This is precisely the partition function of a 2d $BF$~theory \eqref{BF2d}!

You guess the clue: going up in the number of indices -- i.e. building \emph{tensor models}~\cite{Ambjorn:1990ge,DePietri:2000ii} -- we can have the hope of getting skeletons of 3d and 4d manifolds, i.e. spinfoams.

Actually, this is not the case. Tensor models has been proven to fail in this respect. Essentially: 1. they are too simple to incorporate even the 3d $BF$~theory case and 2. they produce both manifolds and pseudo-manifolds\footnote{Sort of triangulation of manifold with singularities~\cite{pseudoman}.}~\cite{Ambjorn:1990ge,DePietri:2000ii}.

The natural way to proceed, is to pass  from matrices to fields,  defined on an appropriate domain space. In GFT -- hence the name -- this domain space is chosen to be an appropriate number of copies of a compact group $G$ 
\be
S^{2d}[\phi]=\f12\int_{G^2}\d g_1\d g_2\ \phi^2(g_1,g_2)+\f{\lambda}{3!}\int_{G^3}\d g_1\d g_2\d g_3\ \phi(g_1,g_2)\phi(g_2,g_3)\phi(g_3,g_1)\ ,
\ee
this is the GFT analogous of the matrix model \eqref{matact}. Instead of analyzing its features, let us move on seeing the key features of a general GFT.
\begin{itemize}
\item
$\phi$ is a (typically) real valued field on $G^n$, with $G$ compact Lie group and $n$ will be the dimension of the `produced' triangulated manifold. We require a gauge invariance under the $G$-(right)action
\be
\phi(g_1,\ldots ,g_n)= \phi(g_1g,\ldots ,g_ng)\ ,\quad \forall g_i,g\in G\ .
\ee
\item
the GFT action (for a real valued field) has the (very) generic structure 
\be
S[\phi]=T[\phi]+\lambda V[\phi]\ ,
\ee
with a ``kinetic'' part which has no dynamical meaning, since at this stage there is nothing as a time variable; and an interaction term which defines how the various group elements are attached with one another: the interaction term is actually the `atom' of hidden spinfoam model, and it determines the kind of $n$-complexes by which the manifold is triangulated. 

\item
The Feynman diagrams are 2-complexes, whose combinatorial structure is entirely given by the interaction term of $S[\phi]$. They are interpreted as dual to a $n$-discretization in terms of $n$-complexes (if the interaction term is particularly simple -- actually the only case I will consider -- these complexes are indeed simplices, and the discretization is a triangulation). 
\item
The value of the Feynman diagram is seen as a spinfoam amplitude, the amplitude associated with the discretization of that specific diagram.
\item
The GFT partition function
\be
\mathcal{Z}=\int\mathcal{D}\phi\ e^{-S[\phi]}=\int\mathcal{D}\phi\ e^{-T[\phi]}e^{-\lambda V[\phi]}= \sum_\Delta w(\Delta)\lambda^V(\Delta)A_\Delta\ ,
\ee
which is the typical perturbative expansion in terms of Feynman diagram: $\Delta$ denote equivalently the Feynman diagram or its associated discretization; $V(\Delta)$ is the number of vertices of the diagram, $w(\Delta)$ is its symmetry factor and $A_\Delta$ is the value of the diagram (the spinfoam amplitude).

Notice that this gives also a very natural way of dealing with triangulation dependence of spinfoam: each triangulation has a different spinfoam amplitude and we simply sum them with the weight given by the coupling and the symmetry factor.

\item
The correlation functions of GFT give the amplitude in going from one state to another, i.e. the (most wanted) $\bek{s}{A}{s'}$
\be
A[\Gamma,\psi]=\sum_{\Delta|\partial\Delta=\Gamma}w(\Delta)\lambda^V(\Delta)A_\Delta=\int\mathcal{D}\phi\ P_\psi(\phi)e^{-S[\phi]}\ ,
\ee
where $\psi$ is a state on the (boundary) graph $\Gamma$ and $P$ is some polynomial function of $\phi$ that is able to render the specific graph $\Gamma$ labeled by variables so to reproduce $\psi$\footnote{Think of $n$-point functions of QFT and you will get the idea. However, I will not go in much more details on this, I refer the reader to~\cite{Livine:2010zx} and references therein.}.

\end{itemize}

Let us now review some GFT models, stating only the results 
\begin{itemize}
 \item \emph{Ponzano-Regge}. The 3d $SU(2)$ GFT model with trivial kinetic term and tetrahedron-like interaction term, generates exactly the 3d $BF$ path integral \eqref{BF3d} for $SU(2)$. The action is
\begin{multline}
S[\phi]=\f12\int\d g_1\d g_2\d g_3\ \phi^2(g_1,g_2,g_3)\\
+\f{\lambda}{4!}\int\prod_{i=1}^6\d g_i\phi(g_1,g_2,g_3)\phi(g_3,g_4,g_6)\phi(g_6,g_2,g_4)\phi(g_5,g_4,g_1)\ .
\label{GFT3d}
\end{multline}
The interaction term is easily understood looking to figure \ref{fig:GFTtetrahedron}. It is indeed a tetrahedron like graph by itself; moreover, if you take the dual in this sense strand $\rightarrow$ edge (of a triangle), field (i.e. 3-strand) $\rightarrow$ triangle, interaction $\rightarrow$ tetrahedron. In this sense the interaction term is precisely telling us how to glue the 3 edges of 4 triangles to produce a tetrahedron.

In this precise sense, gluing interaction terms through propagators you create Feynman diagrams as well as a 3d triangulation. 

Taking the Fourier representation (i.e. the Peter-Weyl decomposition) of the field you get for the vertex the Ponzano-Regge amplitude.

\begin{figure}\centering
\includegraphics{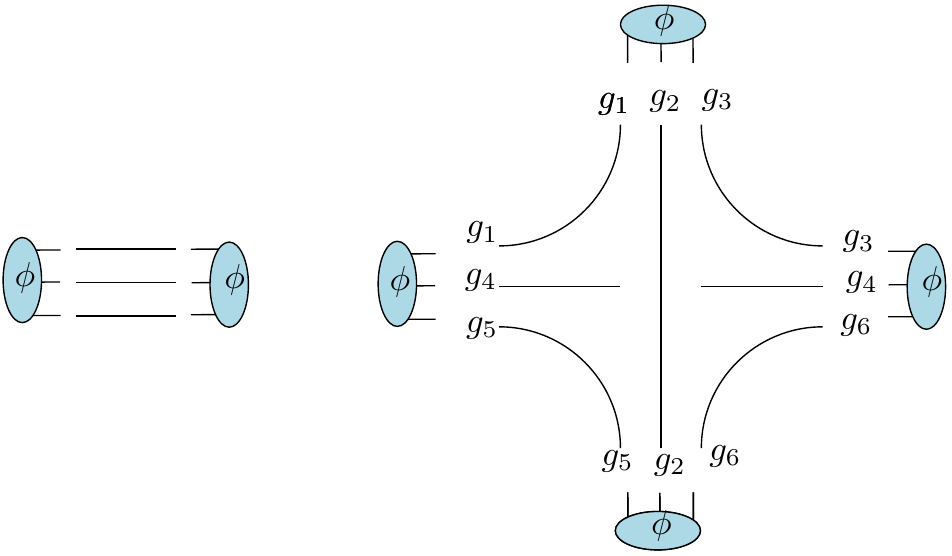}
\caption{Propagator and interaction term of the 3d GFT model \eqref{GFT3d}. The propagator is simply a delta over the group, while the interaction term is defined by a tetrahedron-like contraction.}\label{fig:GFTtetrahedron}
\end{figure}

\item \emph{4d GFT model}. Take the following action
\begin{multline}
S[\phi]=\f12\int\d g_1\d g_2\d g_3\d g_4\ \phi^2(g_1,g_2,g_3,g_4)+\f{\lambda}{5!}\int\prod_{i=1}^{10}\d g_i\phi(g_1,g_2,g_3,g_4)\times\\
\phi(g_4,g_5,g_6,g_7)\phi(g_7,g_3,g_8,g_9)\phi(g_9,g_6,g_2,g_{10})\phi(g_{10},g_8,g_5,g_1)\ .
\label{GFT4d}
\end{multline}
Here the interaction vertex is a 4-simplex (see figure~\ref{fig:GFT4simplex}), thus Feynman diagrams are dual to a 4d triangulation, and the vertex value, once decomposed into irreducible representations of the underlying group, is exactly the 4d $BF$ vertex amplitude, as in equation \eqref{BF4d}.

\begin{figure}\centering
\includegraphics{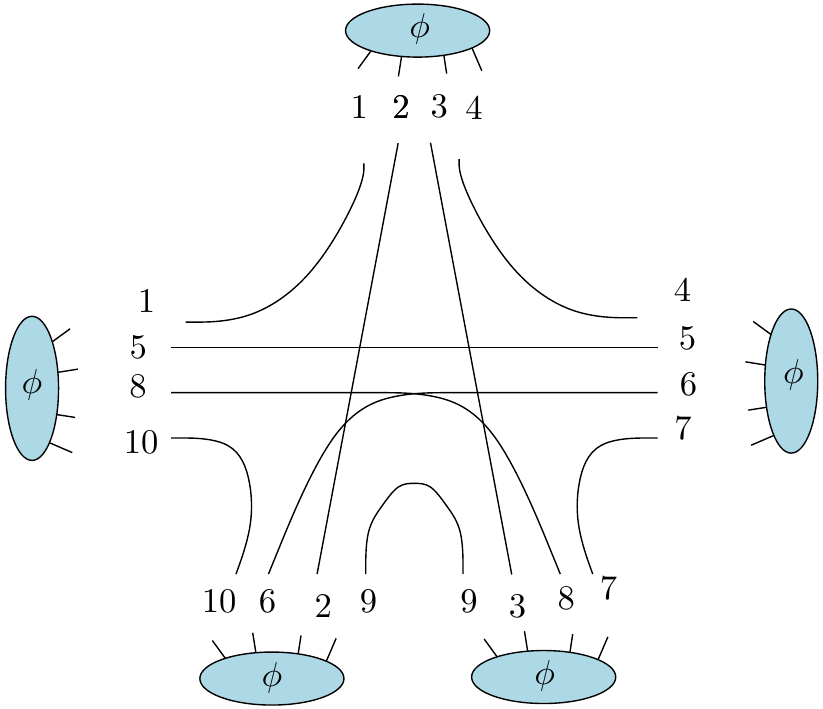}
\caption{Interaction term of the 4d GFT model \eqref{GFT4d}.}\label{fig:GFT4simplex}
\end{figure}

\item \emph{Barrett-Crane}. As a final example I give a sketch of how it is possible to deduce the Barrett-Crane spinfoam model \ref{sec:BC} from a GFT model. The gauge group is of course $G=Spin(4)$. The trick is to introduce a (slightly) more complicated kinetic term in \eqref{GFT4d}, namely
\be
T[\phi]=\f12\int\d^4 g \left(P\phi(g_1,\ldots ,g_4)\right)^2\ ,
\ee
with
\be
P\phi(g_1,\ldots ,g_4)=\int_{\sl}\d g\int_{SU(2)^4}\d^4u\ \phi(g_1gh_1,g_2gh_2,g_3gh_3,g_4gh_4)\ .
\ee
It is matter of calculations to prove that this GFT gives exactly the BC spinfoam amplitude \eqref{ZBC}

\end{itemize}

For a derivation of the EPRL spinfoam model from a GFT, I recommend the reading of~\cite{GFTnow1}.

\paragraph{Bubble divergences}\label{sec:divergences}

A very delicate and important issue of the spinfoam approach to Quantum Gravity is the one regarding divergences. I want here only to inform the reader of the importance of the issue, and give some useful references. Firstly a remark: inside the LQG framework one always talks about \emph{infrared divergences}, i.e. divergences arising when summing over the large-distance-scale degrees of freedom, i.e.  large spins. In LQG there are no ultraviolet divergences, since there is an ultimate minimum length-scale, of the order of the Planck length (see section~\ref{sec:geom}).

Bubble divergences arise in the sum over spins. You can see it as a loop integral of QFT. Simple power counting techniques reveal the presence of divergences depending on the nature of the manifold and of the kind of vertex one is considering (how many edges converge to that vertex, how many faces, and so on). Many studies have been done in this respect (See~\cite{Crane:2001as,Crane:2001qk,Perini:2008pd} and references therein). Moreover, it has been noted that the face amplitude is crucial in determining whether a bubble is or is not  divergent~\cite{Perini:2008pd}. We shall talk about this in chapter~\ref{sec:face}.

Here I want to stress that the GFT approach gives powerful tools to handle and analyze these divergent bubbles and may provide a unified strategy to fix them\footnote{Incidentally, notice that the GFT approach provide with an intriguing duality between UV and IR divergences: indeed the Quantum Gravity IR divergences (namely the large spin divergences) can also be seen as UV divergences on the group.}.  I refer the reader to the recent work~\cite{Rivasseau:2011xg} and to \cite{Geloun:2010nw,Freidel:2009hd}.\\[14pt]

This concludes this brief detour into GFT framework. As I have said at the beginning of this section, GFT is not the focus of my research papers, so this is not the right place to discuss it in more details. What I wanted to give, is the idea of the existence of a (in my opinion) promising unifying framework, inserting Quantum Gravity into a QFT approach while preserving its non-perturbative nature. Indeed recall that the perturbation of the Feynman diagram series in GFT is a perturbation in the coupling $\lambda$, i.e. is a perturbation on the number of vertices of the triangulations, that is on the ``complexity'' of the graph\footnote{Of course we have no clue on the coupling parameter, but this is another problem.} and not a perturbation around a fixed background geometry.   I really encourage the interested reader to look up the bibliography I have referred to during this section, in particular \cite{GFTreviews1,GFTreviews2,GFTreviews3} for good reviews on the subject.

\chapter[A proposal for the face amplitude]{A proposal for fixing the face amplitude in Quantum Gravity spinfoam models}\label{sec:face}

{\it The present chapter is intended to present the content of the paper~\cite{mio_faccia}, in which we propose a way for fixing the face amplitude of a general spinfoam model. The proposal is motivated by the requiring of a sort of ``unitarity'' of time evolution of space geometries, in a sense that will be made clear in what follows. We found that this requirement imposes the face amplitude to be equal to the dimension of the $SU(2)$-projected representation of the $SO(4)$($\sl$) one attached to the face.} 

\section{Introduction and resume of the content of the paper~\texorpdfstring{\cite{mio_faccia}}{on spinfoam face amplitude}}

In this section I present the content of the paper \cite{mio_faccia}, which collects the results of the work I have done in Marseille, in collaboration with Eugenio Bianchi and Carlo Rovelli.

I have repeatedly focused on the fact that, while for $BF$~theory the face amplitude of the spinfoam sum is well determined by the path integral discretization procedure and it is given by the dimension of the representation labeling the face \eqref{BF2d},\eqref{BF3d},\eqref{BF4d}, for Quantum Gravity the situation is much more subtle. One starts with a certain $BF$~theory, obtains a spinfoam sum formula, and then imposes constraints in the vertex amplitude. What happens to the face amplitude? The BC model, for instance, simply let it be the $BF$ face amplitude, i.e. the square of the $SU(2)$ irreducible dimension. In $SO(4)$-models with the Immirzi parameter (such as $SO(4)$-EPRL) one has\footnote{In~\cite{mio_faccia} we have worked in the euclidean ($SO(4)$) case. However, everything can be done in the lorentzian ($\sl$) as well, \emph{mutatis mutandis}. See section~\ref{sec:EPRL} for discussions about this point.} 
\be
   A_{\rho_f} = (2j_++1)(2j_-+1)=(2\gamma_+j_f+1)(2\gamma_-j_f+1).
   \label{bffa}
\ee

However this is not well motivated, indeed doubts can be raised against this argument. For instance, Alexandrov~\cite{Alexandrov:2010pg} has stressed the fact that the implementation of second class constraints into a Feynman path integral in general requires a modification of the measure, and here the face amplitude plays precisely the role of such measure, since $A_v\sim e^{i\,Action}$. Do we have an independent way of fixing the face amplitude?

Let me recall that all the spinfoam models share the same form of partition function, namely the sum 
\be
   \mathcal{Z}_\Delta = \sum_{\rho, \i}\ \prod_f A_{\rho_f}\ \prod_v A_v(\rho_f,\i_e)\ ,
   \label{Z}
\ee
where, as usual (recall section \ref{sec:BF}), that sum is intended to be a sum over all the possible labellings of the faces (edges) of $\Delta^*$ with irreducible representations (intertwiners) of an appropriate group $G$.

In \cite{mio_faccia} we argued that the face amplitude is uniquely determined for any spinfoam sum of the form~\eqref{Z} by three inputs: 1. the choice of the boundary Hilbert space, 2. the requirement that the composition law holds when gluing two-complexes; and 3. a particular ``locality" requirement, or, more precisely, a requirement on the local composition of group elements.  

We argued that these requirements are implemented if the partition function $\mathcal{Z}$ is given by the expression 
\be
\mathcal{Z}_\Delta=\int \d U_f^v\  \prod_v  A_v(U_f^v)\ \prod_f\ \delta(U_f^{v_1}...U_f^{v_k})\ , 
\label{main}
\ee
where $U_f^v\in G$,   $v_1...v_k$ are the vertices surrounding the face $f$, and $A_v(U_f^v)$ is the vertex amplitude $A_v(j_f,i_e)$ expressed in the group element basis \cite{Bianchi:2010}.  Then we showed that this expression leads directly to \eqref{Z}, with arbitrary vertex amplitude, but a fixed choice of face amplitude, which turns out to be the dimension of the representation $j$ of the group $G$,  
\be
           A_j= {\rm dim}(j)\ .
 \label{d}
\ee
In particular, for Quantum Gravity this implies that the $BF$ face amplitude~\eqref{bffa} is ruled out, and should be replaced (both in the Euclidean and in the Lorentzian case) by the $SU(2)$ dimension
\be
           A_j={\rm dim}(j)= 2j+1\ .
\ee

Equation \eqref{main} is the key expression of the whole paper. 

I organize this chapter as the paper from which is taken, i.e.: in section \ref{BFsec} I show that $SO(4)$ $BF$~theory (the prototypical spinfoam model) can be expressed in the form \eqref{main}.  Then I discuss the three requirements above and I show that   \eqref{main} implements these requirements. (Section \ref{inputs}).  
Finally I show that  \eqref{main} gives~\eqref{Z} with the face amplitude \eqref{d} (Section IV).

The problem of fixing the face amplitude has been discussed also by Bojowald and Perez in \cite{Bojowald:2009im}. Bojowald and Perez demand that the amplitude be invariant under suitable refinements of the two-complex. This request is strictly related to the composition law that we considered in \cite{mio_faccia}, and the results we obtained are consistent with those of  \cite{Bojowald:2009im}.

\section{\emph{BF} theory}\label{BFsec}

\begin{figure}
\centering
\includegraphics[width=0.34\textwidth]{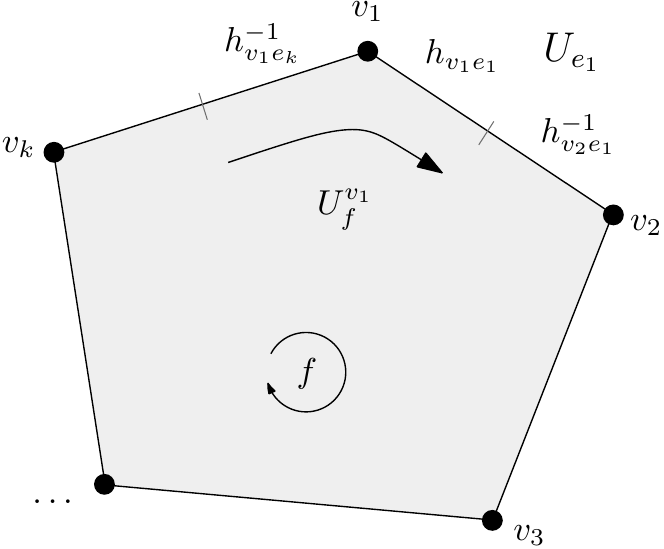}
\caption{Schematic definition of the group elements $h_{ve}$, $U^v_f$ and $U_e$ associated to a portion of a face $f$ of the two-complex.}
\label{fig1}
\end{figure}

Take the general expression of the $BF$ partition function in terms of group elements (see equation~\eqref{bf_path})
\be
\mathcal{Z}_\Delta=\int\prod_e \d U_e\   \prod_f\ \delta(U_{e_1}...U_{e_N})\ , 
\label{ZBF}
\ee
where $U_e$ are group elements associated to the oriented \emph{edges} of $\sigma$, and $(e_1,...,e_N)$ are the edges that surround the face $f$.   Let us introduce group elements $h_{ve}$, labeled by a vertex $v$ and an adjacent edge $e$, such that 
\be
U_e =h_{ve}h_{v'e}^{-1}\ ,
\ee
where $v$ and $v'$ are the source and the target of the edge $e$  (see figure~\ref{fig1}). Then we can trivially rewrite \eqref{ZBF} as 
\be
\mathcal{Z}_\Delta=\int \d h_{ve} \prod_f\ \delta\!\left((h_{v_1e_1} h^{-1}_{v_2 e_1})\ ...\ (h_{v_N e_N} h^{-1}_{v_1 e_N})\right)\ . 
\label{ZBF3}
\ee
Now define the group elements 
\be
U_f^v = h^{-1}_{ve} h_{ve'}
\ee
associated to a single vertex $v$ and two edges $e$ and $e'$ that emerge from $v$ and bound the face $f$ (see figure~\ref{fig1}). Using these, we can rewrite \eqref{ZBF} as 
\be
\mathcal{Z}_\Delta=\int\! \d h_{ve}  \int\! \d U_f^v \,
\prod_{v,f^v} \delta(U_f^v,h^{-1}_{ve} h_{ve'})
\prod_f\ \delta(U_{f}^{v_1}...U_{f}^{v_N})\ ,
\nonumber 
\ee
where the first product is over faces $f^v$ that belong to the vertex $v$, and then a product over all the vertices of the 2-complex.

Notice that this expression has precisely the form \eqref{main}, where the vertex amplitude is
\be
A_v(U_f^v)=\int \d h_{ve}  \prod_{f^v} \delta(U_f^v,h_{ve} h_{ve'}^{-1})\ ,
\ee
which is the well-known expression of the $15j$ Wigner symbol (the vertex amplitude of $BF$ in the spin network basis) in the basis of the group elements (cfr.\eqref{BFatom}). 

We have shown that the $BF$~theory spinfoam amplitude can be put in the form  \eqref{main}. We shall now argue that~\eqref{main} is the \emph{general} form of a local spinfoam model that obeys the composition law.

\section{Three inputs}\label{inputs}

(a) \emph{Hilbert space structure}. Equation \eqref{Z} is a coded expression to define the amplitudes
\be
   A_\Delta(j_l,\i_n) = \sum_{j, \i} \prod_f A_{j_f} \prod_v A_v(j_f,\i_e;j_l,\i_n)\ , 
   \label{W}
\ee
defined for a triangulation $\Delta$ \emph{with boundary}, where the boundary graph $\Gamma$ is formed by links $l$ and nodes $n$. The spins $j_l$ are associated to the links $l$, as well as to the faces $f$ that are bounded by $l$; the intertwiners $\i_n$ are associated to the nodes $n$, as well as to the edges $e$ that are bounded by $n$. The amplitude of the vertices that are adjacent to these boundary faces and edges depend also on the external (thus fixed) variables $(j_l,\i_n)$.  

In a quantum theory, the amplitude $A(j_l,\i_n)$ must be interpreted as a (covariant) vector in a space $H_\Gamma$ of quantum states.\footnote{If $\Gamma$ has two disconnected components interpreted as ``in" and an ''out" spaces, then  $H_\Gamma$ can be identified as the tensor product of the ``in" and an ''out" spaces of non-relativistic quantum mechanics. In the general case,  $H_\Gamma$ is the boundary quantum state in the sense of the boundary formulation of quantum theory \cite{Rovelli_book,Oeckl:2003vu}.} We assume that this space has a Hilbert space structure, which we know.  In particular, we assume that 
\be
 {\cal H}_\Gamma= L_2[G^L,\d U_l]\ ,
\label{n}
\ee
where $L$ is the number of links in $\Gamma$ and $dU_l$ is the Haar measure. Thus we can interpret \eqref{W} as
\be
   A_\Delta(j_l,\i_n) = \bk{j_l,\i_n}{A}\ ,
      \label{Wbraket}
\ee
where $\ket{j_l,\i_n}$ is the spin network function (cfr. equation~\eqref{spin_net})
\be
\bk{U_l}{j_l,\i_n}=\psi_{j_l,\i_n}(U_l)= \bigotimes_l  D^{j_l}(U_l)\cdot \bigotimes_n \i_n\ .
\ee
Using the scalar product defined by \eqref{n}, we have 
\begin{eqnarray}
\bk{j_l,\i_n}{j'_l,\i'_n}&=& \int  \d U_l\ \overline{\psi_{j_l,\i_n}(U_l)}\psi_{j'_l,\i'_n}(U_l)
\nonumber\\
&=&
 \prod_l{\rm dim}(j_l)\; \delta_{j_lj'_l} \ \prod_n \delta_{\i_n\i'_n}\ .
\end{eqnarray}
where ${\rm dim}(j)$ is the dimension of the representation $j$. Therefore the spin-network functions $\psi_{j_l,\i_n}(U_l)$ are not normalized.  (These ${\rm dim}(j)$ normalization factors are due to the convention chosen: they have nothing to do with the dimension of the representation that appears in \eqref{d}.) The resolution of the identity in this basis is 
\be
\id=\sum_{j_l,\i_n}\ \big(\text{\footnotesize $\displaystyle \prod_l$}\, {\rm dim}(j_l)\big)\ \ket{j_l,\i_n}\bra{j_l,\i_n}\ .
\ee

(b) \emph{Composition law.} In non relativistic quantum mechanics, if $U(t_1,t_0)$ is the evolution operator from time $t_0$ to time $t_1$, the composition law reads
\be
U(t_2,t_0)=U(t_2,t_1)U(t_1,t_0)\ . 
\ee
That is, if $\ket{n}$ is an orthonormal basis,
\be
\bek{f}{\!U(t_2,t_0)\!}{i}=
\sum_{n}\bek{f}{\!U(t_2,t_1)\!}{n}\bek{n}{\!U(t_1,t_0)\!}{i}\ .\nonumber
\ee
Let us write an analogous condition of the spinfoam sum. Consider for simplicity a 
two-complex $\sigma=\sigma_1\cup\sigma_2$ without boundary, obtained by gluing two two-complexes $\sigma_1$ and $\sigma_2$ along their common boundary $\Gamma$. Then we require that $W$ satisfies the composition law 
\be
\mathcal{Z}_{\sigma_1\cup\sigma_2}=\bk{A_{\sigma_2}}{A_{\sigma_1}}\ ,
\ee
-- where now I label $\mathcal{Z}$ directly with the 2-skeleton $\sigma\subset\Delta^*$ --
as discussed by Atiyah in \cite{Atiyah:1989vu}. Notice that to formulate this condition we need the Hilbert space structure in the space of the boundary states. 

\begin{figure}
\centering
\includegraphics[width=0.4\textwidth]{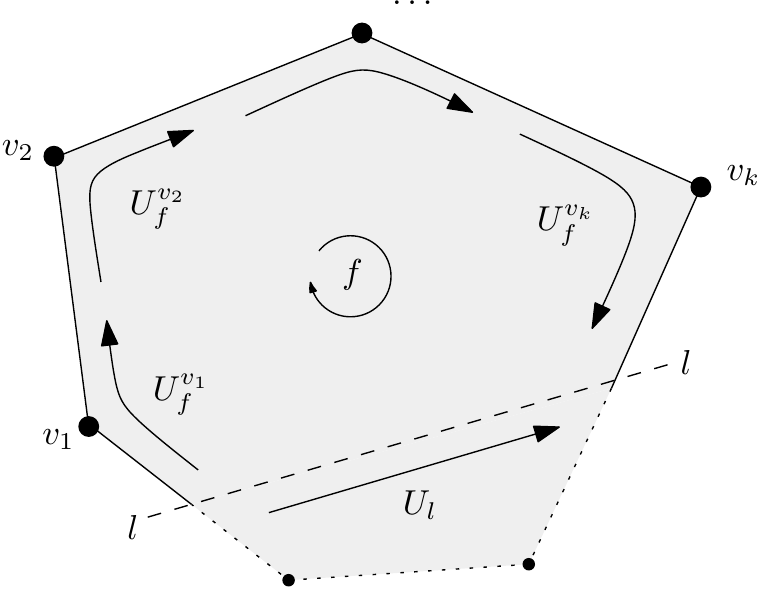}
\caption{Cutting of a face of the 2-skeleton. The holonomy $U_l$ is attached to a link of the boundary spin network and satisfies equation~\eqref{locality}.}
\label{fig2}
\end{figure}

(c) \emph{Locality.} As a vector in $H_\Gamma$, the amplitude $A(j_l,\i_n)$ can be expressed on the group element basis
\be
A(U_l)=\bk{U_l}{A}=\sum_{j_l,\i_n} \; \big(\,\text{\footnotesize $\displaystyle \prod_l$}\, {\rm dim}(j_l)\big)\ \psi_{j_l,\i_n}(U_l) A(j_l,\i_n)\ .
\ee
Similarly, the vertex amplitude can be expanded in the group element basis
\begin{eqnarray}
A_v(U_f^v)&=&\bk{U_f^v}{A_v} \label{va}\\
&=&\sum_{j_f^v,i_n^v} \; \big(\,\text{\footnotesize $\displaystyle \prod_{f^v}$}\, {\rm dim}(j_f^v)\big)\ \psi_{j_f^v,\i_n^v}(U_f^v) A_v(j_f^v,\i_n^v)\ .\nonumber
\end{eqnarray}
Notice that here the group element $U_f^v$ and the spin $j_f^v$ are associated to a vertex $v$ and a face $f$ adjacent to $v$. Similarly, the intertwiner $\i_n^v$ is associated to a vertex $v$ and a node $n$ adjacent to $v$. Consider a boundary link $l$ that bounds a face $f$ (see figure~\ref{fig2}). Let $v_1...v_k$ be the vertices that are adjacent to this face.  We say that the model is local if the relation between the boundary group element $U_l$ and the vertices group elements $U_f^v$ is given by 
\be
U_l=U_f^{v_1}\,...\ U_f^{v_k}\ .
\label{locality}
\ee
In other words: if the boundary group element is simply the product of the group elements around the face. 

\vspace{1em}

Notice that a spinfoam model defined by \eqref{main} is local and satisfies composition law in the sense above.  In fact, \eqref{main} generalizes immediately to 
\begin{eqnarray}
A_\sigma(U_l)&=&\int \d U_f^v\  \prod_v  A_v(U_f^v)\ \prod_{{\rm internal}\ f}\ \delta(U_f^{v_1}...U_f^{v_k})\nonumber\\
&& \ \ \times\ \ \prod_{{\rm external}\ f}\ \delta(U_f^{v_1}...U_f^{v_k}U^{-1}_l)\ .
\label{main2} 
\end{eqnarray}
Here the first product over $f$ is over the (``internal") faces that do not have an external boundary; while the second is over the (``external") faces $f$ that are also bounded by
the vertices $v_1, ..., v_k$ and by the the link $l$.  It is immediate to see that locality is implemented, since the second delta enforces the locality condition \eqref{locality}.

Furthermore, when gluing two amplitudes along a common boundary we have immediately that 
\be
\int \d U_l\ \overline{A_{\sigma_1}(U_l)}\ A_{\sigma_2}(U_l)=\mathcal{Z}_{\sigma_1\cup\sigma_2}\ ,
\ee
because the two delta functions containing $U_l$ collapse into a single delta function associated to the face $l$, which becomes internal.  

Thus, \eqref{main} is a general form of the amplitude where these conditions hold.

In \cite{Bojowald:2009im}, Bojowald and Perez have considered the possibility of fixing the face amplitude by requiring the amplitude of a given spin/intertwiner configuration to be equal to the amplitude of the same spin/intertwiner configuration on a finer two-simplex where additional faces carry the trivial representation.  This requirement imply essentially that the amplitude does not change by splitting a face into two faces. It is easy to see that \eqref{main} satisfies this condition. Therefore \eqref{main} satisfies also the Bojowald-Perez condition.

\section{Face amplitude}

Finally, let us show that \eqref{main} implies \eqref{Z} \emph{and} \eqref{d}. To this purpose, it is sufficient to insert \eqref{va} into  \eqref{main}. This gives
\begin{eqnarray}
\mathcal{Z}_\sigma&=&\!\int \d U_f^v\  \prod_v  \sum_{j_f^v,i_n^v} \big(\,\text{\footnotesize $\displaystyle \prod_{f^v}$}\, {\rm dim}(j_f^v)\big)\ \psi_{j_f^v,\i_n^v}(U_f^v) A_v(j_f^v,\i_n^v)\nonumber\\
&&\ \ \times \   \prod_f\ \delta(U_f^{v_1}...U_f^{v_k})\ .
\label{main22}
\end{eqnarray}
Expand then the delta function in a sum over characters 
\begin{align}\label{calc_faccia}
\mathcal{Z}_\sigma=&\!\int \d U_f^v\  \prod_v  \sum_{j_f^v,\i_n^v}\! \big(\,\text{\footnotesize $\displaystyle \prod_{f^v}$}\, {\rm dim}(j_f^v)\big)\ \psi_{j_f^v,\i_n^v}(U_f^v) A_v(j_f^v,\i_n^v)\\
&\times \   \prod_f\ \sum_{j_f}{\rm dim}(j_f)\ {\rm Tr}(D^{j_f}(U_f^{v_1})\cdots D^{j_f}(U_f^{v_k}))\ .
\nonumber
\end{align}
We can now perform the group integrals.  Each $U_f^v$ appears precisely twice in the integral: once in the sum over $j_f^v$ and the other in the sum over $j_f$. Each integration  over the group, gives a delta function (recall  equation \eqref{id_simple})
\be
\delta_{j_f^v,j_f}=\f{1}{\text{dim}(j_f)}\int\d U^v_f\ D^{j_f^v}(U^v_f)D^{j_f}(U^v_f)\ ,
\ee
which can be used to kill the sum over $j_f^v$ dropping the $v$ subscript. Notice that the dimensional factor involved here, exactly cancels -- once done the product-- the first dimensional factor in \eqref{calc_faccia}, i.e. the one coming from the spin network normalization. Following the contraction path of the indices, it is easy to see that these contract the two intertwiners at the opposite side of each edge. Since intertwiners are orthonormal, this gives a delta function $\delta_{\i_n^v,\i_n^{v'}}$ which reduces the sums over intertwiners to a single sum over $\i_n:=\i_n^v=\i_n^{v'}$. Bringing everything together,  we have 
\be
\mathcal{Z}_\sigma=\sum_{j\, \i}\ \prod_f {\rm dim}(j_f)\ \prod_v  A_v(j_f^v,\i_n^v)\ .
\ee
This is precisely equation \eqref{Z}, with the face amplitude given by \eqref{d}.

Notice that the face amplitude is well defined, in the sense that it cannot be absorbed into the vertex amplitude (as any \emph{edge} amplitude can). The reason is that any factor in the vertex amplitude depending on the spin of the face contributes to the total amplitude at a power $k$, where $k$ is the number of sides of the face. The face amplitude, instead, is a contribution to the total amplitude that does not depend on $k$. This is also the reason why the normalization chosen for the spinfoam basis does not affect the present discussion: it affects the expression for the vertex amplitude, not that for the face amplitude.

By an analogous calculation one can show that the same result holds for the amplitudes $W$: equation \eqref{W} follows from \eqref{main2} expanded on a spin network basis. 

In conclusion, we have shown that the general form \eqref{main} of the partition function, which implements locality and the composition law, implies that the face amplitude of the spinfoam model is given by the dimension of the representation of the group $G$ which appears in the boundary scalar product (\ref{n}). 

In general relativity, in both the Euclidean and the Lorentzian cases, the boundary space is 
\be
 {\cal H}_\Gamma= L_2[SU(2)^L,dU_l]\ ,
\label{n2}
\ee
therefore the face amplitude is $d_j={\rm dim}_{SU(2)}(j)=2j+1$, and not the $SO(4)$ dimension \eqref{bffa}, as previously supposed.  

Notice that such $d_j=2j+1$ amplitude defines a theory that is far less divergent than the theory defined by \eqref{bffa}. In fact, the potential divergence of a bubble is suppressed by a power of $j$ with respect to \eqref{bffa}.  In \cite{Perini:2008pd}, it has been shown that  the $d_j=2j+1$ face amplitude yields a \emph{finite} main radiative correction to a five-valent vertex if all external legs set to zero.

\bookmarksetup{startatroot}
\chapter*{Conclusion\markboth{Conclusion}{Conclusion}}\label{sec:concl}
\addcontentsline{toc}{chapter}{Conclusion}

Let me now briefly review in a bird-flight manner the results which have been achieved. 

We have worked on phantom dark energy models in classical Cosmology. Let me recall that phantom models (i.e. models with some kind of phantom energy) are not ruled out by observations, even if the standard model of Cosmology, $\Lambda$CDM, scientifically still ``wins'', by virtue of Occam's razor. Crossing of the phantom divide line is not excluded as well. We have analyzed what seem to us one of the best candidates to reproduce this crossing: two-field cosmological models with one scalar and one phantom field. 

We have discovered that, starting from an expression for the Hubble variable $h(t)$, there is an infinite number of models (i.e. of potentials) that have, among their solutions, precisely that $h(t)$. This freedom, namely to have infinite different models which are compatible with the same dynamics for the universe, is the most important result presented in~\cite{mio_2field}.

One possible approach to discriminate between such models, is to couple the cosmological fields with other, observable, fields, such as cosmic magnetic fields. The calculations and the numerical simulations have indeed shown that such a coupling gives in principle observable results, capable of select between different models~\cite{mio_magnetic}.

However, we are aware that phantom fields have been heavily criticized, and actually they do deserve the label of `exotic'. The negative sign of the kinetic term is indeed something that disturbs even just to write it. Thus, we tried to find out some mechanism so that the negative sign was only \emph{effective}, while the ``fundamental'' field is a standard one, with no stability problem. Accepting to work in the $PT$~symmetric Quantum Theory framework, we succeeded in this respect, and find a possible way to have an effective phantom field, stable to quantum fluctuations~\cite{mio_PT}.

These were the contributions in the phantom/crossing-of-the-phantom-line branch of Cosmology.
Aside, we analyzed one-field models giving evolutions for the universe with an interesting kind of singularities: sort of Big Bang/Big Crunch with a finite non-zero radius~\cite{mio_soft}. Notice that this is in line with all Quantum Gravity theories, namely to have a minimum length scale. Thus, we thought that the study of non-zero radius singularities could be useful also in perspective of future results from the Quantum Gravity community.  

Quantum Gravity is essential to properly understand what \emph{is} space and what \emph{is} time. Maybe the total absence of experimental results give to these kind of studies an halo of philosophical vagueness. I could agree with this, but still I firmly think that this kind of fundamental research must be carried on. 
Indeed, many results of LQG and spinfoam theory are very suggestive and stimulating, even if on purely theoretical basis. The attempt to find a `sum-over-histories' formulation of Quantum Gravity, based on the LQG approach to quantization, is a recent and promising field of research, where results crowd from many different and far areas of Theoretical Physics. Specifically we have proposed a way of fixing the face amplitude of a general spinfoam model. Up till now, the face amplitude has been taken to be the one of the $BF$~theory underlying the quantization procedure. But we argued that this is not the case, if one want to assure the proper ``gluing'' of spinfoam amplitudes~\cite{mio_faccia}.   

\chapter*{Acknowledgements}
\addcontentsline{toc}{chapter}{Acknowledgements}

Many are the people that deserve my thanks, for different reasons. Honestly, I wouldn't be able to thank properly everybody. So I use this occasion to give my explicit thanks to my thesis advisor, Alexander Kamenshchik, who helped me as a professor, as a research colleague and, most importantly, as a friend, especially in some difficult moments. My only other explicit thank goes to Carlo Rovelli, who guested me in his amazing group in Marseille, where I learned a lot, both from the scientific and personal viewpoint. 

I warmly hope that all the others will feel thanked as well, please. Thank You. Grazie. Merci.

\vfill
\begin{flushright}
 \small 4lab
\end{flushright}
 
\backmatter
\phantomsection
\nocite{*}
\addcontentsline{toc}{chapter}{Bibliography}

\providecommand{\href}[2]{#2}\begingroup\raggedright\endgroup

\end{document}